\documentclass[onecolumn,notitlepage,floats,floatfix,amssymb,aps,prd,showpacs,superscriptaddress,groupedaddress,nofootinbib]{revtex4-2} 
\usepackage{amsmath, amssymb, amsthm, graphicx, epsfig, fancyhdr,epsfig, slashed}

\usepackage[T1]{fontenc}
\usepackage{lmodern}
\usepackage{graphicx}  % needed for figures
\usepackage{dcolumn}   % needed for some tables
\usepackage{bm}        % for math
\usepackage{amssymb}   % for math
\usepackage{amsmath}
\usepackage{bbold}
\usepackage{array}
\usepackage{makecell}
\usepackage{braket}
\usepackage{longtable}
\usepackage{supertabular,booktabs}

\usepackage{titlesec} %for hyperref to detect names of section etc
\usepackage[colorlinks,citecolor=blue,urlcolor=blue,hypertexnames=true]{hyperref}
\usepackage{subfigure}
\usepackage[usenames,dvipsnames,svgnames,table]{xcolor}

\usepackage{tikzsymbols}
\usepackage{natbib}
\usepackage{float}

\usepackage{tikz,xcolor,hyperref}

\usepackage{slashed}

\definecolor{lime}{HTML}{A6CE39}
\DeclareRobustCommand{\orcidicon}{
	\begin{tikzpicture}
	\draw[lime, fill=lime] (0,0) 
	circle [radius=0.2] 
	node[white] {{\fontfamily{qag}\selectfont \tiny ID}};
	\draw[white, fill=white] (-0.0625,0.095) 
	circle [radius=0.007];
	\end{tikzpicture}
	\hspace{-2mm}
}

\foreach \x in {A, ..., Z}{\expandafter\xdef\csname orcid\x\endcsname{\noexpand\href{https://orcid.org/\csname orcidauthor\x\endcsname}
			{\noexpand\orcidicon}}
}
\newcommand{\be}{\begin{equation}}
\newcommand{\ee}{\end{equation}}
\newcommand{\bea}{\begin{eqnarray}}
\newcommand{\eea}{\end{eqnarray}}

\newcommand{\ba}{\begin{eqnarray}}
\newcommand{\ea}{\end{eqnarray}}
\newcommand{\bi}{\begin{itemize}}
\newcommand{\ei}{\end{itemize}}
\newcommand{\x}{\star}

\usepackage{graphicx,slashed,booktabs,xcolor,multirow,float,
amsfonts,bbold,mathtools,sidecap,tikz,bm,enumitem}
\usepackage{multirow}
\usepackage{bbding}
\usepackage{titlesec}
\usepackage{hyperref}
\usepackage{wasysym}
\usepackage{amssymb}% http://ctan.org/pkg/amssymb
\usepackage{pifont}% http://ctan.org/pkg/pifont

\renewcommand{\leq}{\leqslant}

\usepackage[dvipsnames, usenames]{xcolor}

\definecolor{linkcolor}{rgb}{0.55, 0.13, .32}

\definecolor{oucrimsonred}{rgb}{0.6, 0.0, 0.0}
\definecolor{persianblue}{rgb}{0.11, 0.22, 0.73}
\definecolor{forestgreen}{rgb}{0.13,0.35,0.13}
\definecolor{lightgray}{rgb}{0.83, 0.83, 0.83}
 \hypersetup{colorlinks, citecolor=oucrimsonred, linkcolor=persianblue, urlcolor=oucrimsonred}
\definecolor{cornellred}{rgb}{0.7, 0.11, 0.11}
\definecolor{navyblue}{rgb}{0.0, 0.0, 0.5}
\definecolor{amethyst}{rgb}{0.6, 0.4, 0.8}
\definecolor{yellow}{rgb}{1.0, 1.0, 0.0}
\definecolor{firebrick}{rgb}{0.7, 0.13, 0.13}
\definecolor{tangerineyellow}{rgb}{1.0, 0.8, 0.0}
\definecolor{deepfuchsia}{rgb}{0.76, 0.33, 0.76}
\definecolor{amber}{rgb}{1.0, 0.75, 0.0}
\definecolor{VioletRed4}{rgb}{0.55, 0.13, .32}
\definecolor{indiagreen}{rgb}{0.07, 0.53, 0.03}
\definecolor{VioletRed4}{rgb}{0.55, 0.13, .32}

\usepackage{hyperref}

\usepackage{graphics,appendix,afterpage,makecell}

\usepackage{bbold}
\usepackage{tikz}
\usepackage{adjustbox}
\usepackage{tikz-feynman}
\usepackage{tcolorbox}
\usepackage{enumitem}
\usepackage{amsfonts}

\setlist[itemize,1]{label=$\times$}
\setlist[itemize,2]{label=$\checkmark$}
\setlist[itemize,3]{label=$\diamond$}
\setlist[itemize,4]{label=$\bullet$}

\definecolor{oucrimsonred}{rgb}{0.6, 0.0, 0.0}
\definecolor{persianblue}{rgb}{0.11, 0.22, 0.73}
\definecolor{forestgreen}{rgb}{0.13,0.35,0.13}
\definecolor{lightgray}{rgb}{0.83, 0.83, 0.83}
 \hypersetup{colorlinks, citecolor=oucrimsonred, linkcolor=persianblue, urlcolor=oucrimsonred}
\definecolor{cornellred}{rgb}{0.7, 0.11, 0.11}
\definecolor{navyblue}{rgb}{0.0, 0.0, 0.5}
\definecolor{amethyst}{rgb}{0.6, 0.4, 0.8}
\definecolor{yellow}{rgb}{1.0, 1.0, 0.0}
\definecolor{firebrick}{rgb}{0.7, 0.13, 0.13}
\definecolor{tangerineyellow}{rgb}{1.0, 0.8, 0.0}
\definecolor{deepfuchsia}{rgb}{0.76, 0.33, 0.76}
\definecolor{amber}{rgb}{1.0, 0.75, 0.0}
\definecolor{VioletRed4}{rgb}{0.55, 0.13, .32}
\definecolor{indiagreen}{rgb}{0.07, 0.53, 0.03}
\definecolor{VioletRed4}{rgb}{0.55, 0.13, .32}

\definecolor{oucrimsonred}{rgb}{0.6, 0.0, 0.0}
\newcommand\vertarrowbox[3][6ex]{%
  \begin{array}[t]{@{}c@{}} #2 \\
  \left\uparrow\vcenter{\hrule height #1}\right.\kern-\nulldelimiterspace\\
  \makebox[0pt]{\scriptsize#3}
  \end{array}%
}

\definecolor{mtcolor}{rgb}{.8,.3,.1}
\definecolor{violachiaro}{rgb}{1,0.6,1}

\definecolor{gbcolor}{rgb}{.43,.22,.12}
 
\definecolor{gbcolor2}{rgb}{.9,.2,.6}
\definecolor{gbcolor3}{rgb}{.3,.2,.6}

\definecolor{verdechiaro}{rgb}{0.6,1,0.6}
\definecolor{giallochiaro}{rgb}{1,1,0.6}
\definecolor{bluscuro}{rgb}{0.15, 0.2, 0.9}
\definecolor{verdes}{rgb}{0.1, 0.5, 0.1}%
\definecolor{tangerineyellow}{rgb}{1.0, 0.8, 0.0}
\definecolor{smokyblack}{rgb}{0.06, 0.05, 0.03}

\definecolor{americanrose}{rgb}{1.0, 0.01, 0.24}
\definecolor{cobalt}{rgb}{0.0, 0.28, 0.67}
\definecolor{brandeisblue}{rgb}{0.0, 0.44, 1.0}
\definecolor{mycolor}{rgb}{0.0, 0.0, 0.5}%navyblue
\definecolor{oxfordblue}{rgb}{0.0, 0.13, 0.28}
\definecolor{azure}{rgb}{0.0, 0.5, 1.0}
\definecolor{turquoiseblue}{rgb}{0.0, 1.0, 0.94}
\newtcolorbox{mynewbox}[1]{colback=white!5!white,colframe=azure!75!black,fonttitle=\bfseries,title=#1}
\newtcolorbox{mybox}{colback=mycolor!5!white,colframe=azure!75!black}
\newtcolorbox{mynamedbox}[1]{colback=mycolor!5!white,colframe=azure!75!black,title=#1}
\definecolor{venetianred}{rgb}{0.78, 0.03, 0.08}
\newtcolorbox{mynamedbox1}[1]{colback=venetianred!5!white,colframe=venetianred!80!black,title=#1}
\newtcolorbox{mynamedbox2}[1]{colback=azure!5!white,colframe=azure!80!black,title=#1}

\definecolor{rossocorsa}{rgb}{0.83, 0.0, 0.0}

\tikzset{->-/.style={decoration={
  markings,
  mark=at position #1 with {\arrow{>}}},postaction={decorate}}}
\tikzset{-<-/.style={decoration={
  markings,
  mark=at position #1 with {\arrow{<}}},postaction={decorate}}} 
\def\be{\begin{equation}}
\def\ee{\end{equation}}
\def\ba{\begin{eqnarray}}
\def\ea{\end{eqnarray}}

\def\L*{{\cal L}_*}
\def\L{\mathcal{L}}
\def\({\left(}
\def\){\right)}

\def\<{\langle}
\def\>{\rangle}

 \def\neq {\not\equiv}

\def\cs2{c_{s}^{2}}

 \def\be   {\begin{equation}}   \def\ee   {\end{equation}}
 \def\ba   {\begin{array}}      \def\ea   {\end{array}}
 \def\bea  {\begin{eqnarray}}   \def\eea  {\end{eqnarray}}
 \def\bean {\begin{eqnarray*}}  \def\eean {\end{eqnarray*}}
\titleclass{\subsubsubsection}{straight}[\subsection]

\newcounter{subsubsubsection}[subsubsection]
\renewcommand\thesubsubsubsection{\thesubsubsection.\arabic{subsubsubsection}}
\titleformat{\subsubsubsection}
  {\normalfont\normalsize\bfseries}{\thesubsubsubsection}{1em}{}
\titlespacing*{\subsubsubsection}
{0pt}{3.25ex plus 1ex minus .2ex}{1.5ex plus .2ex}

\makeatletter
\renewcommand\paragraph{\@startsection{paragraph}{5}{\z@}%
  {3.25ex \@plus1ex \@minus.2ex}%
  {-1em}%
  {\normalfont\normalsize\bfseries}}
\renewcommand\subparagraph{\@startsection{subparagraph}{6}{\parindent}%
  {3.25ex \@plus1ex \@minus .2ex}%
  {-1em}%
  {\normalfont\normalsize\bfseries}}
\def\toclevel@subsubsubsection{4}
\def\toclevel@paragraph{5}
\def\toclevel@paragraph{6}
\def\l@subsubsubsection{\@dottedtocline{4}{7em}{4em}}
\def\l@paragraph{\@dottedtocline{5}{10em}{5em}}
\def\l@subparagraph{\@dottedtocline{6}{14em}{6em}}
\makeatother

\begin{document}

\title{\Large \textcolor{violet}{Single field inflation in the light of Pulsar Timing Array Data: Quintessential interpretation of blue tilted tensor spectrum through Non-Bunch Davies initial condition}}

\author{{\large  Sayantan Choudhury\orcidA{}}}
\email{sayantan\_ccsp@sgtuniversity.org,  \\ sayanphysicsisi@gmail.com 
}

\affiliation{Centre For Cosmology and Science Popularization (CCSP),\\
        SGT University, Gurugram, Delhi- NCR, Haryana- 122505, India.}

\begin{abstract}
In this work, we present a quintessential interpretation of having a blue-tilted tensor power spectrum for canonical single-field slow-roll inflation to explain the recently observed \textcolor{black}{Pulsar Timing Array (NANOGrav 15-year and EPTA)} signal of Gravitational Waves (GW). We formulate the complete semi-classical description of cosmological perturbation theory in terms of scalar and tensor modes using the Non-Bunch Davies initial condition. We found that the existence of the blue tilt ($n_t$) within the favoured range $1.2<n_t<2.5$ can be explained in terms of a newly derived consistency relation. Further, we compute a new field excursion formula using the Non-Bunch Davies initial condition, that validates the requirement of Effective Field Theory in the sub-Planckian regime, $|\Delta\phi|\ll M_{\rm pl}$ for the allowed value of the tensor-to-scalar ratio, $r<0.06$ from CMB observations. In our study, we refer to this result as Anti Lyth bound as it violates the well-known Lyth bound originally derived for Bunch Davies initial condition. Further, we study the behaviour of the spectral density of GW and the associated abundance with the frequency, which shows that within the frequency domain $10^{-9}{\rm Hz}<f<10^{-7}{\rm Hz}$ the outcome obtained from our analysis is completely consistent with the \textcolor{black}{Pulsar Timing Array (NANOGrav 15-year and EPTA)} signal. Also, we found that the behaviour of GW spectra satisfies the CMB constraints at the low frequency, $f_*\sim 7.7\times 10^{-17}{\rm Hz}$ corresponding to the pivots scale wave number, $k_*\sim 0.05{\rm Mpc}^{-1}$. Finally, the sharp falling behaviour of the GW spectra within the frequency domain $10^{-7}{\rm Hz}<f<1{\rm Hz}$ validates our theory in the comparatively high-frequency regime as well.

\end{abstract}

\maketitle
\tableofcontents
\newpage

 \section{Introduction}

A reliable prediction of many physically motivated scenarios is the presence of a stochastic gravitational wave background (SGWB) spanning a wide frequency range \cite{Caprini:2018mtu,Renzini:2022alw}. There are a number of potential sources, such as merging supermassive black hole binaries (SMBHBs) \cite{Siemens:2006yp,Caprini:2010xv,Ramberg:2019dgi,Caprini:2019egz,Ellis:2020awk,Geller:2023shn,Bi:2023tib} and cosmic remnants linked to phase transitions \cite{Rajagopal:1994zj,Jaffe:2002rt,Wyithe:2002ep,Sesana:2004sp,Burke-Spolaor:2018bvk}. For instance, the incoherent superposition of gravitational radiation generated by all SMBHBs during the slow adiabatic in spiral phase results in a broadband SGWB signal peaking in the ${\cal O}(10^{-9}{\rm Hz})$ frequency range \cite{Burke-Spolaor:2018bvk}. Pulsar timing arrays (PTAs) are a well-known detection method for nHz GWs, taking advantage of the fact that millisecond pulsars exhibit exceptionally steady clock behaviour \cite{Detweiler:1979wn}. When passing GWs travel along the line of sight to a pulsar, they perturb the space-time metric, PTAs look for spatially correlated changes in the pulse arrival-time measurements of the pulsars in a widely dispersed array \cite{Hobbs:2017oam}. In their $12.5$ -year dataset \cite{NANOGrav:2020bcs} of timing residuals from $47$ pulsars, the North American Nanohertz Observatory for Gravitational Waves (NANOGrav) PTA consortium revealed significant evidence in the year $2020$ for a red-stochastic common-spectrum mechanism. Later on, the Parkes PTA (PPTA) \cite{Goncharov:2021oub} and European PTA (EPTA) \cite{Chen:2021rqp,EPTA:2023fyk, EPTA:2023sfo, EPTA:2023akd, EPTA:2023gyr, EPTA:2023xxk, EPTA:2023xiy} partnerships, along with the integrated International PTA (IPTA) \cite{Antoniadis:2022pcn}, verified this signal. While the first expected SGWB signature in a PTA \cite{Romano:2020sxq,NANOGrav:2020spf} is an excess residual power with constant amplitude and spectral shape across all pulsars, a GW origin for such a signal can only be attributed in the presence of phase-coherent inter-pulsar correlations which follow the Hellings-Downs (HD) pattern \cite{Hellings:1983fr}, thereby excluding less interesting possibilities like intrinsic pulsar processes \cite{Goncharov:2022ktc,Zic:2022sxd} or common systematic noise \cite{Tiburzi:2015kqa}. Recent PTA experiments such as NANOGrav, EPTA, PPTA, and the Chinese Pulsar Timing Array (CPTA) have all reported on the analyses of their most recent datasets, which all confirm the presence of excess red common-spectrum signals with strain amplitudes of order ${\cal O}(10^{-15})$ at the reference frequency $f \sim 3.17\times 10^{-8}{\rm Hz}$ \cite{NANOGrav:2023gor,Antoniadis:2023ott,Reardon:2023gzh,Xu:2023wog,NANOGrav:2023hde,NANOGrav:2023ctt,NANOGrav:2023hvm,NANOGrav:2023hfp,NANOGrav:2023tcn,NANOGrav:2023pdq,NANOGrav:2023icp,Antoniadis:2023lym,Antoniadis:2023puu,Antoniadis:2023aac,Antoniadis:2023xlr,Smarra:2023ljf,Reardon:2023zen,Zic:2023gta}. These are the first convincing detections of an SGWB signal in the nHz frequency range because all analyses present evidence (of varying strength) for HD correlations, which denote a genuine GW origin for the signals \footnote{Using the NANOGrav 15-year results \cite{NANOGrav:2023gor,Antoniadis:2023ott,Reardon:2023gzh,Xu:2023wog,NANOGrav:2023hde,NANOGrav:2023ctt,NANOGrav:2023hvm,NANOGrav:2023hfp,NANOGrav:2023tcn,NANOGrav:2023pdq,NANOGrav:2023icp,Antoniadis:2023lym,Antoniadis:2023puu,Antoniadis:2023aac,Antoniadis:2023xlr,Smarra:2023ljf,Reardon:2023zen,Zic:2023gta} as an example, the inferred amplitude of the excess red common-spectrum signal is, ${\cal A}_G = 6.3 \times 10^{-15}$ at the reference frequency $f \sim 3.17\times 10^{-8}{\rm Hz}$, and a model with an HD-correlated power-law SGWB was found to be preferable over one with a spatially uncorrelated common-spectrum power-law SGWB.}. See refs. \cite{Guo:2023hyp,Kitajima:2023cek,Murai:2023gkv,Ashoorioon:2022raz,Athron:2023mer,Li:2023bxy,Li:2023yaj,Oikonomou:2023qfz,Niu:2023bsr} for the various possible theoretical explanations of these signals for GW.

Inflation is a theorized stage of quasi-de Sitter expansion in the very early Universe that was introduced to address the flatness, horizon, and entropy issues \cite{Kazanas:1980tx,Starobinsky:1980te,Sato:1981ds,Guth:1980zm,Mukhanov:1981xt,Linde:1981mu,Albrecht:1982wi}. During inflation, small quantum mechanical fluctuations generated from linearized tensor and scalar perturbations to the metric are stretched outside the Hubble horizon before re-entering much later in the time scale. The inflationary paradigm \cite{Baumann:2009ds,Baumann:2018muz,Senatore:2013roa,Choudhury:2011sq,Choudhury:2011jt,Choudhury:2012yh,Choudhury:2012whm,Choudhury:2013zna,Choudhury:2013jya,Choudhury:2013iaa,Choudhury:2013woa,Choudhury:2014sxa,Choudhury:2014uxa,Choudhury:2014hua,Choudhury:2014kma,Choudhury:2014sua,Choudhury:2014hja,Choudhury:2015pqa,Choudhury:2015hvr,Choudhury:2016wlj,Choudhury:2016cso,Choudhury:2017cos,Naskar:2017ekm,Choudhury:2017glj,Choudhury:2018glz,HosseiniMansoori:2023zop,Geng:2015fla,WaliHossain:2014usl,Hossain:2014coa,Hossain:2014xha} is still in excellent shape which is consistent with a large number of cosmological observations \cite{Martin:2013tda, Benetti:2013cja, Martin:2013nzq,Creminelli:2014oaa,Dai:2014jja,Benetti:2016tvm,Campista:2017ovq,Keeley:2020rmo,Vagnozzi:2020rcz,Vagnozzi:2020dfn,Vagnozzi:2023lwo,Cabass:2022wjy,Cabass:2022ymb}. Further tests of the inflationary paradigm are one of the prime scientific ingredients of future observational probes \cite{CMB-S4:2016ple,SimonsObservatory:2018koc,SimonsObservatory:2019qwx}. However, the detection of the inflationary SGWB has yet to be made. While this signal has typically been desirable at very low frequencies, $f \leq {\cal O}(10^{-15} {\rm Hz}-10^{-16} {\rm Hz})$ \cite{Kamionkowski:2015yta}, based on expectations within the standard framework of inflationary models, such a search need not be restricted to the previously mentioned low-frequency domain, as beyond the canonical single field models of inflation or maybe some interesting version with the canonical one can naturally predict rich phenomenological features at higher frequency domain, including the key frequency range ${\cal O}(10^{-9}{\rm Hz}-10^{-7}{\rm Hz})$ as suggested by very recently observed NANOGrav 15 signal of GW.

The main component and the most important aspects of this essay are summarised below in a brief, point-by-point fashion. This will be beneficial for general readers from a variety of viewpoints:
 \begin{itemize}[label={$\diamond$}]
    \item In this work we start our discussion with the gauge fixing issue of the Cosmological Perturbation Theory using this we will perform the perturbation of scalar and tensor modes dynamical solution of which we have computed from {\it Mukhanov Sasaki equation} along with general quantum initial condition, which we will identify as the {\it Non-Bunch Davies} initial condition \cite{Choudhury:2017glj,Choudhury:2020yaa,Choudhury:2021tuu,Adhikari:2021ked,Akama:2023jsb,Albayrak:2023hie,Choudhury:2022mch,Colas:2022kfu,Aalsma:2022eru,Chapman:2022mqd,Letey:2022hdp,Penna-Lima:2022dmx,Kanno:2022mkx,Fumagalli:2021mpc,Sleight:2021plv,Chen:2010bka,Wang:2014kqa,Ashoorioon:2014nta,Ashoorioon:2013eia}. The parameter space of such new physics is spanned by ${\bf SO(1,4)}$ isometry group of de Sitter space and for this reason, one can expect that the {\it Bunch Davies} initial condition will be represented by a point in the larger parameter space of the {\it Non-Bunch Davies} initial condition. Further, we will compute the impacts of these perturbations in the inflationary observables, such as in the spectrum, spectral tilt, and tensor-to-scalar ratio in the presence of {\it Non-Bunch Davies} initial condition which we believe will show significant deviations compared to the result obtained for the observables with {\it Bunch Davies} initial condition. 
    
    \item From our analysis we will show how one
    can violate the old consistency relation, $r=-8n_t$ with $n_t<0$ derived in the presence of {\it Bunch Davies} initial condition. In our work, we will also point to the preferred range of tensor spectral tilt which is expected to show a blue feature in the tensor power spectrum in the presence of {\it Non-Bunch Davies} initial condition. See refs. \cite{Huang:2015gka,Huang:2015gca,Huang:2017gmp} to know more about tensor spectral tilt and its physical implications in CMB experiments.

    \item Further we will compute the new version of the field excursion formula for the inflaton field and will explicitly show in what amount significant deviation can be achieved from the good old {\it Lyth bound} derived in the same context from {\it Bunch Davies} initial condition. In this computation, our prime objective is to validate the EFT prescription for the canonical single field inflationary paradigm with the help of {\it Non-Bunch Davies} initial condition. Though we will do our analysis in a model-independent fashion i.e. we will not specifically talk about any particular structure of the inflationary potential to derive this bound, but we strongly believe that blue tilted tensor power spectrum and to maintain the theoretical constraint from the new consistency relation derived from {\it Non-Bunch Davies} initial condition the potential should have a certain structure which will satisfy certain physical properties. 

    \item Next, using the derived new consistency relation between tensor-to-scalar ratio and tensor spectral tilt we will study the behaviour of the Gravitational Waves (GW) abundance ($\Omega_{\rm GW}h^2$) with frequency ($f$) in the presence of {\it Non-Bunch Davies} initial condition. Our claim is that the GW abundance computed for {\it Non-Bunch Davies} initial condition can able to justify the various underlying new physical phenomena in the three frequency bands, very low $10^{-20}{\rm Hz}<f<10^{-17}{\rm Hz}$, low $10^{-17}{\rm Hz}<f<10^{-7}{\rm Hz}$ and high $10^{-7}{\rm Hz}<f<1{\rm Hz}$. We will show the status of the {\it Non-Bunch Davies} initial condition generated GW in the light of CMB Planck observation, recently observed NANOGrav 15 signal, LISA \cite{amaro2017laser}, BBO \cite{Kawamura:2011zz}, DECIGO \cite{Crowder:2005nr}, CE \cite{Punturo:2010zz}, ET \cite{Reitze:2019iox}, HLVK and HLV(03) \cite{KAGRA:2018plz,VIRGO:2014yos,LIGOScientific:2014pky}.    

\end{itemize}
The outline of this paper is as follows: In \ref{sec2}, we briefly discuss the canonical single-field model of inflation. In \ref{sec3}, we discuss in detail the generation of scalar and tensor modes using cosmological perturbation theory. In \ref{sec4}, we move on towards a specific and detailed computation of the scalar power spectrum and later discuss its observational impacts. A similar analysis is followed further in \ref{sec5} for the derivation of the tensor power spectrum. In \ref{sec6}, the results derived for the power spectrum in the previous sections are connected with the recent findings of the \textcolor{black}{Pulsar Timing Array (NANOGrav 15-year and EPTA)} Data Set. In \ref{sec7}, a new consistency condition is established to explain the blue-tilted tensor power spectrum based on the observed changes in the amplitude and spectral tilt of the tensor perturbations as discussed in previous sections. In \ref{sec8}, due to the presence of modified Bogoliubov coefficients for the scalar and tensor perturbations, the well-known Lyth bound is also modified and presented with a discussion performed in a model-independent fashion. In \ref{sec9}, we present the numerical analysis required to explain the impact of the blue-tilted power spectrum. After a brief discussion on the generation of stochastic gravitational waves we discuss plots for the tensor-to-scalar ratio based on our new consistency relation and for the abundance of gravitational waves in the presence of Non-Bunch Davies initial condition. Finally, in \ref{sec10} we present the conclusions of our findings and discussions.

\section{The canonical single field model of inflation: The old wine a new glass}
\label{sec2}

To demonstrate the claim of this work let us first write down the fundamental action of the underlying theory which is described by the following equation:
\begin{mynamedbox1}{Canonical single field model minimally coupled to gravity:}
\vspace{-.35cm}
\bea
 S=\frac{1}{2}\int d^4x \sqrt{-g} \left[M^2_{\rm pl} R - (\partial\phi)^2 - 2 V(\phi) \right],
\eea
\end{mynamedbox1}
where $\phi$ is a scalar field that is minimally coupled to gravity and in the present context it describes the single field inflationary paradigm.  In the above action the canonical kinetic term of the scalar field is described by the following shorthand notation:
\bea  (\partial\phi)^2=g^{\mu\nu}\left(\partial_{\mu}\phi\right)\left(\partial_{\nu}\phi\right),\eea and  $V(\phi)$ is the inflationary potential for the scalar field which satisfy the slow-roll conditions during inflation,  $M_{\rm pl}$ is the reduced Planck mass which is $M_{\rm pl}\sim 2.43\times 10^{18}{\rm GeV}$, and last but not the least, the gravitational interaction is described by the Einstein-Hilbert term which is described in terms of $R$, which is the Ricci scalar in the following discussion. Most importantly, it is important to mention that all the terms in the above-mentioned action are minimally coupled with the gravitational sector through the term $\sqrt{-g}$, which physically represents the determinant of the metric in the corresponding space-time.

In this connection, the best possibility is the spatially flat Friedmann-Lemaitre-Robertson-Walker (FLRW) metric which describes our observable Universe, and the corresponding metric of space-time is given by:
\begin{mynamedbox1}{Qusi de Sitter space-time:}
\vspace{-.35cm}
\bea
ds^2=a^2(\tau)\left(-d\tau^2+d{\bf x}^2\right)=a^2(\tau)\left(-d\tau^2+\delta_{ij}dx^{i}dx^{j}\right)\quad\quad{\rm where}\quad\quad a(\tau)=-\frac{1}{H\tau}\quad\quad{\rm with}\quad\quad \tau \in (-\infty,0),
\eea
\end{mynamedbox1}
where $\tau$ is the conformal time coordinate.  The scale factor $a(\tau)$ is expressed in terms of the conformal time coordinate and the corresponding expression describes the quasi-de Sitter solution which is extremely useful for describing the inflationary paradigm. Here $`H'$ represents the Hubble parameter in quasi de Sitter space-time, which is not exactly constant, and deviation from the constancy is treated in terms of slow-roll dynamics which we will discuss in brief now.
To describe the slow-roll dynamics, let us first write down the classical dynamical equation of motions in spatially flat FLRW space-time, which are known as Friedmann equations and Klein-Gordon equation, and given by the following expressions:
\begin{mynamedbox1}{Classical field equations:}
\vspace{-.35cm}
\bea
&&{\cal H}^2=\frac{1}{3M^2_{\rm pl}}\Bigg(\frac{1}{2}\phi^{'2}+a^2V(\phi)\Bigg),\\
%\quad\quad
&&{\cal H}^{'}=-\frac{1}{2M^2_{\rm pl}}\phi^{'2},\\
%\quad\quad
&&\phi^{''}+2{\cal H}\phi^{'}+a^2\frac{dV(\phi)}{d\phi}=0.
\eea
\end{mynamedbox1}
where we have introduced a shorthand notation $'$ which represents the conformal time derivative $d/d\tau$. Also, we use the new definition of the Hubble parameter in conformal time, $\displaystyle {\cal H} \equiv a^{'}/a=aH$ which is going to be extremely useful for further understanding the subject material studied in this paper.

Now, to initiate and end properly the inflationary paradigm at a specific value in the field space one needs to introduce the following parameters in quasi-de Sitter space-time, known as the slow-roll parameters,
\begin{mynamedbox1}{Slow-roll parameters:}
\vspace{-.35cm}
\bea
\epsilon &=&\bigg(1-\frac{{\cal H}^{'}}{{\cal H}^2}\bigg)=\frac{1}{2M^2_{\rm pl}}\frac{\phi^{'2}}{a^2{\cal H}^2},\\
\eta &=&\frac{\epsilon^{'}}{\epsilon {\cal H}}=\bigg(2\epsilon+\frac{\phi^{''}}{\phi^{'} {\cal H}}-1\bigg).
\eea
\end{mynamedbox1}
To validate inflation in the slow-roll region, one needs to satisfy the following conditions,
\bea \epsilon\ll 1,\quad\quad |\eta|\ll 1\quad\quad {\rm and}\quad\quad \left|\frac{\phi^{''}}{\phi^{'} {\cal H}}-1\right|\ll 1.\eea
On the contrary, slow-roll conditions are violated at the end of inflation when we have either $\epsilon=1$ or $|\eta|=1$ or both using which one can explicitly compute the field value at the end of inflation.

 \section{Cosmological Perturbation Theory: Generation of scalar and tensor modes}
\label{sec3}

Let us now have a look at the small perturbations that occur around the spatially flat FLRW background, where the linearized field and the metric perturbations are provided by,
\begin{mynamedbox1}{Linearized field and the metric perturbations:}
\vspace{-.35cm}
\bea
 \phi({\bf x},\tau)&=&\overline{\phi}(\tau)+\delta\phi({\bf x},\tau),\\
    ds^2&=& a^2(\tau)\Bigg[-d\tau^2+\bigg\{\bigg(1+2{\cal R}({\bf x},\tau)\bigg)\delta_{ij}+2h_{ij}({\bf x},\tau)\bigg\}dx^{i}dx^{j}\Bigg]\nonumber\\
   &=&\underbrace{a^2(\tau)\Bigg[-d\tau^2+\delta_{ij}dx^{i}dx^{j}\Bigg]}_{\bf Unperterbed\;background\; metric}+\underbrace{2a^2(\tau)\bigg\{{\cal R}({\bf x},\tau)\delta_{ij}+h_{ij}({\bf x},\tau)\bigg\}dx^{i}dx^{j}}_{\bf Perturbed\;part\;of\;the\;metric}.
\eea
\end{mynamedbox1}
   
   In this description, we have the following important facts that one needs to remember always henceforth:
   \begin{enumerate}
       \item Here $\overline{\phi}(\tau)$ represents the scalar field embedded in a spatially flat background FLRW space-time.

       \item Also, $\delta\phi({\bf x},\tau)$ quantifies small perturbations in the scalar field which can be able to capture the inhomogeneity and anisotropy.

       \item Here ${\cal R}({\bf x},\tau)$ represents the scalar comoving curvature perturbation which is often used to describe the scalar fluctuations in inflation:
       \bea {\cal R}({\bf x},\tau)=\bigg(\zeta({\bf x},\tau)-\frac{{\cal H}}{\bar{\phi}^{'}(\tau)}\delta\phi({\bf x},\tau)\bigg).\eea
       The advantage of this new variable ${\cal R}({\bf x},\tau)$ is that it is constant outside the horizon and invariant under linearized coordinate transformations. But when working directly in the language of ${\cal R}({\bf x},\tau)$, taking the $\bar{\phi}^{'}(\tau)\rightarrow 0$ limit can occasionally be misleading because $\bar{\phi}^{'}(\tau)$ appears in the denominator on the RHS even if the exact de Sitter description should become a good description.

       \item Using two distinct gauges is the most straightforward solution to this problem. 
       
       \item One can work in the gauge where the perturbations are while they are evolving inside the horizon and described by:
       \begin{mynamedbox1}{Gauge A:}
\vspace{-.35cm}
 \bea \zeta({\bf x},\tau)=0.\eea
\end{mynamedbox1}
      
       We refer to this as {\bf gauge A}. The scalar perturbation in this gauge is given by $\delta\phi({\bf x},\tau)$ and exhibits a smooth behaviour, which is described by:
\bea {\cal R}({\bf x},\tau)=-\frac{{\cal H}}{\bar{\phi}^{'}(\tau)}\delta\phi({\bf x},\tau).\eea
      
      \item At this stage one can consider another gauge which is the perfect choice while describing the perturbations outside the horizon and given by:
       \begin{mynamedbox1}{Gauge B:}
\vspace{-.35cm}
\bea \delta\phi({\bf x},\tau)=0.\eea
\end{mynamedbox1}
      
      The correlation functions expressed in terms of $\zeta({\bf x},\tau)$ are time-independent because in this gauge the scalar perturbation is given by:
       \bea {\cal R}({\bf x},\tau)=\zeta({\bf x},\tau),\eea
      which is constant outside the horizon. We refer to this as {\bf gauge B} and in this article, we are going to use this specific gauge to perform the rest of the computation. 

      \item There is an equivalent description using which sometimes one describes the Cosmological Perturbation Theory. This is known as {\it Arnowitt-Deser-Misner (ADM) formalism} and is extremely useful to describe the Hamiltonian formulation of the General Theory of Relativity, which is commonly used in the context of canonical quantum gravity and numerical relativity. In this description, the corresponding ADM metric is described by the following expression:
       \begin{mynamedbox1}{Arnowitt-Deser-Misner (ADM) metric:}
\vspace{-.35cm}
\bea ds^2_{\rm ADM}=-N^2d t^2+g_{ij}\left(dx^{i}+N^{i}dt\right)\left(dx^{j}+N^{j}dt\right),\eea
\end{mynamedbox1}
      
      where instead of using conformal time coordinate the formalism is completely developed in terms of the physical time coordinate $t$ and for this reason, the perturbed metric is not expressed in terms of conformally flat form. Using this metric one needs to choose the following gauge fixing condition:
      \bea N=1,\quad\quad N^{i}=0,\eea
      where $N$ and $N^{i}$ represent the lapse and shift functions. In this context, the spatial component of the metric $g_{ij}$ is described using the following expression:
      \bea g_{ij}=a^2(t)\left(2{\cal R}({\bf x},t)+h_{ij}({\bf x},t)\right).
      \eea
      Here ${\cal R}({\bf x},t)$ and $h_{ij}({\bf x},t)$ represent the comoving curvature perturbation and tensor perturbation and both of them are written in terms of the physical time coordinate $t$. Also, the scale factor is given by, $a(t)=\exp(Ht)$, where $H$ is not exactly constant for quasi-de Sitter space-time. We will talk about the tensor perturbation in detail in the later points, so up to that point please allow us to explain some of the crucial facts regarding the gauge choices and formalism that one follows within the context of Cosmological Perturbation Theory.

      \item It is further important to note that the gauge fixing conditions written in terms of the lapse and shift functions, within the framework of {\it ADM formalism}, do not fix the gauge degrees of freedom completely. After performing the mentioned gauge fixing there is a possibility to have the following {\it spatial reparametrization} which is of the following form:
      \bea x^{i}\rightarrow x^{i}+\xi^{i}({\bf x}),\eea
      and {\it temporal reparametrization} which is of the following form:
       \bea t\rightarrow t+\xi({\bf x}),\quad\quad x^{i}\rightarrow x^{i}+v^{i}({\bf x},t)\quad\quad {\rm where}\quad\quad v^{i}({\bf x},t)=\partial^{i}\xi({\bf x})\int\frac{dt}{a^2(t)}=-\frac{1}{2H}\partial^{i}\xi({\bf x})\exp(-2Ht).\eea
       Here $\xi^{i}({\bf x})$ and $\xi({\bf x})$ represent the corresponding parameters of gauge transformations which describe {\it spatial reparametrization} and {\it temporal reparametrization} respectively. Out of both of these possibilities to write down the Cosmological Perturbation Theory in a correct fashion, one needs to only look into the {\it temporal reparametrization} which provides the correct coordinate transformation where one needs to choose the time-independent parameter  $\xi({\bf x})$ suitably. 

       \item At the linearized level of the perturbation one can note the following relationship among the parameter $\zeta({\bf x},\tau)$ of the {\bf gauge B} and the parameter $\delta\phi({\bf x},\tau)$ corresponding to the {\bf gauge A}, given by:
       \begin{mynamedbox1}{Relationship between gauge A and gauge B:}
\vspace{-.35cm}
\bea \zeta({\bf x},\tau)=-\frac{{\cal H}}{\bar{\phi}^{'}(\tau)}\delta\phi({\bf x},\tau),\eea
\end{mynamedbox1}
       
       which becomes extremely useful for converting the cosmological observables, particularly the correlation functions and other derived parameters, from one gauge to another just by performing a very simple change of variables.

       \item Finally, in the present context $h_{ij}({\bf x},\tau)$ represent the spin-$2$ tensor perturbations which describe the primordial gravitational waves (PGW) that satisfy the following properties at the linearized level of perturbation theory:
         \begin{mynamedbox1}{Properties of PGW:}
\vspace{-.35cm}
\bea &&\textcolor{black}{\bf Symmetric:\Rightarrow}~~h_{ij}=h_{ji},~~~\textcolor{black}{\bf Transverse:\Rightarrow}~~\partial^{i}h_{ij}=0,~~~
       \textcolor{black}{\bf Traceless:\Rightarrow}~~h^{i}_{i}=0.~~~~~~\eea
\end{mynamedbox1}

   \end{enumerate}
   In the upcoming section, we are going to study the impacts of scalar and tensor perturbations in detail and their direct connection with the presently observed NANOGrav 15-year Data Set.

\section{Computation of scalar power spectrum with Non-Bunch Davies initial condition}
\label{sec4}
   
In this section, we are interested in studying the theoretical and observational impacts of the scalar power spectrum. To serve this purpose we begin by expanding the representative canonical scalar field model action up to the second order in the scalar comoving curvature perturbation, ${\cal R}({\bf x},\tau)=\zeta({\bf x},\tau)$, which gives the following simplified form of the action:
 \begin{mynamedbox1}{Second order perturbed action for scalar modes:}
\vspace{-.35cm}
\bea
   S^{(s)}_{(2)}=M^2_{\rm pl}\int d\tau\;  d^3x\;  a^2(\tau)\; \epsilon\; \bigg(\zeta^{'2}({\bf x},\tau)-\left(\partial_i\zeta({\bf x},\tau)\right)^2\bigg).
   \eea
  \end{mynamedbox1} 
    For further computationally simplification purposes we introduce the rescaled space-time dependent field variable, which is written as:
   \bea v({\bf x},\tau)=z(\tau)M_{\rm pl}\zeta({\bf x},\tau),\eea
   using which the second-order action can be further expressed in terms of the canonically normalized form and the corresponding expression is given by:
   \begin{eqnarray}
  \label{ms1} S^{(s)}_{(2)}=\frac{1}{2}\int d\tau\;  d^3x\;  \bigg(v^{'2}({\bf x},\tau)-\left(\partial_iv({\bf x},\tau)\right)^2+\frac{z^{''}(\tau)}{z(\tau)}v^{2}({\bf x},\tau)\bigg),
   \end{eqnarray}
   where we introduce a conformal time-dependent quantity in the quasi-de Sitter space-time:
   \bea z(\tau)=a(\tau)\sqrt{2\epsilon}=-\frac{1}{H\tau}\sqrt{2\epsilon}.\eea
   Here the scale factor and slow-roll parameter $\epsilon$ are defined using the conformal time as mentioned in the previous part of this paper. 
   
   Now instead of performing the rest of the analysis in coordinate space, we will perform the further computations in the momentum space as this is the most cosmologically relevant region from the observational perspective. Also, the coordinate space result involves the Infra Red (IR) cut-off in terms of the underlying length scale of the quasi-de Sitter theory, which is of course not good and one cannot able to fix such cut-offs just from the observational point of view. For this reason, we make use of the following Fourier transform ansatz using which the rescaled field variable in the coordinate space can be written in terms of its Fourier space counterpart by the expression:
   \bea
   v({\bf x},\tau):=\int\frac{d^3{\bf k}}{(2\pi)^3}\; e^{i{\bf k}.{\bf x}}\;v_{\bf k}(\tau).
   \eea
  Using this Fourier transform ansatz the previously mentioned second-order action written in terms of the rescaled scalar perturbation can be recast in the following simplified form:
     \bea
 S^{(s)}_{(2)}=\frac{1}{2}\int \frac{d^3{\bf k}}{(2\pi)^3}\;d\tau\; \bigg(|v^{'}_{\bf k}(\tau)|^{2}-\left(k^2-\frac{z^{''}(\tau)}{z(\tau)}\right)|v_{\bf k}(\tau)|^{2}\bigg).
   \eea
   This action mimics the role of the action of an oscillator having a conformal time-dependent effective frequency in the present discussion. Variation of the aforementioned action with respect to the modes of the rescaled field variable gives us following the equation of motion written as:
    \begin{mynamedbox1}{Mukhanov Sasaki equation for scalar modes:}
\vspace{-.35cm}
 \bea
 v^{''}_{\bf k}(\tau)+\left(k^2-\frac{z^{''}(\tau)}{z(\tau)}\right)v_{\bf k}(\tau)=0\,,
   \eea
  \end{mynamedbox1} 
   which in the corresponding literature is commonly referred to as the {\it Mukhanov Sasaki (MS) equation}. For further computational purposes, it is important to note that in this computation we have the effective, conformal time-dependent, mass which can be expressed in terms of the following simplified language:
   \bea m^2_{\rm eff,(s)}(\tau)=-\frac{z^{''}(\tau)}{z(\tau)}=-\frac{1}{\tau^2}\bigg(\nu^2_s(\tau)-\frac{1}{4}\bigg),\eea
   where we have parameterized the effective mass for the scale mode of perturbation in terms of the conformal time-dependent mass parameter function for the scalar modes by the following relation at the leading order of the slow-roll approximation:
   \bea \nu_s(\tau):\approx \bigg(\frac{3}{2}+3\epsilon-\eta\bigg).\eea
   Here the subscript $s$ corresponds to the contribution computed for the scalar modes only and $\epsilon,\eta$ are the slow-roll parameters that we have explicitly defined in the earlier part of this paper. Using the above-mentioned parameterization and the corresponding slow-roll approximation the most general solution of the second-order MS equation can be written in terms of the following expression:
   \bea v_{\bf k}(\tau)=\sqrt{-\tau}\left[{\alpha}^{(s)}_{\bf k}~H^{(1)}_{\nu_{s}}(-k\tau)+{\beta}^{(s)}_{\bf k}~H^{(2)}_{\nu_{s}}(-k\tau)\right].\eea
   In this solution, $H^{(1)}_{\nu_{s}}(-k\tau)$ and ${\cal H}^{(2)}_{\nu_{s}}(-k\tau)$, represent the Hankel function of the first and second kind which can further be expressed in terms of the Bessel function of the first kind and second kind as:
\bea &&H^{(1)}_{\nu_{s}}(-k\tau)={\cal J}^{(1)}_{\nu_{s}}(-k\tau)+i{\cal Y}^{(1)}_{\nu_{s}}(-k\tau),\\
&&H^{(2)}_{\nu_{s}}(-k\tau)={\cal J}^{(1)}_{\nu_{s}}(-k\tau)-i{\cal Y}^{(1)}_{\nu_{s}}(-k\tau),\eea
which further implies that: 
\bea H^{(1),*}_{\nu_{s}}(-k\tau)=H^{(2)}_{\nu_{s}}(-k\tau), \quad\quad H^{(2),*}_{\nu_{s}}(-k\tau)=H^{(1)}_{\nu_{s}}(-k\tau).\eea
Here, ${\alpha}^{(s)}_{\bf k}$ and ${\beta}^{(s)}_{\bf k}$, are the two Bogoliubov coefficients which are fixed by the proper choice of the quantum initial condition during the slow-roll phase of the inflation. In a deeper sense, the correct choice of the initial condition can able to construct the corresponding quantum mechanical vacuum state. In principle, these Bogoliubov coefficients can be any arbitrary function of the momentum scale which will appear and span within the large classes of ${\bf SO(1,4)}$ de-Sitter isometry group. The generator and the conserved charges of this infinite parameter family should contain the mentioned two Bogoliubov coefficients which describe the corresponding general quantum initial condition and the associated states can be fixed in this description very clearly.

In general, one can consider an arbitrary initial condition described in terms of a general quantum vacuum state which is specified by the two Bogoliubov coefficients ${\alpha}^{(s)}_{\bf k}$ and ${\beta}^{(s)}_{\bf k}$ as mentioned earlier. Within this description of the Quantum Field Theory of de-Sitter space, a general quantum vacuum state is described by these two numbers as $|{\alpha}^{(s)}_{\bf k},{\beta}^{(s)}_{\bf k}\rangle_{\bf NBD}$ and characterized by the following equation: 
\begin{mynamedbox1}{Definition of Non-Buch Davies state for scalar modes:}
\vspace{-.35cm}
  \bea \hat{C}_{\bf k}|{\alpha}^{(s)}_{\bf k},{\beta}^{(s)}_{\bf k}\rangle_{\bf NBD}=0~\forall~{\bf k},\eea
  \end{mynamedbox1} 
  
    where $\hat{C}_{\bf k}$ represents the corresponding annihilation operators for the scalar modes which become extremely important when we canonically quantize the perturbed modes using Cosmological Perturbation Theory. 
    
    In principle, the general quantum state can be expressed in terms of the well-known Euclidean vacuum state, which is commonly referred to as the {\it Bunch Davies} quantum vacuum state in the present description using a well-defined Bogoliubov transformation. Here before going to further computational details in this present section, it is important to mention that in the large parameter space of ${\bf SO(1,4)}$ de-Sitter isometry group {\it Bunch Davies} initial condition is described by a single point, which is actually characterized by the fixing Bogoliubov coefficients as, ${\alpha}^{(s)}_{\bf k}=1$ and ${\beta}^{(s)}_{\bf k}=0$. However if one is interested in utilizing the effect of a general quantum initial condition one should have an infinite number of possibilities where these coefficients are arbitrary functions of the characteristic momentum scale as appearing all over in the present computation. In this corresponding literature, this possibility is often referred to as {\it Non-Bunch Davies} initial condition which we explicitly show to be extremely useful for the present computational purpose. Now let us come back to the previously mentioned Bogoliubov transformation which helps us to express any general {\it Non-Bunch Davies} initial quantum states in terms of the {\it Bunch Davies} initial quantum state and such transformation is described by the following equation:
    \begin{mynamedbox1}{Bogolibov transformation for scalar modes:}
\vspace{-.35cm}
 \bea  |{\alpha}^{(s)}_{\bf k},{\beta}^{(s)}_{\bf k}\rangle_{\bf NBD}&=&\prod_{\bf k}\frac{1}{\sqrt{|{\alpha}^{(s)}_{\bf k}|}}\exp\left[\frac{{\beta}^{(s),*}_{\bf k}}{2{\alpha}^{(s),*}_{\bf k}}~\hat{C}^{\dagger}_{\bf k}\hat{C}^{\dagger}_{-{\bf k}}\right]|1,0\rangle^{(s)}_{\bf BD}\nonumber\\
       &=&\prod_{\bf k}\frac{1}{{\cal N}^{(s)}_{{\bf NBD},{\bf k}}}~\exp\left[\frac{{\beta}^{(s),*}_{\bf k}}{2{\alpha}^{(s),*}_{\bf k}}~\hat{C}^{\dagger}_{\bf k}\hat{C}^{\dagger}_{-{\bf k}}\right]|1,0\rangle^{(s)}_{\bf BD},~~~~~~~~\eea                              
  \end{mynamedbox1} 
    
        where the overall normalization factor of the defined {\it Non-Bunch Davies} initial quantum states are described by the following equation:
        \bea {\cal N}^{(s)}_{{\bf NBD},{\bf k}}:=\sqrt{|{\alpha}^{(s)}_{\bf k}|}.\eea 
        Additionally, it is important for the present description to note that, the initial {\it Bunch Davies} quantum state for the scalar modes is often referred to as:
        \bea |1,0\rangle^{(s)}_{\bf BD}=|0\rangle.\eea
        Here the careful observation clearly shows that we have omitted the superscript $s$ because within the framework of describing {\it Bunch Davies} initial quantum state one cannot able to distinguish between the ground states which describe the generation of scalar and tensor modes i.e. 
        \bea |1,0\rangle^{(s)}_{\bf BD}= |1,0\rangle^{(t)}_{{\bf BD},{\bf k},\lambda}\quad\quad\quad {\rm because}\quad\quad {\alpha}^{(s)}_{\bf k}= {\alpha}^{(t)}_{{\bf k},\lambda}\quad {\rm and}\quad {\beta}^{(s)}_{\bf k}= {\beta}^{(t)}_{{\bf k},\lambda},\eea
        
        On the other hand, once we describe the {\it Non-Bunch Davies} initial quantum states in this context, where $s$ stands for scalar and $t$ for tensor perturbations, such tags cannot be removed as the corresponding states describing the scalar and tensor sectors are not at all same in general:
        \bea |{\alpha}^{(s)}_{\bf k},{\beta}^{(s)}_{\bf k}\rangle_{\bf NBD}\neq |{\alpha}^{(t)}_{{\bf k},\lambda},{\beta}^{(t)}_{{\bf k},\lambda}\rangle_{\bf NBD}\quad\quad\quad {\rm because}\quad\quad {\alpha}^{(s)}_{\bf k}\neq {\alpha}^{(t)}_{{\bf k},\lambda}\quad {\rm and}\quad {\beta}^{(s)}_{\bf k}\neq {\beta}^{(t)}_{{\bf k},\lambda},\eea
        where the quantum states $|{\alpha}^{(t)}_{{\bf k},\lambda},{\beta}^{(t)}_{{\bf k},\lambda}\rangle_{\bf NBD}$ and the corresponding Bogoliubov coefficients ${\alpha}^{(t)}_{{\bf k},\lambda},{\beta}^{(t)}_{{\bf k},\lambda}$ as appearing in the context of tensor perturbations are explicitly defined in the next section. Here the index $\lambda=+,\times$ corresponds to the two polarizations of the tensor modes.
        
        For further better understanding purpose let us also mention the following constraint condition on the {\it Non-Bunch Davies} initial quantum states which needs to be satisfied during performing rest of the computation: 
     \bea \hat{\bf P}^{(s)}_{C}|{\alpha}^{(s)}_{\bf k},{\beta}^{(s)}_{\bf k}\rangle_{\bf NBD}&=&\int \frac{d^3{\bf p}}{(2\pi)^3}~{\bf p}~\hat{C}^{\dagger}_{\bf p}\hat{C}_{\bf p}|{\alpha}^{(s)}_{\bf k},{\beta}^{(s)}_{\bf k}\rangle_{\bf NBD}\nonumber\\
     &=&\prod_{{\bf k}}\int \frac{d^3{\bf p}}{(2\pi)^3}~\frac{{\bf p}~\hat{C}^{\dagger}_{\bf p}\hat{C}_{\bf p}}{\sqrt{|{\alpha}^{(s)}_{\bf k}|}}\exp\left[\frac{{\beta}^{(s),*}_{\bf k}}{2{\alpha}^{(s),*}_{\bf k}}~\hat{C}^{\dagger}_{\bf k}\hat{C}^{\dagger}_{-{\bf k}}\right]|0\rangle\nonumber\\
      &=&0.\eea
    Similarly, the annihilation operators of the {\it Non-Bunch Davies} initial quantum states and {\it Bunch Davies} initial quantum state are related via the following sets of Bogoliubov transformations:
  \bea C_{\bf k}&=&\bigg({\alpha}^{(s),*}_{\bf k}~a_{\bf k}-{\beta}^{(s),*}_{\bf k}~a^{\dagger}_{-{\bf k}}\bigg),\\
    a_{\bf k}&=&\bigg({\alpha}^{(s)}_{\bf k}~C_{\bf k}+{\beta}^{(s)}_{\bf k}~C^{\dagger}_{-{\bf k}}\bigg).\eea 

\textcolor{black}{Further, it is important to note that the signed coordinate transformation has been recently introduced in refs. \cite{Guendelman:2024ezk,Guendelman:2023spd,Guendelman:2023vso} is an interesting and innovative transformation that changes the signature of the Jacobian if considered locally. It resembles the Bogoliubov transformations in a non-trivial curved space background, where it mixes the positive and negative frequency components. Here the non-trivial curved background is important as in the case of the Minkowski flat vacuum there is no mixing of the positive and negative frequency components appearing. As an immediate consequence of this local signed coordinate transformation the Fock space operators of the {\it Non-Bunch Davies} initial quantum state are written in terms of its {\it Bunch Davies} counter-part through Bogolibov transformations in terms of positive and negative frequency components of the creation and annihilation operators. Initially, the inclusion of the {\it Non-Bunch Davies} initial vacuum in the various theoretical frameworks was considered to incorporate the effect of interacting vacuum in different types of computations. Out of various possibilities, $\alpha$- vacua which describes the squeezed vacuum states is very well studied within the framework of Quantum Field Theory of de Sitter space using which the key concepts of the primordial cosmology are established. However, $\alpha$-vacua was introduced in a completely phenomenological fashion in the previous literature and strong physical ground was absent to explain the proper theoretical origin of such concepts to describe the {\it Non-Bunch Davies} initial vacuum. However, the underlying physical connection between the local signed coordinate transformation and the Bogolibov transformation clearly helps to justify the appearance of the {\it Non-Bunch Davies} initial vacuum in the presence of non-trivial curved quasi de Sitter background geometry. It would be interesting if a detailed formal and rigorous derivation of the statement is found and we would certainly keep it in our mind for our future investigations \footnote{We are thankful to the referee for suggesting this important fact.}.} 
    
It is also important to note that the scalar modes satisfy the following normalization condition:
 \bea   \left(v_{\bf k}(\tau),v^{'}_{\bf k}(\tau)\right)_{\bf KG}:=\begin{vmatrix}
     v_{\bf k}(\tau) & v^{*}_{\bf k}(\tau)\\ 
     v^{'}_{\bf k}(\tau) & v^{'*}_{\bf k}(\tau) 
\end{vmatrix}=\bigg(v^{'*}_{\bf k}(\tau)v_{\bf k}(\tau)-v^{'}_{\bf k}(\tau)v^{*}_{\bf k}(\tau)\bigg)=i,\eea
which is referred to as the Klein-Gordon product and can get translated further in terms of the following crucial constraint on the Bogoliubov coefficients describing the {\it Non-Bunch Davies} initial condition:
  \bea \bigg(|{\alpha}^{(s)}_{\bf k}|^2-|{\beta}^{(s)}_{\bf k}|^2\bigg)=1.\eea

Analyzing the derived solutions of the rescaled field and associated canonically conjugate momentum and deriving some physically significant information is challenging. We will therefore take into account the asymptotic limits in the resulting solutions, which will be very beneficial for our research at various cosmic scales.  To determine how Hankel functions of the first and second kind of order $\nu_{s}$ behave, we use the asymptotic limits $-k\tau\rightarrow 0$ and $-k\tau\rightarrow\infty$. After applying the said limits, we obtain the following expressions: 
\bea &&\lim_{k\tau\rightarrow -\infty} H^{(1)}_{\nu_{s}}(-k\tau)=\sqrt{\frac{2}{\pi}}\frac{1}{\sqrt{-k\tau}}\exp(-ik\tau)\exp\left(-\frac{i\pi}{2}\left(\nu_{s}+\frac{1}{2}\right)\right)=-\lim_{k\tau\rightarrow -\infty} H^{(2)}_{\nu_{s}}(-k\tau),\\
 &&\lim_{k\tau\rightarrow 0} H^{(1)}_{\nu_{s}}(-k\tau)=\frac{i}{\pi}\Gamma(\nu_{s})\left(-\frac{k\tau}{2}\right)^{-\nu_{s}}=-\lim_{k\tau\rightarrow 0} H^{(2)}_{\nu_{s}}(-k\tau),\eea 
Further, imposing the above-mentioned asymptotic limits applicable to the super-horizon ($-k\tau\ll 1$) and sub-horizon  ($-k\tau\gg 1$) the expressions for rescaled field variable can be obtained by the following expressions:
 \bea \lim_{k\tau\rightarrow 0}v_{\bf k}(\tau)&=&\sqrt{\frac{2}{k}}\frac{i}{\pi}\Gamma(\nu_{s})\left(-\frac{k\tau}{2}\right)^{\frac{1}{2}-\nu_{s}}\left({\alpha}^{(s)}_{\bf k}-{\beta}^{(s)}_{\bf k}\right),\\
 \lim_{k\tau\rightarrow -\infty}v_{\bf k}(\tau)&=&\sqrt{\frac{2}{\pi k}}\left[{\alpha}^{(s)}_{\bf k}~\exp\left(-i\left\{k\tau+\frac{\pi}{2}\left(\nu_{s}+\frac{1}{2}\right)\right\}\right)-{\beta}^{(s)}_{\bf k}~\exp\left(i\left\{k\tau+\frac{\pi}{2}\left(\nu_{s}+\frac{1}{2}\right)\right\}\right)\right]. ~~~~~~~~\eea
Utilizing this fact the most general solution for rescaled field variable for any general {\it Non-Bunch Davies} initial condition can be expressed as: 
\bea&& v_{\bf k}(\tau)=-\frac{1}{\sqrt{2k}}2^{\nu_{s}-\frac{3}{2}}(-k\tau)^{\frac{3}{2}-\nu_{s}}\left|\frac{\Gamma(\nu_{s})}{\Gamma\left(\frac{3}{2}\right)}\right|\times\left[{\alpha}^{(s)}_{\bf k}~\left(1-\frac{i}{k\tau}\right)~\exp\left(-i\left\{k\tau+\frac{\pi}{2}\left(\nu_{s}+\frac{1}{2}\right)\right\}\right)\right.\nonumber\\
&& \left.~~~~~~~~~~~~~~~~~~~~~~~~~~~~~~~~~~~~~~~~~~~~~~~~~~~~~~~~~~~~~~~+{\beta}^{(s)}_{\bf k}~\left(1+\frac{i}{k\tau}\right)~\exp\left(i\left\{k\tau+\frac{\pi}{2}\left(\nu_{s}+\frac{1}{2}\right)\right\}\right)\right].~~~~~~~~~
\eea
Using the above-mentioned expression for the rescaled scalar modes finally the expression for the comoving curvature perturbation in {\bf gauge B} is given by the following simplified expression:
\begin{mynamedbox1}{Solution for scalar modes in gauge B:}
\vspace{-.35cm}
 \bea\zeta_{\bf k}(\tau)&=&\frac{v_{\bf k}(\tau)}{zM_{\rm pl}}\nonumber\\
&=&-\frac{iH}{2\sqrt{\epsilon}M_{\rm pl}}\frac{2^{\nu_{s}-\frac{3}{2}}(-k\tau)^{\frac{3}{2}-\nu_{s}}}{k^{3/2}}\left|\frac{\Gamma(\nu_{s})}{\Gamma\left(\frac{3}{2}\right)}\right|\times\left[{\alpha}^{(s)}_{\bf k}~\left(1+ik\tau\right)~\exp\left(-i\left\{k\tau+\frac{\pi}{2}\left(\nu_{s}+\frac{1}{2}\right)\right\}\right)\right.\nonumber\\
&& \left.~~~~~~~~~~~~~~~~~~~~~~~~~~~~~~~~~~~~~~~~~~~~~~~~~~~-{\beta}^{(s)}_{\bf k}~\left(1-ik\tau\right)~\exp\left(i\left\{k\tau+\frac{\pi}{2}\left(\nu_{s}+\frac{1}{2}\right)\right\}\right)\right],~~~~~~~~~
\eea                            
  \end{mynamedbox1} 

which represents the most general solution of the {\it MS equation} for scalar modes in the presence of any general {\it Non-Bunch Davies} initial condition.

Now we require the quantization of the comoving scalar curvature perturbation to compute the expression for the power spectrum. For this purpose, we use previously defined {\it Non-Bunch Davies} quantum vacuum state. The canonical quantization of the scalar mode and its associated conjugate momenta should satisfy the necessary equal-time commutation relation:
\bea \left[\hat{\zeta}_{\bf k}(\tau),\hat{\zeta}^{'}_{{\bf k}^{'}}(\tau)\right]=i\;\delta^{3}\left({\bf k}+{\bf k}^{'}\right)\quad \quad\quad{\rm where}\quad\quad \hat{\zeta}_{\bf k}(\tau)&=&\bigg[{\zeta}_{\bf k}(\tau)\hat{C}_{\bf k}+{\zeta}^{*}_{\bf k}(\tau)\hat{C}^{\dagger}_{-{\bf k}}\bigg],\eea 
using which one can get the following commutation relations between the creation and annihilation operators of the {\it Non-Bunch Davies} quantum vacuum state for the scalar modes:
\bea \left[\hat{C}_{\bf k},\hat{C}^{\dagger}_{{\bf k}^{'}}\right]&=&(2\pi)^{3}\;\delta^{3}\left({\bf k}+{\bf k}^{'}\right),\quad
 \left[\hat{C}_{\bf k},\hat{C}_{{\bf k}^{'}}\right]=0,\quad
  \left[\hat{C}^{\dagger}_{\bf k},\hat{C}^{\dagger}_{{\bf k}^{'}}\right]=0.\eea
Then the two-point correlation function for the comoving curvature perturbation can be expressed as follows,
\bea \langle \hat{\zeta}_{\bf k}\hat{\zeta}_{{\bf k}^{'}}\rangle &=&(2\pi)^{3}\;\delta^{3}\left({\bf k}+{\bf k}^{'}\right)P_{\zeta}(k) \quad\quad{\rm where}\quad\quad P_{\zeta}(k)=|\zeta_{\bf k}(\tau)|^{2},\eea
and the associated dimensionless power spectrum can be computed as:
\bea \Delta^{2}_{\zeta}(k)&=&\frac{k^{3}}{2\pi^{2}}P_{\zeta}(k)\nonumber\\
&=&\frac{H^2}{8\pi^2M^2_{\rm pl}\epsilon}2^{2\nu_{s}-3}(-k\tau)^{3-2\nu_{s}}\left|\frac{\Gamma(\nu_{s})}{\Gamma\left(\frac{3}{2}\right)}\right|^2\times\Bigg|{\alpha}^{(s)}_{\bf k}~\left(1+ik\tau\right)~\exp\left(-i\left\{k\tau+\frac{\pi}{2}\left(\nu_{s}+\frac{1}{2}\right)\right\}\right)\nonumber\\
&& ~~~~~~~~~~~~~~~~~~~~~~~~~~~~~~~~~~~~~~~~~~~~~~~~~~~~~~~~~~~~~~~-{\beta}^{(s)}_{\bf k}~\left(1-ik\tau\right)~\exp\left(i\left\{k\tau+\frac{\pi}{2}\left(\nu_{s}+\frac{1}{2}\right)\right\}\right)\Bigg|^2.\quad\quad\eea
Finally, for super-horizon scales ($-k\tau\ll 1$), the dimensionless power spectrum can be further simplified as:
\begin{mynamedbox1}{Scalar power spectrum in super-horizon scale:}
\vspace{-.35cm}
\bea \Delta^{2}_{\zeta}(k)
=\frac{H^2}{8\pi^2M^2_{\rm pl}\epsilon}2^{2\nu_{s}-3}(-k\tau)^{3-2\nu_{s}}\left|\frac{\Gamma(\nu_{s})}{\Gamma\left(\frac{3}{2}\right)}\right|^2\times\Bigg|{\alpha}^{(s)}_{\bf k}~\exp\left(-\frac{i\pi}{2}\left(\nu_{s}+\frac{1}{2}\right)\right)-{\beta}^{(s)}_{\bf k}~\exp\left(\frac{i\pi}{2}\left(\nu_{s}+\frac{1}{2}\right)\right)\Bigg|^2.\quad\eea
\end{mynamedbox1}
At horizon exit, $-k\tau=1$ condition is imposed in the above expression which is used for further estimation of the amplitude and connection with the constraints from observations.

\section{Computation of tensor power spectrum with Non-Bunch Davies initial condition}
\label{sec5}

In this section, our prime objective will be to study the theoretical and observational impacts of the tensor power spectrum. To serve this purpose, further expanding the representative Einstein-Hilbert classical gravitational action up to the second order gives the following simplified result:
\begin{mynamedbox1}{Second order perturbed action for tensor modes:}
\vspace{-.35cm}
 \bea S^{(t)}_{(2)}=\frac{M^2_{\rm pl}}{8}\int d\tau~d^3x~a^2(\tau)~\left[(h^{'}_{ij}({\bf x},\tau))^2-\left(\nabla h_{ij}({\bf x},\tau)\right)^2\right],\eea
  \end{mynamedbox1} 

Further, we consider the following rescaling in tensor mode perturbation field variable which will be useful for the further computational purpose:
\bea h_{ij}({\bf x},\tau):&=&\sum_{\lambda=+,\times}h^{(\lambda)}({\bf x},\tau)~e^{(\lambda)}_{ij}({\bf x})\nonumber\\
&=&\frac{2}{M_{\rm pl}}\sum_{\lambda=+,\times}\frac{f_{\lambda}({\bf x},\tau)}{a(\tau)}~e^{(\lambda)}_{ij}({\bf x})\nonumber\\
&=&\frac{1}{a(\tau)}\frac{2}{M_{\rm pl}}\frac{1}{\sqrt{2}}\begin{pmatrix}
~f_{(+)}({\bf x},\tau))~ & ~f_{(\times)}({\bf x},\tau))~ &~ 0~\\
~f_{(\times)}({\bf x},\tau))~ & ~-f_{(+)}({\bf x},\tau))~ &~ 0~\\
0 & 0 & 0\
\end{pmatrix}\quad\quad\quad{\rm where}\quad\quad a(\tau)=-\frac{1}{H\tau},\eea
where, $e^{(\lambda)}_{ij}({\bf x})$ represents polarization tensor for the tensor modes for two helicities, $\lambda=+,\times$, respectively.

Using the above-mentioned field redefinition the second-order perturbed action for the tensor modes can be written as:
\bea S^{(t)}_{(2)}=\frac{1}{2}\sum_{\lambda=+,\times}\int d\tau~d^3x~\left[(f^{'}_{\lambda}({\bf x},\tau)))^2-\left(\nabla f_{\lambda}({\bf x},\tau))\right)^2+\frac{a^{''}(\tau)}{a(\tau)}\left(f_{\lambda}({\bf x},\tau))\right)^2\right].\eea
Now, making use of the following Fourier transform ansatz which helps to convert the above-mentioned second order action for the tensor modes in the momentum space:
\bea f_{\lambda}(\tau,{\bf x}):=\int \frac{d^3{\bf k}}{(2\pi)^3}~f_{\lambda,{\bf k}}(\tau)~e^{i{\bf k}.{\bf x}}~~~~~\forall ~~\lambda=+,\times.\eea
Using this Fourier transform ansatz the previously mentioned second-order action written in terms of the tensor modes of rescaled perturbation can be recast in the given simplified form:
\bea S^{(t)}_{(2)}=\frac{1}{2}\sum_{\lambda=+,\times}\int d\tau~d^3{\bf k}~\left[|f^{'}_{\lambda,{\bf k}}(\tau)|^2-\left(k^2-\frac{a^{''}(\tau)}{a(\tau)}\right)\left|f_{\lambda,{\bf k}}(\tau)\right|^2\right],\eea
   This action mimics the role of the action of an oscillator having a conformal time-dependent effective frequency in the present context of the discussion. Varying the action with respect to the rescaled field variable written in Fourier space the MS equation for the tensor mode can be written as:
   \begin{mynamedbox1}{Mukhanov Sasaki equation for tensor modes:}
\vspace{-.35cm}
 \bea f^{''}_{\lambda,{\bf k}}(\tau)+\left(k^2-\frac{a^{''}(\tau)}{a(\tau)}\right)f_{\lambda,{\bf k}}(\tau)=0.\eea
  \end{mynamedbox1} 
   For the further computational purpose, it becomes important to remember that here we have obtained an effective, conformal time-dependent, mass which can further be expressed in the following simplified language:
   \bea m^2_{\rm eff,(t)}(\tau)=-\frac{a^{''}(\tau)}{a(\tau)}=-\frac{2}{\tau^2}.\eea
   Here the subscript $t$ corresponds to the contribution computed for the tensor modes. Utilizing these facts, the most general solution of the second-order MS equation can be written using the following expression:
 \bea f_{\lambda,{\bf k}}(\tau)=\sqrt{-\tau}\left[{\alpha}^{(t)}_{{\bf k},\lambda}~H^{(1)}_{3/2}(-k\tau)+{\beta}^{(t)}_{{\bf k},\lambda}~H^{(2)}_{3/2}(-k\tau)\right]~~~~~\forall ~~\lambda=+,\times.\eea
   In this solution, $H^{(1)}_{3/2}(-k\tau)$ and $H^{(2)}_{3/2}(-k\tau)$, represent the Hankel function of the first and second kind. Here, ${\alpha}^{(t)}_{{\bf k},\lambda}=\left({\alpha}^{(t)}_{{\bf k},+},{\alpha}^{(t)}_{{\bf k},\times})\right)$ and ${\beta}^{(t)}_{{\bf k},\lambda}=\left({\beta}^{(s)}_{{\bf k},+},{\beta}^{(t)}_{{\bf k},\times}\right)$, are the two sets of Bogoliubov coefficients describing the two helicity dependent sectors of tensor modes which can be fixed by the proper choice of the quantum initial condition during the slow-roll phase of the inflation. In general, one can consider an arbitrary initial condition described in terms of a general quantum vacuum state which is characterized by the following equation: 
    \begin{mynamedbox1}{Definition of Non-Bunch Davies state for tensor modes:}
\vspace{-.35cm}
   \bea \hat{D}_{{\bf k},\lambda}|{\alpha}^{(t)}_{{\bf k},\lambda},{\beta}^{(t)}_{{\bf k},\lambda}\rangle_{\bf NBD}=0~\forall~{\bf k},\quad\quad{\rm where}\quad\quad \lambda=+,\times,\eea
  \end{mynamedbox1} 
 
    where $\hat{D}_{{\bf k},\lambda}$ represents the corresponding annihilation operators for the two helicity-dependent sectors of the tensor modes.
    
    Further, it is important to mention that in the large parameter space of ${\bf SO(1,4)}$ de Sitter isometry group {\it Bunch Davies} initial condition for the tensor modes is described by a single point for each helicity, which is characterized by the fixing Bogoliubov coefficients as, ${\alpha}^{(t)}_{{\bf k},\lambda}=1$ i.e $\left({\alpha}^{(t)}_{{\bf k},+},{\alpha}^{(t)}_{{\bf k},\times})\right)=(1,1)$ and ${\beta}^{(t)}_{{\bf k},\lambda}=0$ i.e. $\left({\beta}^{(s)}_{{\bf k},+},{\beta}^{(t)}_{{\bf k},\times}\right)=(0,0)$. However if one is interested to utilize the effect of a general quantum initial condition one should have an infinite number of possibilities where these coefficients are arbitrary functions of the characteristic momentum scale as appearing in both helicity-dependent sectors of the tensor modes and often referred to as {\it Non-Bunch Davies} initial condition. Here any general {\it Non-Bunch Davies} initial quantum states can be expressed in terms of the well-known {\it Bunch Davies} initial quantum state and such Bogoliubov transformation is described by the following equation:
    \begin{mynamedbox1}{Bogoliubov transformation  for tensor modes:}
\vspace{-.35cm}
  \bea  |{\alpha}^{(t)}_{{\bf k},\lambda},{\beta}^{(t)}_{{\bf k},\lambda}\rangle_{\bf NBD}&=&\prod_{\bf k}\frac{1}{\sqrt{|{\alpha}^{(t)}_{{\bf k},\lambda}|}}\exp\left[\frac{{\beta}^{(t),*}_{{\bf k},\lambda}}{2{\alpha}^{(t),*}_{{\bf k},\lambda}}~\hat{D}^{\dagger}_{{\bf k},\lambda}\hat{D}^{\dagger}_{-{{\bf k},\lambda}}\right]|1,0\rangle^{(t)}_{\bf BD}\nonumber\\
       &=&\prod_{\bf k}\frac{1}{{\cal N}^{(t)}_{{\bf NBD},{\bf k},\lambda}}~\exp\left[\frac{{\beta}^{(t),*}_{{\bf k},\lambda}}{2{\alpha}^{(t),*}_{{\bf k},\lambda}}~\hat{D}^{\dagger}_{{\bf k},\lambda}\hat{D}^{\dagger}_{-{{\bf k},\lambda}}\right]|1,0\rangle^{(t)}_{\bf BD}\quad\quad\quad\forall \lambda=+,\times,~~~~~~~~\eea      
  \end{mynamedbox1} 
                            
        where the overall normalization factor of the defined {\it Non-Bunch Davies} initial quantum states for the two helicity-dependent tensor modes is described using the following equation:
        \bea {\cal N}^{(t)}_{{\bf NBD},{\bf k},\lambda}:=\sqrt{|{\alpha}^{(t)}_{{\bf k},\lambda}|}.\eea 
        On top of this, it is important to note that in the present description, the {\it Bunch Davies} initial quantum state for the tensor modes is often referred to as, $|1,0\rangle^{(t)}_{\bf BD}=|0\rangle$. The similar constraints which we have discussed in the case of scalar modes holds good in the case of describing tensor modes with two helicities as well.

Further utilizing the asymptotic properties of the Hankel function in the sub-horizon and super-horizon region, as discussed for scalar modes in the previous section, the most general solution for the rescaled field variable for the tensor perturbation modes for any general {\it Non-Bunch Davies} initial condition can be expressed as: 
\bea&& f_{\lambda,{\bf k}}(\tau)=\frac{1}{\sqrt{2k}}\left[{\alpha}^{(t)}_{{\bf k},\lambda}~\left(1-\frac{i}{k\tau}\right)~\exp\left(-ik\tau\right)+{\beta}^{(t)}_{{\bf k},\lambda}~\left(1+\frac{i}{k\tau}\right)~\exp\left(ik\tau\right)\right]\quad\quad\quad\forall \lambda=+,\times.~~~~~~~~~
\eea
These tensor modes satisfy the following necessary normalization condition:
 \bea   \sum_{\lambda=+,\times}\left(f_{\lambda,{\bf k}}(\tau),f^{'}_{\lambda,{\bf k}}(\tau)\right)_{\bf KG}:=\sum_{\lambda=+,\times}\begin{vmatrix}
     f_{\lambda,{\bf k}}(\tau) & f^{*}_{\lambda,{\bf k}}(\tau)\\ 
     f^{'}_{\lambda,{\bf k}}(\tau) & f^{'*}_{\lambda,{\bf k}}(\tau) 
\end{vmatrix}=\sum_{\lambda=+,\times}\bigg(f^{'*}_{\lambda,{\bf k}}(\tau) f_{\lambda,{\bf k}}(\tau)-f^{'}_{\lambda,{\bf k}}(\tau)f^{*}_{\lambda,{\bf k}}(\tau)\bigg)=i,\eea
which is referred to as the Klein-Gordon product and this further translates into the following crucial constraint on the Bogoliubov coefficients describing the {\it Non-Bunch Davies} initial condition:
  \bea \sum_{\lambda=+,\times}\bigg(|{\alpha}^{(t)}_{{\bf k},\lambda}|^2-|{\beta}^{(t)}_{{\bf k},\lambda}|^2\bigg)=1.\eea

Using the above-mentioned expression for the rescaled field variable finally, the expression for the tensor mode for two helicities is given by the following combined format:
 \begin{mynamedbox1}{Solution for tensor modes:}
\vspace{-.35cm}
  \bea h_{\lambda,{\bf k}}(\tau)&=&\frac{f_{\lambda,{\bf k}}(\tau)}{aM_{\rm pl}}\nonumber\\
&=&\frac{iH}{\sqrt{2}}\frac{2}{M_{\rm pl}}\frac{1}{k^{3/2}}\left[{\alpha}^{(t)}_{{\bf k},\lambda}~\left(1+ik\tau\right)~\exp\left(-ik\tau\right)-{\beta}^{(t)}_{{\bf k},\lambda}~\left(1-ik\tau\right)~\exp\left(ik\tau\right)\right],~~~~~~~~~
\eea    
  \end{mynamedbox1} 

which represents the most general solution of the {\it MS equation} for scalar modes in the presence of any general {\it Non-Bunch Davies} initial condition.

Now we need to quantize the tensor perturbation to compute the expression for the power spectrum. For this purpose, we use previously defined {\it Non-Bunch Davies} quantum vacuum state. The canonical quantization between the tensor mode and its associated conjugate momenta has to satisfy the following equal-time commutation relation:
\bea \left[\hat{h}_{\lambda,{\bf k}}(\tau),\hat{h}^{'}_{\lambda^{'},{\bf k}^{'}}(\tau)\right]=i\;\delta_{\lambda\lambda^{'}}\delta^{3}\left({\bf k}+{\bf k}^{'}\right)\quad \quad{\rm where}\quad\quad \hat{h}_{\lambda,{\bf k}}(\tau)&=&\bigg[h_{\lambda,{\bf k}}(\tau)\hat{D}_{\lambda,{\bf k}}+h^{*}_{\lambda,{\bf k}}(\tau)\hat{D}^{\dagger}_{\lambda,-{\bf k}}\bigg]\quad\forall\lambda=+,\times,\quad\eea 
using which one can get the following commutation relations between the creation and annihilation operators of the {\it Non-Bunch Davies} quantum vacuum state for the tensor modes:
\bea \left[\hat{D}_{\lambda,{\bf k}},\hat{D}^{\dagger}_{\lambda^{'},{\bf k}^{'}}\right]&=&(2\pi)^{3}\;\delta_{\lambda\lambda^{'}}\delta^{3}\left({\bf k}+{\bf k}^{'}\right),\quad
 \left[\hat{D}_{\lambda,{\bf k}},\hat{D}_{\lambda^{'},{\bf k}^{'}}\right]=0,\quad
  \left[\hat{D}^{\dagger}_{\lambda,{\bf k}},\hat{D}^{\dagger}_{\lambda^{'},{\bf k}^{'}}\right]=0\quad\forall\lambda,\lambda^{'}=+,\times.\eea
  
  Then the two-point correlation function for the tensor perturbation can be expressed as follows,
\bea \langle \hat{h}_{\lambda,{\bf k}}\hat{h}_{\lambda^{'},{\bf k}^{'}}\rangle &=&(2\pi)^{3}\;\delta_{\lambda\lambda^{'}}\delta^{3}\left({\bf k}+{\bf k}^{'}\right)P_{h}(k) \quad\quad{\rm where}\quad\quad P_{h}(k)=\sum_{\lambda=+,\times}|h_{\lambda,{\bf k}}(\tau)|^{2},\eea
and the associated dimensionless power spectrum can be computed as:
\bea \Delta^{2}_{t}(k)&=&\frac{k^{3}}{2\pi^{2}}\times 4P_{h}(k)\nonumber\\
&=&\frac{2H^2}{\pi^2M^2_{\rm pl}}\times\sum_{\lambda=+,\times}\Bigg|{\alpha}^{(t)}_{{\bf k},\lambda}~\left(1+ik\tau\right)~\exp\left(-ik\tau\right)-{\beta}^{(t)}_{{\bf k},\lambda}~\left(1-ik\tau\right)~\exp\left(ik\tau\right)\Bigg|^2.\quad\quad\eea
Finally, for super-horizon scales ($-k\tau\ll 1$), the dimensionless power spectrum can be further simplified as:
\begin{mynamedbox1}{Tensor power spectrum in super-horizon scale:}
\vspace{-.35cm}
\bea \Delta^{2}_{t}(k)
=\frac{2H^2}{\pi^2M^2_{\rm pl}}\times\sum_{\lambda=+,\times}\Bigg|{\alpha}^{(t)}_{{\bf k},\lambda}-{\beta}^{(t)}_{{\bf k},\lambda}\Bigg|^2.\quad\eea
\end{mynamedbox1}
At horizon exit, $-k\tau=1$ condition is imposed in the above expression which is used for further estimation of the amplitude and connection with the constraints from observations.

\section{Inflationary observables from inflation in presence Non-Bunch Davies initial condition}
\label{sec6}

In this section, our prime objective is to connect the derived results for {\it Non-Bunch Davies} initial condition to describe the inflationary paradigm with the recent findings of the NANOGrav 15-year and the EPTA Data Set. Most importantly, in this section, we will provide a quintessential explanation of accommodating the blue-tilted tensor power spectrum within the context of canonical single-field slow-roll models of inflation in a model-independent fashion. We will write down the expressions for each observable and some other important quantities using {\it Non-Bunch Davies} initial condition in the super-horizon region which we believe will be extremely helpful to establish our key point in the rest half of this paper:
\begin{enumerate}
    \item \underline{\bf Scalar power spectrum:}\\
    The previously derived expression for the dimensionless scalar power spectrum in the presence of {\it Non-Bunch Davies} initial condition, in the super-horizon region, can be written in terms of a simplified formula as:
    \begin{mynamedbox1}{Scalar power spectrum in super-horizon scale:}
\vspace{-.35cm}
 \bea \Delta^{2}_{\zeta}(k)
=\Delta^{2}_{\zeta}(k_*)\left(\frac{k}{k_*}\right)^{n_s-1}.\quad\eea
\end{mynamedbox1}
where at the pivot scale $k=k_*=0.05{\rm Mpc}^{-1}$, the amplitude of the scalar power spectrum is given by the following expression:
 \begin{mynamedbox1}{Scalar power spectrum at the pivot scale:}
\vspace{-.35cm}
  \bea \Delta^{2}_{\zeta}(k_*)
=\left(\frac{H^2}{8\pi^2M^2_{\rm pl}\epsilon}\right)_*2^{2\nu_{s}-3}\left|\frac{\Gamma(\nu_{s})}{\Gamma\left(\frac{3}{2}\right)}\right|^2\times\Bigg|{\alpha}^{(s)}_{{\bf k}_*}~\exp\left(-\frac{i\pi}{2}\left(\nu_{s}+\frac{1}{2}\right)\right)-{\beta}^{(s)}_{{\bf k}_*}~\exp\left(\frac{i\pi}{2}\left(\nu_{s}+\frac{1}{2}\right)\right)\Bigg|^2.\quad\eea
\end{mynamedbox1}

\item \underline{\bf Scalar spectral tilt:}\\
The scalar spectral tilt from the aforementioned form of the scalar power spectrum is computed to be as:
\begin{mynamedbox1}{Scalar spectral tilt at the pivot scale:}
\vspace{-.35cm}
  \bea n_s-1&=&\left(\frac{d\ln \Delta^{2}_{\zeta}(k)}{d\ln k}\right)_{k=k_*}\nonumber\\
  &=&3-2\nu_s+\Bigg(\frac{d}{d\ln k}\Bigg|{\alpha}^{(s)}_{{\bf k}}~\exp\left(-\frac{i\pi}{2}\left(\nu_{s}+\frac{1}{2}\right)\right)-{\beta}^{(s)}_{{\bf k}}~\exp\left(\frac{i\pi}{2}\left(\nu_{s}+\frac{1}{2}\right)\right)\Bigg|^2\Bigg)_{k=k_*}\nonumber\\
&\approx&\eta-3\epsilon\approx 2\eta_V-6\epsilon_V.\eea
\end{mynamedbox1}
Here we fix the {\it Non-Bunch Davies} initial condition for the scalar counterpart of the Bogoliubov coefficients in such a way that the following constraint condition hast to be obeyed throughout our analysis:
\bea \Bigg(\frac{d}{d\ln k}\Bigg|{\alpha}^{(s)}_{{\bf k}}~\exp\left(-\frac{i\pi}{2}\left(\nu_{s}+\frac{1}{2}\right)\right)-{\beta}^{(s)}_{{\bf k}}~\exp\left(\frac{i\pi}{2}\left(\nu_{s}+\frac{1}{2}\right)\right)\Bigg|^2\Bigg)_{k=k_*}\ll 1,\eea
which means that the spectral tilt for scalar perturbation is not very sensitive to the choices of the respective Bogoliubov coefficients used to describe them. 
Here, first of all, it is important to note that both the slow-roll parameters $\epsilon$ and $\eta$ are evaluated at the pivot scale and to avoid any cumbersome notation we will not be using the $*$ notation for the these parameters. We must also mention the important fact that the following approximations are applicable in the slow-roll regime of inflation which helps us to write everything using the effective potential during inflation and its derivatives:
\bea \epsilon\approx \epsilon_V, \quad \eta\approx 2\eta_V-3\epsilon_V\quad\quad \epsilon_V=\frac{M^2_{\rm pl}}{2}\left(\frac{V^{'}}{V}\right)^2,\quad{\rm and}\quad\eta_V=M^2_{\rm pl}\left(\frac{V^{''}}{V}\right).\eea
Here the $'$ is used to describe the derivative with respect to the inflaton field.

\item \underline{\bf Tensor power spectrum:}\\
    The previously derived expression for the dimensionless tensor power spectrum in the presence of {\it Non-Bunch Davies} initial condition in the super-horizon region can be expressed in terms of the following simplified formula:
    \begin{mynamedbox1}{Tensor power spectrum in super-horizon scale:}
\vspace{-.35cm}
  \bea \Delta^{2}_{t}(k)
=\Delta^{2}_{t}(k_*)\left(\frac{k}{k_*}\right)^{n_t}.\quad\eea
\end{mynamedbox1}
where at the pivot scale $k=k_*=0.05{\rm Mpc}^{-1}$, the amplitude of the tensor power spectrum is given by the following expression:
    \begin{mynamedbox1}{Tensor power spectrum at the pivot  scale:}
\vspace{-.35cm}
  \bea \Delta^{2}_{t}(k_*)=\left(\frac{2H^2}{\pi^2M^2_{\rm pl}}\right)_*\times\sum_{\lambda=+,\times}\Bigg|{\alpha}^{(t)}_{{\bf k_*},\lambda}-{\beta}^{(t)}_{{\bf k_*},\lambda}\Bigg|^2.\quad\quad\eea
\end{mynamedbox1}

\item \underline{\bf Tensor spectral tilt:}\\
The tensor spectral tilt from the derived above-mentioned form of the tensor power spectrum is computed as:
\begin{mynamedbox1}{Tensor spectral tilt at the pivot scale:}
\vspace{-.35cm}
  \bea\label{nt} n_t&=&\left(\frac{d\ln \Delta^{2}_{t}(k)}{d\ln k}\right)_{k=k_*}=-2\epsilon+{\cal G}\approx -2\epsilon_V+{\cal G}.\eea
\end{mynamedbox1}
Here we fix the {\it Non-Bunch Davies} initial condition for the tensor counterpart of Bogoliubov coefficients in such a way that the following constraint condition hast to be obeyed throughout our analysis:
\bea \underbrace{{\cal G}:=\Bigg(\frac{d}{d\ln k}\sum_{\lambda=+,\times}\Bigg|{\alpha}^{(t)}_{{\bf k},\lambda}-{\beta}^{(t)}_{{\bf k},\lambda}\Bigg|^2\Bigg)_{k=k_*}}_{\sim {\cal O}(1)}\gg 2\epsilon_V,\eea
which means that the spectral tilt for tensor perturbation is extremely sensitive to the choices of respective Bogoliubov coefficients used to describe them. Also, this implies that the tensor spectral tilt is highly blue tilted 
as during the slow-roll phase of inflation, the slow-roll parameter $\epsilon$ is very small for almost all classes of inflationary effective potentials. One needs to always satisfy the above-mentioned constraint on the specific functional form of the Bogoliubov coefficients which describe the tensor perturbation. 

\textcolor{black}{Now, it is important to provide a warning to the general readers that the tensor spectral tilt described in equation (\ref{nt}) represents the first term in the following Taylor series expansion $\ln \Delta^{2}_{t}(k)$ in terms of $\ln k$, which is given by:}
\bea \ln \Delta^{2}_{t}(k)&=&\sum^{\infty}_{n=0}\frac{1}{n!}\left(\frac{d^n\ln \Delta^{2}_{t}(k)}{d\ln k^n}\right)_{k=k_*}\left(\ln\left(\frac{k}{k_*}\right)\right)^n\nonumber\\
&=&\ln \Delta^{2}_{t}(k_*)+\left(\frac{d\ln \Delta^{2}_{t}(k)}{d\ln k}\right)_{k=k_*}\ln\left(\frac{k}{k_*}\right)+\frac{1}{2!}\left(\frac{d^2\ln \Delta^{2}_{t}(k)}{d\ln k^2}\right)_{k=k_*}\ln^2\left(\frac{k}{k_*}\right)\nonumber\\
&&\quad\quad\quad\quad\quad\quad\quad\quad\quad\quad\quad\quad+\frac{1}{3!}\left(\frac{d^3\ln \Delta^{2}_{t}(k)}{d\ln k^3}\right)_{k=k_*}\ln^3\left(\frac{k}{k_*}\right)+\cdots\nonumber\\
&=&\ln \Delta^{2}_{t}(k_*)+n_t\ln\left(\frac{k}{k_*}\right)+\frac{1}{2}\alpha_t\ln^2\left(\frac{k}{k_*}\right)+\frac{1}{6}\beta_t\ln^3\left(\frac{k}{k_*}\right)+\cdots\eea
\textcolor{black}{which can be further simplified as:}
\bea \Delta^{2}_{t}(k)&=&\Delta^{2}_{t}(k_*)\left(\frac{k}{k_*}\right)^{n_t+\frac{1}{2}\alpha_t\ln\left(\frac{k}{k_*}\right)+\frac{1}{6}\beta_t\ln\left(\frac{k}{k_*}\right)+\cdots}.\eea
\textcolor{black}{In the above computation, we define:}
\bea n_t&=&\left(\frac{d\ln \Delta^{2}_{t}(k)}{d\ln k}\right)_{k=k_*}=n_t^{\bf BD} +{\cal G},\\
\alpha_t &=&\left(\frac{d^2\ln \Delta^{2}_{t}(k)}{d\ln k^2}\right)_{k=k_*}=\left(\frac{dn_{t}(k)}{d\ln k}\right)_{k=k_*}=\alpha_t^{\bf BD}+\left(\frac{d{\cal G}}{d\ln k}\right),\\
\beta_t &=&\left(\frac{d^3\ln \Delta^{2}_{t}(k)}{d\ln k^3}\right)_{k=k_*}=\left(\frac{d^2n_{t}(k)}{d\ln k^2}\right)_{k=k_*}=\beta_t^{\bf BD}+\left(\frac{d^2{\cal G}}{d\ln k^2}\right),\\
\eea
\textcolor{black}{where the {\it Bunch-Davies} counterpart satisfies the slow-roll limiting condition given by the following expression:}
\bea n_t^{\bf BD}&=&-2\epsilon_V,\\
    \alpha_t^{\bf BD}&=&\left(4\epsilon_V\eta_V-8\epsilon^2_V\right),\\ 
    \beta_t^{\bf BD}&=&\left(56\epsilon^2_V\eta_V-64\epsilon^3_V-8\eta^2_V\epsilon_V-4\epsilon_V\xi^2_V\right).\eea
 \textcolor{black}{Now, if the quantity ${\cal G}$ is of order unity or larger, that would call for including higher terms in the above-mentioned series expansion. The series can only be truncated iff its coefficients $n_t$, $\alpha_t$, $\beta_t$, etc are small, i.e. in the slow-roll limit for the Bunch Davis vacuum when it takes the values $n_t^{\bf BD}$, $\alpha_t^{\bf BD}$ and $\beta_t^{\bf BD}$ \footnote{\textcolor{black}{We are thankful to the Editor to suggesting us to explain this issue in this context. }}. The non-negligible values of the additional terms ${\cal G}$, $\left(\frac{d{\cal G}}{d\ln k}\right)$ and $\left(\frac{d^2{\cal G}}{d\ln k^2}\right)$ demand to include higher order terms and strictly pointing towards the violation of standard slow-roll consistency conditions, which is only valid in presence of {\it Bunch Davies} vacuum.
In this article, we utilize this violation of slow-roll condition in the presence of {\it Non-Bunch Davies} vacuum to achieve our final desired result established at the end of the paper for gravitational waves.}

Recently in the NANOGrav 15-year and the EPTA Data Set, it is reported that the tensor power spectrum is highly blue-tilted and described using the following suggested fitting formula:
\begin{mynamedbox1}{NANOGrav 15-year Data Set fitting formula for tensor spectral tilt:}
\vspace{-.35cm}
  \bea n_t=-0.14\log_{10}r+0.58,\eea
\end{mynamedbox1}
where $r$ is the tensor-to-scalar ratio which we will compute from our present prescription.
The above-suggested form clearly suggests the existence of the spectral blue tilt for the tensor spectrum.\textcolor{black}{ Further, this formula gives hints towards having a violation of the standard slow-roll condition, $n_t=-r/8$, which is only true for {\it Bunch Davies} vacuum. In this article, through our established methodology we have suggested that inclusion of the {\it Non-Bunch Davies} vacuum is one of the most promising options using which one can violate the standard slow-roll consistency condition.} We will discuss this fact in a more detailed fashion in the next section where we will justify that our findings strongly support the recently observed result from the NANOGrav 15-year and EPTA Data Set. \textcolor{black}{Additionally, we will provide a new consistency relationship in the next section which violates the standard slow-roll consistency condition in the presence of {\it Non-Bunch Davies} vacuum.}

\item \underline{\bf Tensor-to-scalar ratio:}\\
Further, in the super-horizon region, the tensor-to-scalar ratio in the presence of {\it Non-Bunch Davies} initial condition can be expressed as:
 \begin{mynamedbox1}{Tensor-to-scalar ratio in super-horizon scale:}
\vspace{-.35cm}
 \bea r(k)
=r(k_*)\left(\frac{k}{k_*}\right)^{n_t-n_s+1}.\quad\eea
\end{mynamedbox1}
where at the pivot scale $k=k_*=0.05{\rm Mpc}^{-1}$, the amplitude of the tensor-to-scalar ratio is given by the following expression:
 \begin{mynamedbox1}{Tensor-to-scalar ratio at the pivot scale:}
\vspace{-.35cm}
  \bea r(k_*)
=\frac{\Delta^{2}_{t}(k_*)}{\Delta^{2}_{\zeta}(k_*)}=16\epsilon_V{\cal F},\quad\eea
\end{mynamedbox1}
where in presence of {\it Non Bunch Davies} initial condition the modification factor ${\cal F}$ is defined as:
\bea {\cal F}:=2^{3-2\nu_{s}}\left|\frac{\Gamma\left(\frac{3}{2}\right)}{\Gamma(\nu_{s})}\right|^2 \times\frac{\displaystyle \sum_{\lambda=+,\times}\Bigg|{\alpha}^{(t)}_{{\bf k_*},\lambda}-{\beta}^{(t)}_{{\bf k_*},\lambda}\Bigg|^2}{\displaystyle\Bigg|{\alpha}^{(s)}_{{\bf k}_*}~\exp\left(-\frac{i\pi}{2}\left(\nu_{s}+\frac{1}{2}\right)\right)-{\beta}^{(s)}_{{\bf k}_*}~\exp\left(\frac{i\pi}{2}\left(\nu_{s}+\frac{1}{2}\right)\right)\Bigg|^2}.\eea
\end{enumerate}
We will now mention the immediate important outcomes from the inflationary paradigm in the presence of Non-
Bunch Davies initial condition.

\section{New consistency relation from Non-Bunch Davies initial condition}
\label{sec7}

Due to having changes in the amplitude and spectral tilt of the tensor perturbation in the presence of {\it Non-Bunch Davies} initial condition, it is expected to have a violation of old consistency relation, $r=-8n_t$, derived for the canonical single field slow-roll models of inflation in presence of {\it Bunch Davies} initial condition. Here we found that the new consistency relation between the amplitude and spectral tilt of the tensor perturbation is given by the following simplified formula:  
\begin{mynamedbox1}{New consistency relation:}
\vspace{-.35cm}
 \bea n_t=\bigg(-\frac{r}{8{\cal F}}+{\cal G}\bigg),\eea
\end{mynamedbox1}
where the modification factors ${\cal F}$ and ${\cal G}$ are already defined before in terms of the Bogoliubov coefficients of {\it Non-Bunch Davies} vacuum which describe the scalar and tensor perturbations respectively. In the next section, we will suggest the functional forms of both of the modification factors which can be able to explain the blue-titled tensor spectrum through Non-Bunch Davies initial condition in the light of PTA (NANOGrav 15-year and EPTA) Data Set. One important fact of the above relation is in the case of {\it Bunch Davies} initial condition where we have ${\cal F}=1$ and ${\cal G}=0$, which gives the old consistency relation and is only able to explain red-tilted tensor perturbation in the power spectrum. The PTA Data Set discarded this possibility and pointed towards having a blue-tilted tensor power spectrum which is almost impossible to explain via {\it Bunch Davies} initial condition. Though there exists a very small number of counterexamples using which one can produce a very small blue-titled tensor power spectrum, which is, a quintessential inflationary paradigm \cite{Hossain:2014xha,Hossain:2014coa,Hossain:2014ova,WaliHossain:2014usl,Geng:2015fla}, phantom models of inflation \cite{Piao:2004tq,Liu:2010dh}, alternatives to inflation, such as string gas cosmology \cite{Brandenberger:2008nx,Brandenberger:2011et}, ekpyrotic \cite{Khoury:2001wf,Lehners:2008vx} and bouncing scenarios \cite{Brandenberger:2016vhg,Brandenberger:2023wtd,Koehn:2015vvy,Lehners:2015mra,Ijjas:2018qbo,Bhargava:2020fhl} and many more in the list \cite{Agullo:2016tjh,Bojowald:2005epg,Bojowald:1999tr,Koshelev:2022olc,Koshelev:2020foq}. But the PTA Data Set is suggested to have large spectral blue tilt which cannot be explained by the above-mentioned examples without incorporating the details of reheating phenomenology. Recently in ref \cite{Vagnozzi:2023lwo} the author showed that large blue tilt favours very low reheating temperature, for example, the numerical analysis suggests that to have $n_t\sim 2$ one should have reheating temperature within the range $10{\rm GeV}<T_{\rm reh}<1{\rm TeV}$, which is again almost impossible to generate from the standard known models of canonical single field inflation. However, we strongly believe once the microphysics of reheating will be known, which is till date completely unknown to all of us, some of the ambiguities and problems associated with reheating phenomenology related to the present work can be addressed in a more crystal clear fashion. Some of the efforts have been made in refs. \cite{Choudhury:2020yaa,Choudhury:2021tuu,Choudhury:2018rjl,Choudhury:2018bcf} to address this mentioned issue, which we believe will be useful for the purpose of a better understanding of the general readers. In this work instead of talking about such possibilities, we have pointed out that the {\it Non-Bunch Davies} initial conditions are one of the best possibilities using which one can reliably explain the origin of the blue tilted tensor power spectrum as observed by the PTA (NANOGrav 15-year and EPTA) Data Set. Not only can {\it Non-Bunch Davies} initial conditions explain the blue-tilted tensor spectrum in the low-frequency domain, but also one can able to show using a bit of detailed numerical analysis that it is consistent with the constraints obtained from the high-frequency probes of the gravitation waves spectrum. In the next section, we are going to analyze all of these possibilities together in a detailed fashion. 

\section{Anti Lyth bound: New formula for field excursion using with Non-Bunch Davies initial condition}
\label{sec8}

Since the {\it Non-Bunch Davies} initial condition is the key ingredient of this paper then due to having modifications in the presence of different Bogoliubov coefficients which explain the scalar and tensor perturbations correctly the well-known field excursion formula for the inflaton field derived in the presence of the {\it Bunch Davies} initial condition, commonly known as the {\it Lyth bound}, will also be going to modify. Though we perform our analysis in a completely model-independent fashion, i.e. without using any specific structure of inflationary effective potentials in the slow-roll regime, the result derived in this section will be going to further constrain the behaviour of such effective potentials in the presence of the {\it Non-Bunch Davies} initial condition. In this connection, it is important to point out that the {\it Lyth bound} is the determining factor that tells whether inflation occurs in the super-Planckian or in the sub-Planckian regime. The old version of the {\it Lyth bound} always suggests having a super-Planckian inflationary paradigm, which is against the general notion of the basics of the Effective Field Theory (EFT) prescription. However, enormous efforts have been made in the past to evade such strong outcomes of {\it Lyth bound} which renders the EFT prescription trustworthy within the framework of inflationary paradigm governed by {\it Bunch Davies} initial condition. See refs. \cite{Choudhury:2015hvr,Choudhury:2015pqa,Choudhury:2014sua,Choudhury:2014kma,Choudhury:2013iaa,Hossain:2014ova} more details on this aspects. Here we are digging out this good old issue again in this paper again as we believe the {\it Non-Bunch Davies} initial condition is one of the promising possibilities using which one can able to evade the good old {\it Lyth bound} derived for {\it Bunch Davies} initial condition. This means that the existence of {\it Non-Bunch Davies} initial condition clearly validates the inflationary paradigm in a region where the basic understandings and general notions of EFT hold good perfectly. Not only validating the EFT framework but also our computation performed in this section is going to be extremely helpful to know how exactly the observed tensor blue tilt by the PTA (NANOGrav 15-year and EPTA) Data Set can be able to constrain the structure of the inflationary potential to validate EFT prescription in the present context of the discussion. Here we start our derivation by the following conversion formula which connects the underlying momentum scale with the inflaton field:
\bea \frac{d}{d\ln k}=-M_{\rm pl}\sqrt{2\epsilon_V}\frac{d}{d\phi},\eea
which is also used during the computation of running of the scalar and tensor spectral tilt of the primordial spectrum as discussed before in the earlier part of this paper. Using the above-mentioned conversion formula one can further write down the expression for the first slow-roll parameter $\epsilon_V$ in the following simplified form:
\bea \epsilon_V=\frac{1}{2M^2_{\rm pl}}\left(\frac{d\phi}{d\ln k}\right)^2.\eea
Now through the use of the expression mentioned above, in the super-horizon region one can go on to write down the expression for the tensor-to-scalar ratio as a function of momentum scale in the presence of {\it Non-Bunch Davies} initial condition:
\bea r(k)=\frac{8}{M^2_{\rm pl}}\left(\frac{d\phi}{d\ln k}\right)^2\left(-k\tau\right)^{2\nu_s-3}{\cal F}=r(k_*)\left(\frac{k}{k_*}\right)^{n_t-n_s+1},\eea
using which one can further compute the following expression for the field excursion formula:
\bea \frac{|\Delta\phi|}{M_{\rm pl}}&=&\Bigg|\int^{k_*}_{k_{\rm end}}\frac{dk}{k}\;\sqrt{\frac{r(k)}{8{\cal F}}}\left(-k\tau\right)^{\frac{3}{2}-\nu_s}\Bigg|\nonumber\\
&=&\frac{1}{k^{\frac{n_t-n_s+1}{2}}_*}\sqrt{\frac{r(k_*)}{8{\cal F}}}\left(-\tau\right)^{\frac{3}{2}-\nu_s}\Bigg|\int^{k_*}_{k_{\rm end}}dk\;k^{\frac{1}{2}-\nu_s+\frac{n_t-n_s+1}{2}}\Bigg|\nonumber\\
&=&\sqrt{\frac{r(k_*)}{8{\cal F}}}\frac{2}{\left(n_t+n_s-1\right)}\Bigg|\Bigg\{1-\left(\frac{k_{\rm end}}{k_*}\right)^{\frac{n_t+n_s-1}{2}}\Bigg\}\Bigg|,\eea
where the field excursion in terms of the inflaton field is defined as:
\bea |\Delta\phi|:=|\phi_{\rm end}-\phi_*|,\eea
where the field value at the end of inflation and at the horizon crossing scale is described by, $\phi_{\rm end}$ and $\phi_*$.

Now for the further simplification purpose we introduce the number of e-foldings $\Delta {\cal N}$ which in the present context helps us to make the connection with momentum-dependent scale and is described by the following expression:
 \bea \Delta {\cal N}=\left({\cal N}_{\rm *}-{\cal N}_{\rm end}\right)=\ln\left(\frac{k_*}{k_{\rm end}}\right).\eea
 Using the above-mentioned result, the new field excursion formula for the inflaton field  can be recast in the following form:
 \begin{mynamedbox1}{New field excursion formula: Anti Lyth bound}
\vspace{-.35cm}
 \bea \frac{|\Delta\phi|}{M_{\rm pl}}&=&\sqrt{\frac{r(k_*)}{8{\cal F}}}\frac{2}{\left(n_t+n_s-1\right)}\Bigg|\Bigg\{1-\exp\left(-\Delta {\cal N}\left(\frac{n_t+n_s-1}{2}\right)\right)\Bigg\}\Bigg|.\eea
\end{mynamedbox1}
For the blue-tilted tensor spectrum let us fix, $n_t\sim 2$ and for the red tilted scalar spectrum, we consider $n_s\sim 0.96$. Additionally, we fix the number of e-foldings at $\Delta{\cal N}\sim 60$. We also use the relation $r(k_*)=16\epsilon_V{\cal F}$, which further gives the following simplified result in terms of order of magnitude:
\bea \frac{|\Delta\phi|}{M_{\rm pl}}&=&1.44\times \sqrt{\epsilon_V}\sim 0.096 \ll 1,\eea
where in the last step during the simplification we use the constraint on the potential dependent first slow-roll parameter as, $\epsilon_V\sim 0.0044$, which is consistent with the Planck data \cite{Planck:2018jri}. This implies that blue tensor spectral tilt with {\it Non-Bunch Davies} initial condition describes sub-Plancking EFT description of the canonical single field inflation. It is then  expected from the above-mentioned estimation that if we push the blue tilt of the tensor power spectrum to the lower values, say $n_t\sim 0.22$ then we get $|\Delta\phi|\sim {\cal O}(1)M_{\rm pl}$ where the EFT prescription breaks and the super-Planckian physics start dominating over the sub-Planckian behaviour of the inflationary potential. This means that at, $n_t\leq 0.22$, EFT prescription breaks even if we consider a small blue tilt in the mentioned spectrum. This analysis also tells us that if we consider the red tilt in the tensor power spectrum as obtained from {\it Bunch Davies} initial condition, then the field excursion formula suggests that EFT description breaks and super-Planckian physics play a very significant role in the present computation. After doing this analysis we have found that the findings of {\it Non-Bunch Davies} initial condition are more appropriate than the result obtained from {\it Bunch Davies} initial condition as we always want that the EFT prescription holds good perfectly.

\section{Numerical results: Explanation of blue titled tensor spectrum through Non-Bunch Davies initial condition in the light of Pulsar Timing Array Data Set}
\label{sec9}

In this section, our prime objective is to explain the impact of the blue titled tensor spectrum through the Non-Bunch Davies initial condition in the light of the \textcolor{black}{Pulsar Timing Array (NANOGrav 15-year and EPTA)} Data Set which we believe is one of the promising possibilities to defend the canonical single-field slow roll inflationary paradigm. To explain the prescribed scenario we emphasize the violation of the consistency conditions as derived in the earlier section. In the upcoming discussion, we suggest some frequency-dependent and tensor spectral tilt-dependent specific features of the Bogoliubov coefficients describing the tensor sector of the perturbation, which will be going to fix some of the unknown quantities as appearing in the expression for the derived new consistency relation in the presence of {\it Non-Bunch Davies} initial condition. Before going into further details on the related discussion, below we will first discuss some of the basic details about the generation of stochastic gravitational waves and related constraints using the \textcolor{black}{Pulsar Timing Array (NANOGrav 15-year and EPTA)} Data Set.

Let us start with the following conversion formula which helps us to convert the underlying cosmological momentum scale in terms of the GW frequency:
\bea f=1.6\times 10^{-9}{\rm Hz}\times\left(\frac{k}{10^6{\rm Mpc}^{-1}}\right),\eea
where we must point that, $1{\rm nHz} \equiv 10^{-9}{\rm Hz}$. Using this conversion formula the amplitude of the tensor power spectrum can be recast in the following simplified formula:
\bea \Delta^{2}_{t}(f)
=\Delta^{2}_{t}\left(\frac{f}{f_*}\right)^{n_t}=r\Delta^{2}_{\zeta}\left(\frac{f}{f_*}\right)^{n_t}.\quad\eea
Using this relation in the region $k\gg k_{\rm eq}$, the spectral energy density of GW can be expressed as \cite{Zhao:2013bba}:
\begin{mynamedbox1}{Spectral density of GW from theory of inflation:}
\vspace{-.35cm}
\bea \Omega_{\rm GW}(f)\approx \frac{15}{16}\frac{\Omega^2_m}{H^2_0\eta^4_0}r\Delta^{2}_{\zeta}\left(\frac{f}{f_*}\right)^{n_t}.\eea
\end{mynamedbox1}
where $f_*\sim 7.7\times 10^{-17}{\rm Hz}$ represent the frequency of the pivot scale which is $k_*\sim 0.05 {\rm Mpc}^{-1}$. Here $\eta_0$ is the conformal time scale at the present day, $H_0$ is the present epoch Hubble parameter, $k_{\rm eq}$ is the wave number associated with matter radiation equality, which is fixed by $k_{\rm eq}\sim 7.3\times 10^{-2}\times \Omega_m h^2 {\rm Mpc}^{-1}\sim {\cal O}(10^{-2}){\rm Mpc}^{-1}$ corresponding to the frequency scale, $f_{\rm eq}\sim 1.54\times 10^{-17}{\rm Hz}$. Here $\Omega_m$ is the density parameter for matter and $h$ represents the reduced Hubble parameter in this discussion. From Planck 2018 \cite{Planck:2018vyg} one can take the best-fit values of the cosmological parameters for the further analysis purpose, which are $\Omega_m=0.315$, $h=0.673$ for $\eta_0\sim 1.4\times 10^{4}{\rm Mpc}$, $\Delta^{2}_{\zeta}\sim 2.2\times 10^{-9}$, where the pivot scale and matter-radiation equality scale are given by, $k_*=0.05{\rm Mpc}^{-1}$ and $k_{\rm eq}\sim 0.01{\rm Mpc}^{-1}$.

Now we will talk about the GW produced by PTA where the spectral density is given by the following expression:
\begin{mynamedbox1}{Spectral density of GW from PTA portal:}
\vspace{-.35cm}
\bea \Omega^{\rm PTA}_{\rm GW}(f)=\frac{2\pi^2}{3H^2_0}f^2 h^2_G(f),\eea 
\end{mynamedbox1}
where $h(f)$ is the frequency-dependent GW power spectrum, which appears at the characteristic frequency of PTA, which is $f_{\rm PTA}\sim 10^{-9}{\rm Hz}=1{\rm nHz}$. For the practical purpose in the mentioned characteristic frequency domain one can use power law parametrization using which the GW power spectrum can be expressed by the following formula:
\bea h_G(f)={\cal A}_G\left(\frac{f}{f_{\rm PTA}}\right)^{2\alpha},\eea
where ${\cal A}_G$ is the amplitude of the PTA-based GW and the corresponding spectral tilt is $\alpha$, which is related to the spectral tilt $\gamma$ of the PTA residual cross-power spectral density by, $\alpha= \frac{1}{2}(3-\gamma)$. Using this specific power law form of the spectrum the spectral density for GW from PTA can be further expressed by the following simplified relationship:
\bea \Omega^{\rm PTA}_{\rm GW}(f)=\frac{2\pi^2}{3H^2_0}{\cal A}^2_G \times\left(\frac{f^{2(1+\alpha)}}{f^{2\alpha}_{\rm PTA}}\right)=\frac{2\pi^2}{3H^2_0}{\cal A}^2_G \times\left(\frac{f^{5-\gamma}}{f^{3-\gamma}_{\rm PTA}}\right).\eea
Now at the pivot scale $k_*=0.05{\rm Mpc}^{-1}$ one can explicitly compute the amplitude of the GW amplitude from PTA, which is given by the following expression:
\begin{mynamedbox1}{Constraint on GW amplitude from PTA portal:}
\vspace{-.35cm}
\bea {\cal A}_G=\sqrt{\frac{45}{32\pi^2}\frac{\Omega^2_m\Delta^2_{\zeta}}{\left(k_{\rm eq}\eta_0\right)^2}}\times \frac{1}{f_{\rm PTA}\eta_0}\times\left(\frac{f_{\rm PTA}}{f_*}\right)^{\frac{n_t}{2}}\times \sqrt{r}\sim 2.14\times 10^{4.3n_t-19}\times \sqrt{\frac{r}{0.06}},\eea
\end{mynamedbox1}
where we use $n_t=5-\gamma$ in this computation. This relation suggests that for the allowed upper bound of the tensor-to-scalar ratio $r\lesssim 0.06$, if one can provide a large blue tilt from a fundamental theoretical framework then only with PTA portal in the ${\rm nHz}$ frequency domain one can able to explain the production of GW amplitude. As we have pointed already out in the previous part of the discussion in this paper, {\it Non-Bunch Davies } initial condition is one of the interesting and promising possibilities using which one can able to explain the generation of a large blue tilted tensor spectrum. For this reason, it is highly possible to detect the imprints of such theoretical framework in the GW spectrum in the light of observed by the Pulsar Timing Array (NANOGrav 15-year and EPTA) Data.

%%%%%3-\%%%%%%%%%%%%%%%%%FIGURE%%%%%%%%%%%%%%%%%%%%%%%%%%%%%%%%%%%%%
    \begin{figure*}[htb!]
    	\centering
   {
      	\includegraphics[width=18cm,height=9cm] {
      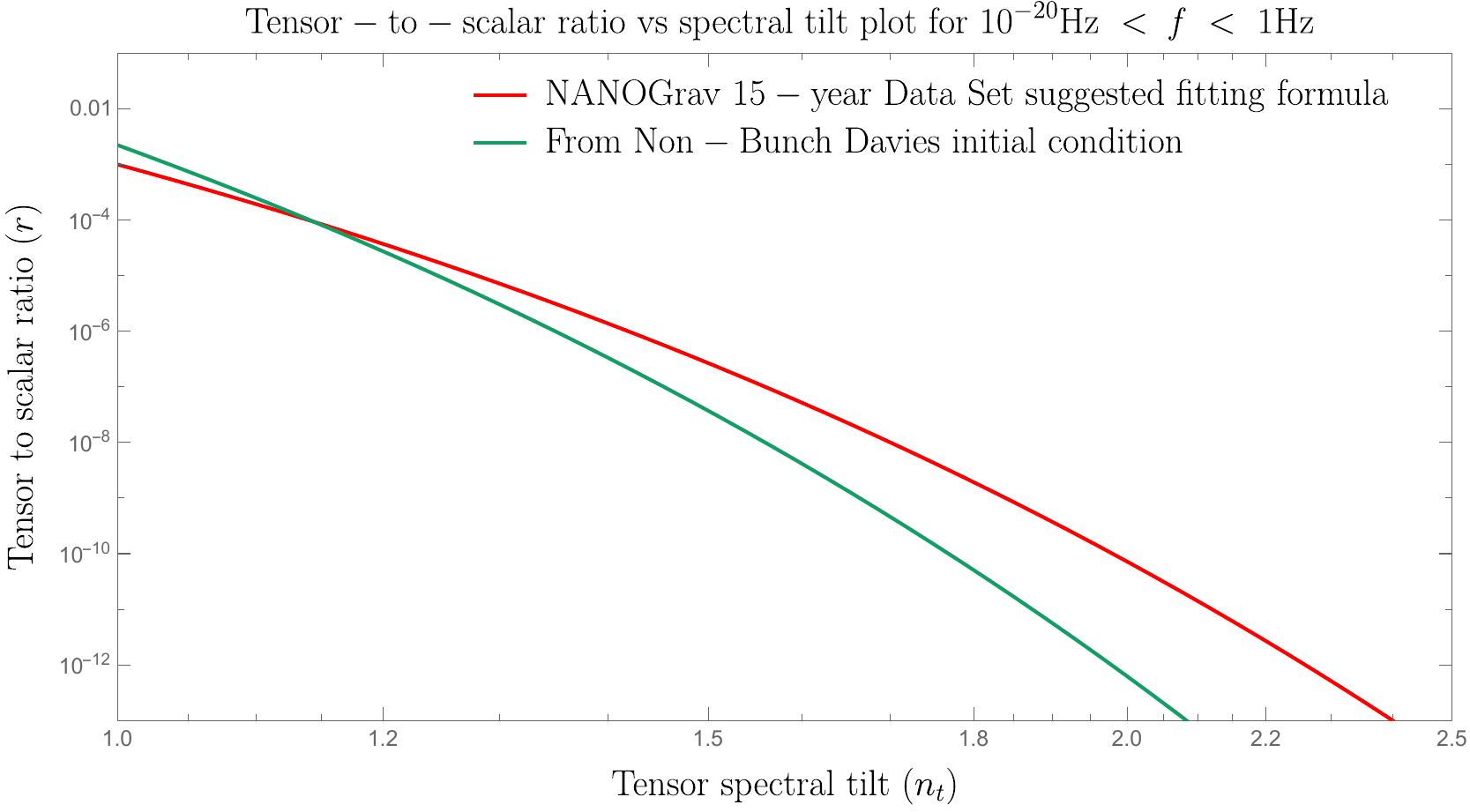}
        \label{OUTPBH}
    }
    	\caption[Optional caption for list of figures]{Tensor-to-scalar ratio ($r$) vs tensor spectral tilt ($n_t$) plot using new consistency relation for {\it Non-Bunch Davies} initial condition. We have also compared our result with the newly proposed fitting formula as suggested in NANOGrav 15 year result.} 
        \label{fig1}
    \end{figure*}
    %%%%%%%%%%%%%%%%%%%%%%%%%%%%%%%%%%%%%

In the context of {\it Non-Bunch Davies } initial condition, we parameterize the Bogoliubov coefficients associated with scalar and tensor 
perturbations in such a fashion that the following new consistency relation is maintained in all frequency ranges:
\bea r=8{\cal F}\left({\cal G}-n_t\right),\eea
where we suggest that the factor ${\cal F}$ is very small and approximately a constant and ${\cal G}$ is a function of the tensor spectral tilt $n_t$. We suggest the following functional forms from the numerical best-fitting perspective when considering all frequency domains by the following expressions:
\bea  {\cal F}\sim {\cal O}(10^{-22})\quad\quad\quad {\rm for}\quad 10^{-20}{\rm Hz}<f<1{\rm Hz}\eea
and 
\bea  {\cal G}
&=& \displaystyle
\displaystyle\left\{
	\begin{array}{ll}
		\displaystyle 0& \mbox{for}\quad 10^{-20}{\rm Hz}< f < 10^{-17}{\rm Hz} \;  \\  
			\displaystyle 
			\displaystyle 10^{28}\exp(-22n_t)+n_t \quad\quad\quad\quad\quad\quad\quad\quad& \mbox{for}\quad  10^{-17}{\rm Hz}< f < 10^{-7}{\rm Hz} \; \\
   \displaystyle 
			\displaystyle 10^{28+6.89n_t}\exp(22n_t)-n_t \quad\quad\quad\quad\quad\quad\quad\quad& \mbox{for}\quad  10^{-7}{\rm Hz}< f < 1{\rm Hz} \;. 
	\end{array}
\right. \eea
Consequently, the mathematical form of the new consistency relations in the above-mentioned three important frequency domains can be expressed as:
\begin{mynamedbox1}{New consistency relation from Non-Bunch Davies initial condition:}
\vspace{-.35cm}
\bea  r
&=& 8\times 10^{-22}\times\displaystyle
\displaystyle\left\{
	\begin{array}{ll}
		\displaystyle n_t & \mbox{for}\quad 10^{-20}{\rm Hz}< f < 10^{-17}{\rm Hz} \;  \\  
			\displaystyle 
			\displaystyle 10^{28}\exp(-22n_t) \quad\quad\quad\quad\quad\quad\quad\quad& \mbox{for}\quad  10^{-17}{\rm Hz}< f < 10^{-7}{\rm Hz} \; \\
   \displaystyle 
			\displaystyle 10^{28+6.89n_t}\exp(22n_t) \quad\quad\quad\quad\quad\quad\quad\quad& \mbox{for}\quad  10^{-7}{\rm Hz}< f < 1{\rm Hz} \;. 
	\end{array}
\right. \eea
\end{mynamedbox1}
In figure (\ref{fig1}), we have plotted the behaviour of tensor-to-scalar ratio ($r$) vs tensor spectral tilt ($n_t$) plot using new consistency relation for {\it Non-Bunch Davies} initial condition. We have also compared our result with the newly proposed fitting formula as suggested in the PTA (NANOGrav 15-year and EPTA) result. This representative plot clearly suggests that our proposed new consistency relation conforms well with the PTA result.

Now keeping the above parametrization in mind the spectral energy density of GW can be further written in the following simplified form:
\begin{mynamedbox1}{Sepctral density of GW amplitude for Non-Bunch Davies initial condition:}
\vspace{-.35cm}
\bea  \Omega_{\rm GW}(f)\approx \frac{15}{2}\frac{\Omega^2_m}{H^2_0\eta^4_0}{\cal F}\Delta^{2}_{\zeta}
\times \displaystyle
\displaystyle\left\{
	\begin{array}{ll}
		\displaystyle n_t\left(\frac{f}{f_*}\right)^{-n_t}& \mbox{for}\quad 10^{-20}{\rm Hz}< f < 10^{-17}{\rm Hz} \;  \\  
			\displaystyle 
			\displaystyle 10^{28}\exp(-22n_t)\left(\frac{f}{f_*}\right)^{n_t}\quad\quad\quad\quad\quad\quad\quad\quad& \mbox{for}\quad  10^{-17}{\rm Hz}< f < 10^{-7}{\rm Hz} \; \\
   \displaystyle 
			\displaystyle 10^{28+6.89n_t}\exp(22n_t)\left(\frac{f}{f_*}\right)^{-n_t} \quad\quad\quad\quad\quad\quad\quad\quad& \mbox{for}\quad  10^{-7}{\rm Hz}< f < 1{\rm Hz} \;.
	\end{array}
\right. \eea
\end{mynamedbox1}

%%%%%3-feature

The suggested feature of the spectral energy density of GW in the three frequency domain for {\it Non-Bunch Davies} initial condition with $n_t>0$ is actually motivated by the scenario of analyzing slow-roll (SRI), ultra slow-roll (USR) and then a slow-roll (SRII) phase one after another to study the generation of large amplitude fluctuation in presence of instantaneous transition necessarily required to produce PBHs. See refs. \cite{Kristiano:2022maq,Kristiano:2023scm,Choudhury:2023vuj,Choudhury:2023jlt,Choudhury:2023rks,Choudhury:2023hvf,Choudhury:2023kdb,Riotto:2023hoz,Riotto:2023gpm,Firouzjahi:2023aum,Motohashi:2023syh,Firouzjahi:2023ahg,Franciolini:2023lgy,Tasinato:2023ukp,Cheng:2023ikq,Choudhury:2023hfm,Bhattacharya:2023ysp,Choudhury:2023fwk,Choudhury:2023fjs,Choudhury:2024one} for more details on this issue. However, through deeper analysis, we have understood in ref \cite{Choudhury:2023jlt,Choudhury:2023rks,Choudhury:2023hvf} that quantum loop effects dominate in between the frequency domain $10^{-10}{\rm Hz}<f<10^{-9}{\rm Hz}$ corresponding to the momentum scale, $10^{5}{\rm Mpc}^{-1}<k<10^{6}{\rm Mpc}^{-1}$ and one cannot able to produce a sufficient number of e-foldings (we have found ${\cal N}\sim 25$) to produce $M_{\rm PBH}\sim M_{\odot}$, where $M_{\odot}\sim 10^{31}{\rm kg}$. Renormalization and DRG resummation of the one-loop scalar power spectrum put such strong constraints on the final result. For this reason, one can only able to produce a small mass PBHs ($M_{\rm PBH}\sim 10^2{\rm gm}$) at very high frequency $10^{6}{\rm Hz}<f<10^{7}{\rm Hz}$ corresponding to the momentum scale, $10^{21}{\rm Mpc}^{-1}<k<10^{22}{\rm Mpc}^{-1}$. In this case, a sufficient number of e-foldings can be achieved i.e. ${\cal N}\sim 60$ even after applying strong constraints from renormalization and DRG resummation in the one-loop corrected result of the scalar power spectrum. This problem can be evaded in the case of Galileon inflation which we have explicitly shown in ref \cite{Choudhury:2023hvf,Choudhury:2023kdb,Choudhury:2023hfm,Choudhury:2023fwk} due to having a strong non-renormalization theorem. In this article, our motivation is not PBHs production but rather violating the well-known slow-roll consistency condition $r=-8n_t$, with red tilt $n<0$ derived for canonical single field slow-roll models of inflation with {\it Bunch Davies} initial condition. Though {\it Bunch Davies} initial condition seems like a natural choice as it asymptotically goes to the {\it Minkowski vacuum}, but not capable enough to resolve all the above-mentioned crucial issues. Also, the corresponding red-tilted spectrum could not able to show enhancement to confront the findings of the recently pointed \textcolor{black}{Pulsar Timing Array (NANOGrav 15-year and EPTA)} result. Though it is not fully very naturally originated but {\it Non-Bunch Davies} initial condition helps us to violate the consistency condition. Instead of inserting a slow-roll (SRI), ultra slow-roll (USR), and then a slow-roll (SRII) phase one equivalently performs the analysis by violating the consistency condition in the presence of {\it Non-Bunch Davies} initial condition. Previous analysis also points towards the fact that the transition from SRI to USR and USR to SRII shifts the corresponding vacuum state to a {\it Non-Bunch Davies} one if we initially start with the {\it Bunch Davies} condition at the starting point of SRI phase. Such a newly created vacua in USR and SRII phases not only deviates from its {\it Bunch Davies} counterpart but also makes the corresponding Bogoliubov coefficients of the states momentum and time scale-dependent. In the above-mentioned parametrization of the new consistency relation, all of this information is not only maintained but also helps us to go beyond our general notion, which was established in the ground of very specific choice of {\it Bunch Davies} initial condition. 

%%%%%%%%%%%%%%%%FIGURE%%%%%%%%%%%%%%%%%%%%%%%%%%%%%%%%%%%%%
    \begin{figure*}[htb!]
    	\centering
   {
      	\includegraphics[width=18cm,height=9.5cm] {
      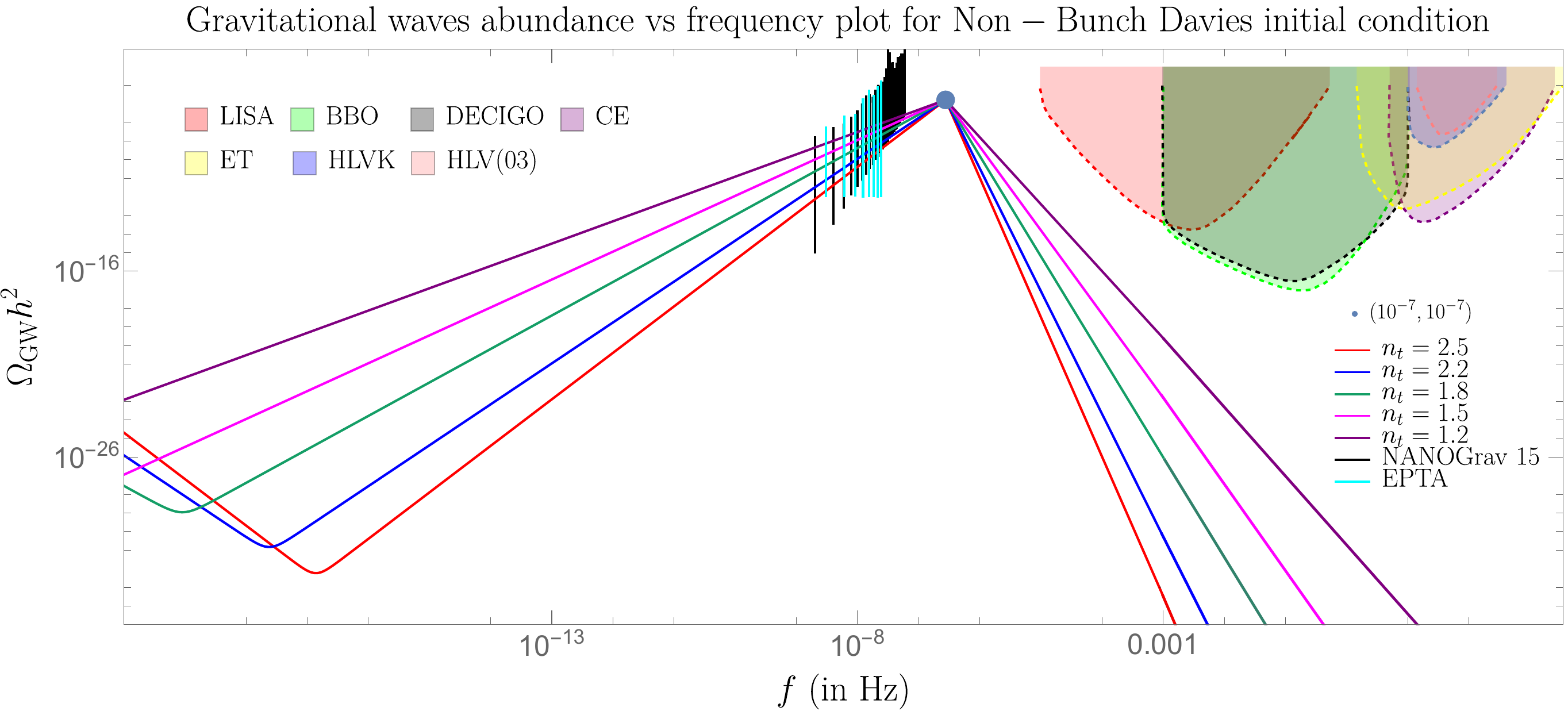}
        \label{OUTPBH}
    }
    	\caption[Optional caption for list of figures]{Representative plot showing the behaviour of the Gravitational Waves (GWs) abundance ($\Omega_{\rm GW}h^2$) with frequency ($f$) in the presence of {\it Non-Bunch Davies} initial condition. This plot clearly shows that the blue tilted behaviour of the spectrum in the frequency range $10^{-17}{\rm Hz}<f<10^{-7}{\rm Hz}$ is perfectly consistent with the NANOGrav 15 signal (in black bands) and the EPTA signal (in cyan bands). This plot is also consistent with the CMB observation (Planck) in the very low-frequency range, at the pivot value $f_*\sim 7.7\times 10^{-17}{\rm Hz}$ corresponding to momentum scale $k=0.05{\rm Mpc}^{-1}$. For the visualization purpose and to show the applicability of our result here we have taken five benchmark points, which are $n_t=1.2$ for $r=2.3\times 10^{-5}$, $n_t=1.5$ for $r=3.6\times 10^{-8}$,
        $n_t=1.8$ for $r=4.3\times 10^{-11}$,
        $n_t=2.2$ for $r=7.3\times 10^{-15}$,
        $n_t=2.5$ for $r=7.9\times 10^{-18}$. All of these values suggest that blue-tilted tensor spectra obtained from {\it Non-Bunch Davies} initial condition is more favoured than the red-tilted spectra as originally derived from the canonical single field model of inflation in the presence of {\it Bunch Davies} initial condition. In addition to this, it is important to note that all the benchmark values of the tensor-to-scalar ratio are consistent with the upper bound $r<0.06$ as obtained from Planck observation. } 
        \label{fig2}
    \end{figure*}
    %%%%%%%%%%%%%%%%%%%%%%%%%%%%%%%%%%%%%
  
    In figure (\ref{fig2}) we have depicted the behaviour of the Gravitational Waves (GWs) abundance ($\Omega_{\rm GW}h^2$) with frequency ($f$) in the presence of {\it Non-Bunch Davies} initial condition. By following the new structure of the modification factors in the three frequency bands, very low $10^{-20}{\rm Hz}<f<10^{-17}{\rm Hz}$, low $10^{-17}{\rm Hz}<f<10^{-7}{\rm Hz}$ and high $10^{-7}{\rm Hz}<f<1{\rm Hz}$, the behaviour of the spectral density of the GW and consequently the GW abundance changes its features with respect to the tensor spectral tilt $n_t$ which in turn fixes the value of the tensor-to-scalar ratio by making use of the newly proposed consistent relations for {\it Non-Bunch Davies} initial condition. Throughout the analysis, we have maintained the positive signature of the tensor spectral tilt $n_t>0$. The suggested new consistency relation is applicable to the mentioned three frequency regions and the above-mentioned plot clearly shows that in the intermediate low region, $10^{-17}{\rm Hz}<f<10^{-7}{\rm Hz}$, the blue-tilted behaviour of the power spectrum is dominating which finally gives rise to the enhancement in the GW abundance and using such prescription we can easily observe that it is able to explain the signals obtained from the recent \textcolor{black}{Pulsar Timing Array (NANOGrav 15-year and EPTA)} results. On the other hand, in the very low-frequency domain, $10^{-20}{\rm Hz}<f<10^{-17}{\rm Hz}$, the spectrum is behaving like a red-tilted though we have maintained the signature of $n_t>0$. It means that in this region the frequency-dependent behaviour of the spectral density of GW and the corresponding GW abundance is such that the outcome gives a red-tilted spectrum. Careful observation tells us that this is the region where CMB observation plays a very crucial role and one has to respect the constraints from the Planck observation as well in this domain, particularly at the pivot scale $f_*\sim 7.7\times 10^{-17}{\rm Hz}$ which corresponds to the wave number $k_*\sim 0.05{\rm Mpc}^{-1}$. We have obtained the red-tilted behaviour of the tensor spectrum in this particular domain which is also consistent with the Planck observation. Further, it is important to note that in this particular region with the help of {\it Bunch Davies} initial condition one can able to explain the existence of the red tilted tensor power spectrum which supports CMB observation. That means, in our analysis, the mentioned low-frequency domain is the {\it Bunch Davies} limiting condition of the proposed {\it Non-Bunch Davies} initial condition where everything seems perfect and the outcomes conform to the Planck observation as well. This is a good news that the proposed {\it Non-Bunch Davies} condition captured the information of the good old {\it Bunch Davies} initial condition in the very low-frequency domain which validates our analysis from the perspective of the CMB observation as well. On top of that, in the mentioned intermediate low-frequency domain due to having the blue tilted feature in the spectrum it also satisfies the constraint obtained from Pulsar Timing Array (NANOGrav 15-year and EPTA) data with high accuracy. Now we discuss the comparatively high-frequency domain $10^{-7}{\rm Hz}<f<1{\rm Hz}$, where the behaviour of the spectrum shows rapidly-falling behaviour compared to the behaviour observed from the very low-frequency domain. This particular feature points towards the fact that in this region the spectrum is different but shows a red-tilted feature, which is different from the feature obtained from the very low-frequency domain GW spectrum. This high-frequency domain behaviour of the spectrum is also consistent with other observational constraints obtained from LISA \cite{amaro2017laser}, BBO \cite{Kawamura:2011zz}, DECIGO \cite{Crowder:2005nr}, CE \cite{Punturo:2010zz}, ET \cite{Reitze:2019iox}, HLVK and HLV(03) \cite{KAGRA:2018plz,VIRGO:2014yos,LIGOScientific:2014pky}. These mentioned probes have not observed any signature from inflation and as the obtained spectrum from the {\it Non-Bunch Davies} initial condition falls in a rapid fashion and does not intersect the scanning parameter space of any of them, the high-frequency domain behaviour of the spectrum is also consistent. In figure (\ref{fig2}) for the visualization purpose and to show the applicability of our result here we have taken five benchmark points, which are $n_t=1.2$ for $r=2.3\times 10^{-5}$, $n_t=1.5$ for $r=3.6\times 10^{-8}$,
        $n_t=1.8$ for $r=4.3\times 10^{-11}$,
        $n_t=2.2$ for $r=7.3\times 10^{-15}$,
        $n_t=2.5$ for $r=7.9\times 10^{-18}$. All of these values suggest that blue-tilted tensor spectra obtained from {\it Non-Bunch Davies} initial condition is more favourable than the red-tilted spectra as originally derived from the canonical single field model of inflation in the presence of {\it Bunch Davies} initial condition. Furthermore, it is important to note that all the benchmark values of the tensor-to-scalar ratio are consistent with the upper bound $r<0.06$ as obtained from Planck observation.

\section{Conclusion}
\label{sec10}
 We finally conclude our discussion with the following key highlighting points from our analysis:
 \begin{itemize}[label={\checkmark}]
    \item In this work we start our discussion with the gauge fixing issue of the Cosmological Perturbation Theory and mention clearly the usefulness of each of the choices in the context of primordial cosmology.

    \item Next, using a preferred gauge condition we have performed the perturbation of scalar and tensor modes dynamical solution of which we have computed from {\it Mukhanov Sasaki equation} along with general quantum initial condition, which we have identified as the {\it Non-Bunch Davies} initial condition in this paper. The parameter space of this new initial condition is spanned by ${\bf SO(1,4)}$ isometry group of de Sitter space and for this reason one can expect that the {\it Bunch Davies} initial condition will represent a point in the large parameter space of the {\it Non-Bunch Davies} initial condition. 

    \item Further we have computed impacts of these purterbations in the inflationary observables, such as in the spectrum, spectral tilt and tensor-to-scalar ratio in presence of {\it Non-Bunch Davies} initial condition. The obtained result show significant modifications/deviations compared to the result obtained for the observables with {\it Bunch Davies} initial condition. 
    
    \item From our analysis we expect that the Bogoliubov coefficients representing the scalar and tensor sector of the cosmological perturbations are different. For this reason it will show immediately its impact in the consistency relation between the tensor-to-scalar ratio ($r$) and tensor spectral tilt ($n_t$). We have found from our analysis that the old consistency relation, $r=-8n_t$ with $n_t<0$ derived in presence of {\it Bunch Davies} initial condition is violated in the present context. We have explicitly derived the expressions for the new consistency relation in the presence of {\it Non-Bunch Davies} initial condition where the significant deviation is pointed in the derived result.

    \item Further with the help of newly derived results for inflationary observables derived from {\it Non-Bunch Davies} initial condition we have computed the new version of the field excursion formula for the inflaton field and show significant deviation from the good old {\it Lyth bound} derived in the same context from {\it Bunch Davies} initial condition. This newly derived result in our paper suggests that the field excursion in the red-tilted regime of the tensor spectrum is controlled by the tensor-to-scalar ratio and the total number of e-foldings necessarily required for inflation. On the other hand, the blue-tilted regime of the tensor power spectrum is controlled by the tensor-to-scalar ratio and the values of the tensor spectral tilt and scalar spectral tilt. We have found from our analysis that in the allowed region of tensor-to-scalar ratio, $r<0.06$ from CMB observation one is able to achieve sub-Planckian field excursion which is the fundamental requirement to validate EFT prescription within the framework of inflation. Though during the derivation of this bound we have not used any specific structure of the inflationary potential but the present analysis suggests that to get blue tilted tensor power spectrum and to maintain the theoretical constraint from the new consistency relation derived from {\it Non-Bunch Davies} initial condition the potential should have a certain structure and physical properties which will respect both the newly derived consistency condition and field excursion formula. 

    \item Next, we have suggested certain structure of the modification factors as appearing in the new consistency relation, which are made up of certain combinations of the Bogoliubov coefficients computed from scalar and tensor perturbations. This suggested form is incorporated so that it is able to respect the red tilted behaviour in the frequency ranges, $10^{-20}{\rm Hz}<f<10^{-17}{\rm Hz}$ and $10^{-7}{\rm Hz}<f<1{\rm Hz}$ and the blue tilted behaviour in the domain $10^{-17}{\rm Hz}<f<10^{-7}{\rm Hz}$. Though we are currently not interested in PBH formation from the present setup but it is important to note that the present set up somewhat mimics the transient roll inflation where the scenario is describe in terms of SRI, USR and SRII phases and the corresponding transition from one phase to other phases. In near future we will investigate this issue in great detail and establish such underlying connection more clarity. 

    \item Finally we have depicted the behaviour of the abundance of Gravitational Waves (GW) ($\Omega_{\rm GW}h^2$) with frequency ($f$) in the presence of {\it Non-Bunch Davies} initial condition. By following the new structure of the modification factors applicable in the three frequency bands, very low $10^{-20}{\rm Hz}<f<10^{-17}{\rm Hz}$, low $10^{-17}{\rm Hz}<f<10^{-7}{\rm Hz}$ and high $10^{-7}{\rm Hz}<f<1{\rm Hz}$ the behaviour of the spectral density of the GW and consequently the GW abundance changes its features with respect to the tensor spectral tilt $n_t$ which in turn fixes the value of the tensor-to-scalar ratio by making use of the newly proposed consistent relations for {\it Non-Bunch Davies} initial condition. 
    We have found that this GW spectra computed from {\it Non-Bunch Davies} initial condition is consistent with CMB observation within the frequency range, $10^{-20}{\rm Hz}<f<10^{-17}{\rm Hz}$, where the pivot scale is $f_*\sim 7.7\times 10^{-17}{\rm Hz}$. Most importantly, the blue tilted behaviour of this GW spectrum conforms well with \textcolor{black}{Pulsar Timing Array (NANOGrav 15-year and EPTA)} signal in the frequency range $10^{-9}{\rm Hz}<f<10^{-7}{\rm Hz}$, where the peak of the spectrum appears at the frequency scale, $f\sim 10^{-7}{\rm Hz}$. In the comparatively high-frequency domain $10^{-7}{\rm Hz}<f<1{\rm Hz}$ behaviour of the spectrum is also consistent with other observational constraints obtained from  LISA \cite{amaro2017laser}, BBO \cite{Kawamura:2011zz}, DECIGO \cite{Crowder:2005nr}, CE \cite{Punturo:2010zz}, ET \cite{Reitze:2019iox}, HLVK and HLV(03) \cite{KAGRA:2018plz,VIRGO:2014yos,LIGOScientific:2014pky}. These mentioned probes did not have observed any signature from inflation and as the obtained spectrum from the {\it Non-Bunch Davies} initial condition falls in a rapid fashion in this regime and does not intersect the scanning parameter space of any of them, the high-frequency domain behaviour of the spectrum is also consistent. Our analysis suggests blue tilted tensor spectra obtained from {\it Non-Bunch Davies} initial condition is more favoured than the red tilted spectra as originally derived from the canonical single field model of inflation in the presence of {\it Bunch Davies} initial condition. Finally, we have found that the values of the tensor-to-scalar ratio are consistent with the upper bound $r<0.06$ as obtained from Planck observation.

\end{itemize}

	\subsection*{Acknowledgements}

 SC would like to thank The National Academy of Sciences (NASI), Prayagraj, India for being elected as a member
of the academy. SC would like to thank 
Siddharth Kumar Tiwari and Ahaskar Karde for their enormous help and discussions during this project. SC sincerely thanks Md. Sami for various useful discussions which helps to improve the analysis performed in this paper. SC would like to thank the work-friendly environment of The Thanu Padmanabhan Centre For Cosmology and Science Popularization (CCSP), SGT University, Gurugram, for providing tremendous support in research. SC would also like to thank all the members of Quantum Aspects of the Space-Time \& Matter (QASTM) for elaborative discussions. Last but not least, we would like to acknowledge our debt to the people belonging to the various parts of the
world for their generous and steady support for research in natural sciences.
%\newpage

%\clearpage

%\newpage
%\phantomsection
%\addcontentsline{toc}{section}{References}
\bibliographystyle{utphys}
\bibliography{Ref}

\providecommand{\href}[2]{#2}\begingroup\raggedright\begin{thebibliography}{100}

\bibitem{Caprini:2018mtu}
C.~Caprini and D.~G. Figueroa, ``{Cosmological Backgrounds of Gravitational
  Waves},'' \href{http://dx.doi.org/10.1088/1361-6382/aac608}{{\em Class.
  Quant. Grav.} {\bfseries 35} no.~16, (2018) 163001},
  \href{http://arxiv.org/abs/1801.04268}{{\ttfamily arXiv:1801.04268
  [astro-ph.CO]}}.

\bibitem{Renzini:2022alw}
A.~I. Renzini, B.~Goncharov, A.~C. Jenkins, and P.~M. Meyers, ``{Stochastic
  Gravitational-Wave Backgrounds: Current Detection Efforts and Future
  Prospects},'' \href{http://dx.doi.org/10.3390/galaxies10010034}{{\em
  Galaxies} {\bfseries 10} no.~1, (2022) 34},
  \href{http://arxiv.org/abs/2202.00178}{{\ttfamily arXiv:2202.00178 [gr-qc]}}.

\bibitem{Siemens:2006yp}
X.~Siemens, V.~Mandic, and J.~Creighton, ``{Gravitational wave stochastic
  background from cosmic (super)strings},''
  \href{http://dx.doi.org/10.1103/PhysRevLett.98.111101}{{\em Phys. Rev. Lett.}
  {\bfseries 98} (2007) 111101},
  \href{http://arxiv.org/abs/astro-ph/0610920}{{\ttfamily
  arXiv:astro-ph/0610920}}.

\bibitem{Caprini:2010xv}
C.~Caprini, R.~Durrer, and X.~Siemens, ``{Detection of gravitational waves from
  the QCD phase transition with pulsar timing arrays},''
  \href{http://dx.doi.org/10.1103/PhysRevD.82.063511}{{\em Phys. Rev. D}
  {\bfseries 82} (2010) 063511},
  \href{http://arxiv.org/abs/1007.1218}{{\ttfamily arXiv:1007.1218
  [astro-ph.CO]}}.

\bibitem{Ramberg:2019dgi}
N.~Ramberg and L.~Visinelli, ``{Probing the Early Universe with Axion Physics
  and Gravitational Waves},''
  \href{http://dx.doi.org/10.1103/PhysRevD.99.123513}{{\em Phys. Rev. D}
  {\bfseries 99} no.~12, (2019) 123513},
  \href{http://arxiv.org/abs/1904.05707}{{\ttfamily arXiv:1904.05707
  [astro-ph.CO]}}.

\bibitem{Caprini:2019egz}
C.~Caprini {\em et~al.}, ``{Detecting gravitational waves from cosmological
  phase transitions with LISA: an update},''
  \href{http://dx.doi.org/10.1088/1475-7516/2020/03/024}{{\em JCAP} {\bfseries
  03} (2020) 024}, \href{http://arxiv.org/abs/1910.13125}{{\ttfamily
  arXiv:1910.13125 [astro-ph.CO]}}.

\bibitem{Ellis:2020awk}
J.~Ellis, M.~Lewicki, and J.~M. No, ``{Gravitational waves from first-order
  cosmological phase transitions: lifetime of the sound wave source},''
  \href{http://dx.doi.org/10.1088/1475-7516/2020/07/050}{{\em JCAP} {\bfseries
  07} (2020) 050}, \href{http://arxiv.org/abs/2003.07360}{{\ttfamily
  arXiv:2003.07360 [hep-ph]}}.

\bibitem{Geller:2023shn}
M.~Geller, S.~Ghosh, S.~Lu, and Y.~Tsai, ``{Challenges in Interpreting the
  NANOGrav 15-Year Data Set as Early Universe Gravitational Waves Produced by
  ALP Induced Instability},'' \href{http://arxiv.org/abs/2307.03724}{{\ttfamily
  arXiv:2307.03724 [hep-ph]}}.

\bibitem{Bi:2023tib}
Y.-C. Bi, Y.-M. Wu, Z.-C. Chen, and Q.-G. Huang, ``{Implications for the
  Supermassive Black Hole Binaries from the NANOGrav 15-year Data Set},''
  \href{http://arxiv.org/abs/2307.00722}{{\ttfamily arXiv:2307.00722
  [astro-ph.CO]}}.

\bibitem{Rajagopal:1994zj}
M.~Rajagopal and R.~W. Romani, ``{Ultralow frequency gravitational radiation
  from massive black hole binaries},''
  \href{http://dx.doi.org/10.1086/175813}{{\em Astrophys. J.} {\bfseries 446}
  (1995) 543--549}, \href{http://arxiv.org/abs/astro-ph/9412038}{{\ttfamily
  arXiv:astro-ph/9412038}}.

\bibitem{Jaffe:2002rt}
A.~H. Jaffe and D.~C. Backer, ``{Gravitational waves probe the coalescence rate
  of massive black hole binaries},''
  \href{http://dx.doi.org/10.1086/345443}{{\em Astrophys. J.} {\bfseries 583}
  (2003) 616--631}, \href{http://arxiv.org/abs/astro-ph/0210148}{{\ttfamily
  arXiv:astro-ph/0210148}}.

\bibitem{Wyithe:2002ep}
J.~S.~B. Wyithe and A.~Loeb, ``{Low - frequency gravitational waves from
  massive black hole binaries: Predictions for LISA and pulsar timing
  arrays},'' \href{http://dx.doi.org/10.1086/375187}{{\em Astrophys. J.}
  {\bfseries 590} (2003) 691--706},
  \href{http://arxiv.org/abs/astro-ph/0211556}{{\ttfamily
  arXiv:astro-ph/0211556}}.

\bibitem{Sesana:2004sp}
A.~Sesana, F.~Haardt, P.~Madau, and M.~Volonteri, ``{Low - frequency
  gravitational radiation from coalescing massive black hole binaries in
  hierarchical cosmologies},'' \href{http://dx.doi.org/10.1086/422185}{{\em
  Astrophys. J.} {\bfseries 611} (2004) 623--632},
  \href{http://arxiv.org/abs/astro-ph/0401543}{{\ttfamily
  arXiv:astro-ph/0401543}}.

\bibitem{Burke-Spolaor:2018bvk}
S.~Burke-Spolaor {\em et~al.}, ``{The Astrophysics of Nanohertz Gravitational
  Waves},'' \href{http://dx.doi.org/10.1007/s00159-019-0115-7}{{\em Astron.
  Astrophys. Rev.} {\bfseries 27} no.~1, (2019) 5},
  \href{http://arxiv.org/abs/1811.08826}{{\ttfamily arXiv:1811.08826
  [astro-ph.HE]}}.

\bibitem{Detweiler:1979wn}
S.~L. Detweiler, ``{Pulsar timing measurements and the search for gravitational
  waves},'' \href{http://dx.doi.org/10.1086/157593}{{\em Astrophys. J.}
  {\bfseries 234} (1979) 1100--1104}.

\bibitem{Hobbs:2017oam}
G.~Hobbs and S.~Dai, ``{Gravitational wave research using pulsar timing
  arrays},'' \href{http://dx.doi.org/10.1093/nsr/nwx126}{{\em Natl. Sci. Rev.}
  {\bfseries 4} no.~5, (2017) 707--717},
  \href{http://arxiv.org/abs/1707.01615}{{\ttfamily arXiv:1707.01615
  [astro-ph.IM]}}.

\bibitem{NANOGrav:2020bcs}
{\bfseries NANOGrav} Collaboration, Z.~Arzoumanian {\em et~al.}, ``{The
  NANOGrav 12.5 yr Data Set: Search for an Isotropic Stochastic
  Gravitational-wave Background},''
  \href{http://dx.doi.org/10.3847/2041-8213/abd401}{{\em Astrophys. J. Lett.}
  {\bfseries 905} no.~2, (2020) L34},
  \href{http://arxiv.org/abs/2009.04496}{{\ttfamily arXiv:2009.04496
  [astro-ph.HE]}}.

\bibitem{Goncharov:2021oub}
B.~Goncharov {\em et~al.}, ``{On the Evidence for a Common-spectrum Process in
  the Search for the Nanohertz Gravitational-wave Background with the Parkes
  Pulsar Timing Array},''
  \href{http://dx.doi.org/10.3847/2041-8213/ac17f4}{{\em Astrophys. J. Lett.}
  {\bfseries 917} no.~2, (2021) L19},
  \href{http://arxiv.org/abs/2107.12112}{{\ttfamily arXiv:2107.12112
  [astro-ph.HE]}}.

\bibitem{Chen:2021rqp}
S.~Chen {\em et~al.}, ``{Common-red-signal analysis with 24-yr high-precision
  timing of the European Pulsar Timing Array: inferences in the stochastic
  gravitational-wave background search},''
  \href{http://dx.doi.org/10.1093/mnras/stab2833}{{\em Mon. Not. Roy. Astron.
  Soc.} {\bfseries 508} no.~4, (2021) 4970--4993},
  \href{http://arxiv.org/abs/2110.13184}{{\ttfamily arXiv:2110.13184
  [astro-ph.HE]}}.

\bibitem{EPTA:2023fyk}
{\bfseries EPTA} Collaboration, J.~Antoniadis {\em et~al.}, ``{The second data
  release from the European Pulsar Timing Array III. Search for gravitational
  wave signals},'' \href{http://arxiv.org/abs/2306.16214}{{\ttfamily
  arXiv:2306.16214 [astro-ph.HE]}}.

\bibitem{EPTA:2023sfo}
{\bfseries EPTA} Collaboration, J.~Antoniadis {\em et~al.}, ``{The second data
  release from the European Pulsar Timing Array I. The dataset and timing
  analysis},'' \href{http://arxiv.org/abs/2306.16224}{{\ttfamily
  arXiv:2306.16224 [astro-ph.HE]}}.

\bibitem{EPTA:2023akd}
{\bfseries EPTA} Collaboration, J.~Antoniadis {\em et~al.}, ``{The second data
  release from the European Pulsar Timing Array II. Customised pulsar noise
  models for spatially correlated gravitational waves},''
  \href{http://arxiv.org/abs/2306.16225}{{\ttfamily arXiv:2306.16225
  [astro-ph.HE]}}.

\bibitem{EPTA:2023gyr}
{\bfseries EPTA} Collaboration, J.~Antoniadis {\em et~al.}, ``{The second data
  release from the European Pulsar Timing Array IV. Search for continuous
  gravitational wave signals},''
  \href{http://arxiv.org/abs/2306.16226}{{\ttfamily arXiv:2306.16226
  [astro-ph.HE]}}.

\bibitem{EPTA:2023xxk}
{\bfseries EPTA} Collaboration, J.~Antoniadis {\em et~al.}, ``{The second data
  release from the European Pulsar Timing Array: V. Implications for massive
  black holes, dark matter and the early Universe},''
  \href{http://arxiv.org/abs/2306.16227}{{\ttfamily arXiv:2306.16227
  [astro-ph.CO]}}.

\bibitem{EPTA:2023xiy}
{\bfseries EPTA} Collaboration, C.~Smarra {\em et~al.}, ``{The second data
  release from the European Pulsar Timing Array: VI. Challenging the ultralight
  dark matter paradigm},'' \href{http://arxiv.org/abs/2306.16228}{{\ttfamily
  arXiv:2306.16228 [astro-ph.HE]}}.

\bibitem{Antoniadis:2022pcn}
J.~Antoniadis {\em et~al.}, ``{The International Pulsar Timing Array second
  data release: Search for an isotropic gravitational wave background},''
  \href{http://dx.doi.org/10.1093/mnras/stab3418}{{\em Mon. Not. Roy. Astron.
  Soc.} {\bfseries 510} no.~4, (2022) 4873--4887},
  \href{http://arxiv.org/abs/2201.03980}{{\ttfamily arXiv:2201.03980
  [astro-ph.HE]}}.

\bibitem{Romano:2020sxq}
J.~D. Romano, J.~S. Hazboun, X.~Siemens, and A.~M. Archibald,
  ``{Common-spectrum process versus cross-correlation for gravitational-wave
  searches using pulsar timing arrays},''
  \href{http://dx.doi.org/10.1103/PhysRevD.103.063027}{{\em Phys. Rev. D}
  {\bfseries 103} no.~6, (2021) 063027},
  \href{http://arxiv.org/abs/2012.03804}{{\ttfamily arXiv:2012.03804 [gr-qc]}}.

\bibitem{NANOGrav:2020spf}
{\bfseries NANOGrav} Collaboration, N.~S. Pol {\em et~al.}, ``{Astrophysics
  Milestones for Pulsar Timing Array Gravitational-wave Detection},''
  \href{http://dx.doi.org/10.3847/2041-8213/abf2c9}{{\em Astrophys. J. Lett.}
  {\bfseries 911} no.~2, (2021) L34},
  \href{http://arxiv.org/abs/2010.11950}{{\ttfamily arXiv:2010.11950
  [astro-ph.HE]}}.

\bibitem{Hellings:1983fr}
R.~w. Hellings and G.~s. Downs, ``{UPPER LIMITS ON THE ISOTROPIC GRAVITATIONAL
  RADIATION BACKGROUND FROM PULSAR TIMING ANALYSIS},''
  \href{http://dx.doi.org/10.1086/183954}{{\em Astrophys. J. Lett.} {\bfseries
  265} (1983) L39--L42}.

\bibitem{Goncharov:2022ktc}
B.~Goncharov {\em et~al.}, ``{Consistency of the Parkes Pulsar Timing Array
  Signal with a Nanohertz Gravitational- wave Background},''
  \href{http://dx.doi.org/10.3847/2041-8213/ac76bb}{{\em Astrophys. J.}
  {\bfseries 932} no.~2, (2022) L22},
  \href{http://arxiv.org/abs/2206.03766}{{\ttfamily arXiv:2206.03766 [gr-qc]}}.

\bibitem{Zic:2022sxd}
A.~Zic {\em et~al.}, ``{Evaluating the prevalence of spurious correlations in
  pulsar timing array data sets},''
  \href{http://dx.doi.org/10.1093/mnras/stac2100}{{\em Mon. Not. Roy. Astron.
  Soc.} {\bfseries 516} no.~1, (2022) 410--420},
  \href{http://arxiv.org/abs/2207.12237}{{\ttfamily arXiv:2207.12237
  [astro-ph.HE]}}.

\bibitem{Tiburzi:2015kqa}
C.~Tiburzi, G.~Hobbs, M.~Kerr, W.~Coles, S.~Dai, R.~Manchester, A.~Possenti,
  R.~Shannon, and X.~You, ``{A study of spatial correlations in pulsar timing
  array data},'' \href{http://dx.doi.org/10.1093/mnras/stv2143}{{\em Mon. Not.
  Roy. Astron. Soc.} {\bfseries 455} no.~4, (2016) 4339--4350},
  \href{http://arxiv.org/abs/1510.02363}{{\ttfamily arXiv:1510.02363
  [astro-ph.IM]}}.

\bibitem{NANOGrav:2023gor}
{\bfseries NANOGrav} Collaboration, G.~Agazie {\em et~al.}, ``{The NANOGrav 15
  yr Data Set: Evidence for a Gravitational-wave Background},''
  \href{http://dx.doi.org/10.3847/2041-8213/acdac6}{{\em Astrophys. J. Lett.}
  {\bfseries 951} no.~1, (2023) L8},
  \href{http://arxiv.org/abs/2306.16213}{{\ttfamily arXiv:2306.16213
  [astro-ph.HE]}}.

\bibitem{Antoniadis:2023ott}
J.~Antoniadis {\em et~al.}, ``{The second data release from the European Pulsar
  Timing Array III. Search for gravitational wave signals},''
  \href{http://arxiv.org/abs/2306.16214}{{\ttfamily arXiv:2306.16214
  [astro-ph.HE]}}.

\bibitem{Reardon:2023gzh}
D.~J. Reardon {\em et~al.}, ``{Search for an Isotropic Gravitational-wave
  Background with the Parkes Pulsar Timing Array},''
  \href{http://dx.doi.org/10.3847/2041-8213/acdd02}{{\em Astrophys. J. Lett.}
  {\bfseries 951} no.~1, (2023) L6},
  \href{http://arxiv.org/abs/2306.16215}{{\ttfamily arXiv:2306.16215
  [astro-ph.HE]}}.

\bibitem{Xu:2023wog}
H.~Xu {\em et~al.}, ``{Searching for the Nano-Hertz Stochastic Gravitational
  Wave Background with the Chinese Pulsar Timing Array Data Release I},''
  \href{http://dx.doi.org/10.1088/1674-4527/acdfa5}{{\em Res. Astron.
  Astrophys.} {\bfseries 23} no.~7, (2023) 075024},
  \href{http://arxiv.org/abs/2306.16216}{{\ttfamily arXiv:2306.16216
  [astro-ph.HE]}}.

\bibitem{NANOGrav:2023hde}
{\bfseries NANOGrav} Collaboration, G.~Agazie {\em et~al.}, ``{The NANOGrav 15
  yr Data Set: Observations and Timing of 68 Millisecond Pulsars},''
  \href{http://dx.doi.org/10.3847/2041-8213/acda9a}{{\em Astrophys. J. Lett.}
  {\bfseries 951} no.~1, (2023) L9},
  \href{http://arxiv.org/abs/2306.16217}{{\ttfamily arXiv:2306.16217
  [astro-ph.HE]}}.

\bibitem{NANOGrav:2023ctt}
{\bfseries NANOGrav} Collaboration, G.~Agazie {\em et~al.}, ``{The NANOGrav 15
  yr Data Set: Detector Characterization and Noise Budget},''
  \href{http://dx.doi.org/10.3847/2041-8213/acda88}{{\em Astrophys. J. Lett.}
  {\bfseries 951} no.~1, (2023) L10},
  \href{http://arxiv.org/abs/2306.16218}{{\ttfamily arXiv:2306.16218
  [astro-ph.HE]}}.

\bibitem{NANOGrav:2023hvm}
{\bfseries NANOGrav} Collaboration, A.~Afzal {\em et~al.}, ``{The NANOGrav 15
  yr Data Set: Search for Signals from New Physics},''
  \href{http://dx.doi.org/10.3847/2041-8213/acdc91}{{\em Astrophys. J. Lett.}
  {\bfseries 951} no.~1, (2023) L11},
  \href{http://arxiv.org/abs/2306.16219}{{\ttfamily arXiv:2306.16219
  [astro-ph.HE]}}.

\bibitem{NANOGrav:2023hfp}
{\bfseries NANOGrav} Collaboration, G.~Agazie {\em et~al.}, ``{The NANOGrav
  15-year Data Set: Constraints on Supermassive Black Hole Binaries from the
  Gravitational Wave Background},''
  \href{http://arxiv.org/abs/2306.16220}{{\ttfamily arXiv:2306.16220
  [astro-ph.HE]}}.

\bibitem{NANOGrav:2023tcn}
{\bfseries NANOGrav} Collaboration, G.~Agazie {\em et~al.}, ``{The NANOGrav
  15-year Data Set: Search for Anisotropy in the Gravitational-Wave
  Background},'' \href{http://arxiv.org/abs/2306.16221}{{\ttfamily
  arXiv:2306.16221 [astro-ph.HE]}}.

\bibitem{NANOGrav:2023pdq}
{\bfseries NANOGrav} Collaboration, G.~Agazie {\em et~al.}, ``{The NANOGrav
  15-year Data Set: Bayesian Limits on Gravitational Waves from Individual
  Supermassive Black Hole Binaries},''
  \href{http://arxiv.org/abs/2306.16222}{{\ttfamily arXiv:2306.16222
  [astro-ph.HE]}}.

\bibitem{NANOGrav:2023icp}
{\bfseries NANOGrav} Collaboration, A.~D. Johnson {\em et~al.}, ``{The NANOGrav
  15-year Gravitational-Wave Background Analysis Pipeline},''
  \href{http://arxiv.org/abs/2306.16223}{{\ttfamily arXiv:2306.16223
  [astro-ph.HE]}}.

\bibitem{Antoniadis:2023lym}
J.~Antoniadis {\em et~al.}, ``{The second data release from the European Pulsar
  Timing Array I. The dataset and timing analysis},''
  \href{http://arxiv.org/abs/2306.16224}{{\ttfamily arXiv:2306.16224
  [astro-ph.HE]}}.

\bibitem{Antoniadis:2023puu}
J.~Antoniadis {\em et~al.}, ``{The second data release from the European Pulsar
  Timing Array II. Customised pulsar noise models for spatially correlated
  gravitational waves},'' \href{http://arxiv.org/abs/2306.16225}{{\ttfamily
  arXiv:2306.16225 [astro-ph.HE]}}.

\bibitem{Antoniadis:2023aac}
J.~Antoniadis {\em et~al.}, ``{The second data release from the European Pulsar
  Timing Array IV. Search for continuous gravitational wave signals},''
  \href{http://arxiv.org/abs/2306.16226}{{\ttfamily arXiv:2306.16226
  [astro-ph.HE]}}.

\bibitem{Antoniadis:2023xlr}
J.~Antoniadis {\em et~al.}, ``{The second data release from the European Pulsar
  Timing Array: V. Implications for massive black holes, dark matter and the
  early Universe},'' \href{http://arxiv.org/abs/2306.16227}{{\ttfamily
  arXiv:2306.16227 [astro-ph.CO]}}.

\bibitem{Smarra:2023ljf}
C.~Smarra {\em et~al.}, ``{The second data release from the European Pulsar
  Timing Array: VI. Challenging the ultralight dark matter paradigm},''
  \href{http://arxiv.org/abs/2306.16228}{{\ttfamily arXiv:2306.16228
  [astro-ph.HE]}}.

\bibitem{Reardon:2023zen}
D.~J. Reardon {\em et~al.}, ``{The Gravitational-wave Background Null
  Hypothesis: Characterizing Noise in Millisecond Pulsar Arrival Times with the
  Parkes Pulsar Timing Array},''
  \href{http://dx.doi.org/10.3847/2041-8213/acdd03}{{\em Astrophys. J. Lett.}
  {\bfseries 951} no.~1, (2023) L7},
  \href{http://arxiv.org/abs/2306.16229}{{\ttfamily arXiv:2306.16229
  [astro-ph.HE]}}.

\bibitem{Zic:2023gta}
A.~Zic {\em et~al.}, ``{The Parkes Pulsar Timing Array Third Data Release},''
  \href{http://arxiv.org/abs/2306.16230}{{\ttfamily arXiv:2306.16230
  [astro-ph.HE]}}.

\bibitem{Guo:2023hyp}
S.-Y. Guo, M.~Khlopov, X.~Liu, L.~Wu, Y.~Wu, and B.~Zhu, ``{Footprints of
  Axion-Like Particle in Pulsar Timing Array Data and JWST Observations},''
  \href{http://arxiv.org/abs/2306.17022}{{\ttfamily arXiv:2306.17022
  [hep-ph]}}.

\bibitem{Kitajima:2023cek}
N.~Kitajima, J.~Lee, K.~Murai, F.~Takahashi, and W.~Yin, ``{Nanohertz
  Gravitational Waves from Axion Domain Walls Coupled to QCD},''
  \href{http://arxiv.org/abs/2306.17146}{{\ttfamily arXiv:2306.17146
  [hep-ph]}}.

\bibitem{Murai:2023gkv}
K.~Murai and W.~Yin, ``{A Novel Probe of Supersymmetry in Light of Nanohertz
  Gravitational Waves},'' \href{http://arxiv.org/abs/2307.00628}{{\ttfamily
  arXiv:2307.00628 [hep-ph]}}.

\bibitem{Ashoorioon:2022raz}
A.~Ashoorioon, K.~Rezazadeh, and A.~Rostami, ``{NANOGrav signal from the end of
  inflation and the LIGO mass and heavier primordial black holes},''
  \href{http://dx.doi.org/10.1016/j.physletb.2022.137542}{{\em Phys. Lett. B}
  {\bfseries 835} (2022) 137542},
  \href{http://arxiv.org/abs/2202.01131}{{\ttfamily arXiv:2202.01131
  [astro-ph.CO]}}.

\bibitem{Athron:2023mer}
P.~Athron, A.~Fowlie, C.-T. Lu, L.~Morris, L.~Wu, Y.~Wu, and Z.~Xu, ``{Can
  Supercooled Phase Transitions explain the Gravitational Wave Background
  observed by Pulsar Timing Arrays?},''
  \href{http://arxiv.org/abs/2306.17239}{{\ttfamily arXiv:2306.17239
  [hep-ph]}}.

\bibitem{Li:2023bxy}
S.-P. Li and K.-P. Xie, ``{A collider test of nano-Hertz gravitational waves
  from pulsar timing arrays},''
  \href{http://arxiv.org/abs/2307.01086}{{\ttfamily arXiv:2307.01086
  [hep-ph]}}.

\bibitem{Li:2023yaj}
Y.~Li, C.~Zhang, Z.~Wang, M.~Cui, Y.-L.~S. Tsai, Q.~Yuan, and Y.-Z. Fan,
  ``{Primordial magnetic field as a common solution of nanohertz gravitational
  waves and Hubble tension},''
  \href{http://arxiv.org/abs/2306.17124}{{\ttfamily arXiv:2306.17124
  [astro-ph.HE]}}.

\bibitem{Oikonomou:2023qfz}
V.~K. Oikonomou, ``{Flat Energy Spectrum of Primordial Gravitational Waves vs
  Peaks and the NANOGrav 2023 Observation},''
  \href{http://arxiv.org/abs/2306.17351}{{\ttfamily arXiv:2306.17351
  [astro-ph.CO]}}.

\bibitem{Niu:2023bsr}
X.~Niu and M.~H. Rahat, ``{NANOGrav signal from axion inflation},''
  \href{http://arxiv.org/abs/2307.01192}{{\ttfamily arXiv:2307.01192
  [hep-ph]}}.

\bibitem{Kazanas:1980tx}
D.~Kazanas, ``{Dynamics of the Universe and Spontaneous Symmetry Breaking},''
  \href{http://dx.doi.org/10.1086/183361}{{\em Astrophys. J. Lett.} {\bfseries
  241} (1980) L59--L63}.

\bibitem{Starobinsky:1980te}
A.~A. Starobinsky, ``{A New Type of Isotropic Cosmological Models Without
  Singularity},'' \href{http://dx.doi.org/10.1016/0370-2693(80)90670-X}{{\em
  Phys. Lett. B} {\bfseries 91} (1980) 99--102}.

\bibitem{Sato:1981ds}
K.~Sato, ``{Cosmological Baryon Number Domain Structure and the First Order
  Phase Transition of a Vacuum},''
  \href{http://dx.doi.org/10.1016/0370-2693(81)90805-4}{{\em Phys. Lett. B}
  {\bfseries 99} (1981) 66--70}.

\bibitem{Guth:1980zm}
A.~H. Guth, ``{The Inflationary Universe: A Possible Solution to the Horizon
  and Flatness Problems},''
  \href{http://dx.doi.org/10.1103/PhysRevD.23.347}{{\em Phys. Rev. D}
  {\bfseries 23} (1981) 347--356}.

\bibitem{Mukhanov:1981xt}
V.~F. Mukhanov and G.~V. Chibisov, ``{Quantum Fluctuations and a Nonsingular
  Universe},'' {\em JETP Lett.} {\bfseries 33} (1981) 532--535.

\bibitem{Linde:1981mu}
A.~D. Linde, ``{A New Inflationary Universe Scenario: A Possible Solution of
  the Horizon, Flatness, Homogeneity, Isotropy and Primordial Monopole
  Problems},'' \href{http://dx.doi.org/10.1016/0370-2693(82)91219-9}{{\em Phys.
  Lett. B} {\bfseries 108} (1982) 389--393}.

\bibitem{Albrecht:1982wi}
A.~Albrecht and P.~J. Steinhardt, ``{Cosmology for Grand Unified Theories with
  Radiatively Induced Symmetry Breaking},''
  \href{http://dx.doi.org/10.1103/PhysRevLett.48.1220}{{\em Phys. Rev. Lett.}
  {\bfseries 48} (1982) 1220--1223}.

\bibitem{Baumann:2009ds}
D.~Baumann,
  \href{http://dx.doi.org/10.1142/9789814327183_0010}{``{Inflation},''} in {\em
  {Theoretical Advanced Study Institute in Elementary Particle Physics}:
  {Physics of the Large and the Small}}, pp.~523--686.
\newblock 2011.
\newblock \href{http://arxiv.org/abs/0907.5424}{{\ttfamily arXiv:0907.5424
  [hep-th]}}.

\bibitem{Baumann:2018muz}
D.~Baumann, ``{Primordial Cosmology},''
  \href{http://dx.doi.org/10.22323/1.305.0009}{{\em PoS} {\bfseries TASI2017}
  (2018) 009}, \href{http://arxiv.org/abs/1807.03098}{{\ttfamily
  arXiv:1807.03098 [hep-th]}}.

\bibitem{Senatore:2013roa}
L.~Senatore, \href{http://dx.doi.org/10.1142/9789814525220_0006}{``{TASI 2012
  Lectures on Inflation},''} in {\em {Theoretical Advanced Study Institute in
  Elementary Particle Physics}: {Searching for New Physics at Small and Large
  Scales}}, pp.~221--302.
\newblock 2013.

\bibitem{Choudhury:2011sq}
S.~Choudhury and S.~Pal, ``{Brane inflation in background supergravity},''
  \href{http://dx.doi.org/10.1103/PhysRevD.85.043529}{{\em Phys. Rev. D}
  {\bfseries 85} (2012) 043529},
  \href{http://arxiv.org/abs/1102.4206}{{\ttfamily arXiv:1102.4206 [hep-th]}}.

\bibitem{Choudhury:2011jt}
S.~Choudhury and S.~Pal, ``{Fourth level MSSM inflation from new flat
  directions},'' \href{http://dx.doi.org/10.1088/1475-7516/2012/04/018}{{\em
  JCAP} {\bfseries 04} (2012) 018},
  \href{http://arxiv.org/abs/1111.3441}{{\ttfamily arXiv:1111.3441 [hep-ph]}}.

\bibitem{Choudhury:2012yh}
S.~Choudhury and S.~Pal, ``{DBI Galileon inflation in background SUGRA},''
  \href{http://dx.doi.org/10.1016/j.nuclphysb.2013.05.010}{{\em Nucl. Phys. B}
  {\bfseries 874} (2013) 85--114},
  \href{http://arxiv.org/abs/1208.4433}{{\ttfamily arXiv:1208.4433 [hep-th]}}.

\bibitem{Choudhury:2012whm}
S.~Choudhury and S.~Pal, ``{Primordial non-Gaussian features from DBI Galileon
  inflation},'' \href{http://dx.doi.org/10.1140/epjc/s10052-015-3452-3}{{\em
  Eur. Phys. J. C} {\bfseries 75} no.~6, (2015) 241},
  \href{http://arxiv.org/abs/1210.4478}{{\ttfamily arXiv:1210.4478 [hep-th]}}.

\bibitem{Choudhury:2013zna}
S.~Choudhury, T.~Chakraborty, and S.~Pal, ``{Higgs inflation from new K\"ahler
  potential},'' \href{http://dx.doi.org/10.1016/j.nuclphysb.2014.01.002}{{\em
  Nucl. Phys. B} {\bfseries 880} (2014) 155--174},
  \href{http://arxiv.org/abs/1305.0981}{{\ttfamily arXiv:1305.0981 [hep-th]}}.

\bibitem{Choudhury:2013jya}
S.~Choudhury, A.~Mazumdar, and S.~Pal, ``{Low \& High scale MSSM inflation,
  gravitational waves and constraints from Planck},''
  \href{http://dx.doi.org/10.1088/1475-7516/2013/07/041}{{\em JCAP} {\bfseries
  07} (2013) 041}, \href{http://arxiv.org/abs/1305.6398}{{\ttfamily
  arXiv:1305.6398 [hep-ph]}}.

\bibitem{Choudhury:2013iaa}
S.~Choudhury and A.~Mazumdar, ``{An accurate bound on tensor-to-scalar ratio
  and the scale of inflation},''
  \href{http://dx.doi.org/10.1016/j.nuclphysb.2014.03.005}{{\em Nucl. Phys. B}
  {\bfseries 882} (2014) 386--396},
  \href{http://arxiv.org/abs/1306.4496}{{\ttfamily arXiv:1306.4496 [hep-ph]}}.

\bibitem{Choudhury:2013woa}
S.~Choudhury and A.~Mazumdar, ``{Primordial blackholes and gravitational waves
  for an inflection-point model of inflation},''
  \href{http://dx.doi.org/10.1016/j.physletb.2014.04.050}{{\em Phys. Lett. B}
  {\bfseries 733} (2014) 270--275},
  \href{http://arxiv.org/abs/1307.5119}{{\ttfamily arXiv:1307.5119
  [astro-ph.CO]}}.

\bibitem{Choudhury:2014sxa}
S.~Choudhury, A.~Mazumdar, and E.~Pukartas, ``{Constraining ${\cal N}=1$
  supergravity inflationary framework with non-minimal K\"ahler operators},''
  \href{http://dx.doi.org/10.1007/JHEP04(2014)077}{{\em JHEP} {\bfseries 04}
  (2014) 077}, \href{http://arxiv.org/abs/1402.1227}{{\ttfamily arXiv:1402.1227
  [hep-th]}}.

\bibitem{Choudhury:2014uxa}
S.~Choudhury, ``{Constraining N = 1 supergravity inflation with non-minimal
  Kaehler operators using $\delta$N formalism},''
  \href{http://dx.doi.org/10.1007/JHEP04(2014)105}{{\em JHEP} {\bfseries 04}
  (2014) 105}, \href{http://arxiv.org/abs/1402.1251}{{\ttfamily arXiv:1402.1251
  [hep-th]}}.

\bibitem{Choudhury:2014hua}
S.~Choudhury, ``{Inflamagnetogenesis redux: Unzipping sub-Planckian inflation
  via various cosmoparticle probes},''
  \href{http://dx.doi.org/10.1016/j.physletb.2014.06.029}{{\em Phys. Lett. B}
  {\bfseries 735} (2014) 138--145},
  \href{http://arxiv.org/abs/1403.0676}{{\ttfamily arXiv:1403.0676 [hep-th]}}.

\bibitem{Choudhury:2014kma}
S.~Choudhury and A.~Mazumdar, ``{Reconstructing inflationary potential from
  BICEP2 and running of tensor modes},''
  \href{http://arxiv.org/abs/1403.5549}{{\ttfamily arXiv:1403.5549 [hep-th]}}.

\bibitem{Choudhury:2014sua}
S.~Choudhury, ``{Can Effective Field Theory of inflation generate large
  tensor-to-scalar ratio within Randall\textendash{}Sundrum single
  braneworld?},'' \href{http://dx.doi.org/10.1016/j.nuclphysb.2015.02.024}{{\em
  Nucl. Phys. B} {\bfseries 894} (2015) 29--55},
  \href{http://arxiv.org/abs/1406.7618}{{\ttfamily arXiv:1406.7618 [hep-th]}}.

\bibitem{Choudhury:2014hja}
S.~Choudhury, B.~K. Pal, B.~Basu, and P.~Bandyopadhyay, ``{Quantum Gravity
  Effect in Torsion Driven Inflation and CP violation},''
  \href{http://dx.doi.org/10.1007/JHEP10(2015)194}{{\em JHEP} {\bfseries 10}
  (2015) 194}, \href{http://arxiv.org/abs/1409.6036}{{\ttfamily arXiv:1409.6036
  [hep-th]}}.

\bibitem{Choudhury:2015pqa}
S.~Choudhury, ``{Reconstructing inflationary paradigm within Effective Field
  Theory framework},'' \href{http://dx.doi.org/10.1016/j.dark.2015.11.003}{{\em
  Phys. Dark Univ.} {\bfseries 11} (2016) 16--48},
  \href{http://arxiv.org/abs/1508.00269}{{\ttfamily arXiv:1508.00269
  [astro-ph.CO]}}.

\bibitem{Choudhury:2015hvr}
S.~Choudhury and S.~Panda, ``{COSMOS-e\textquoteright{}-GTachyon from string
  theory},'' \href{http://dx.doi.org/10.1140/epjc/s10052-016-4072-2}{{\em Eur.
  Phys. J. C} {\bfseries 76} no.~5, (2016) 278},
  \href{http://arxiv.org/abs/1511.05734}{{\ttfamily arXiv:1511.05734
  [hep-th]}}.

\bibitem{Choudhury:2016wlj}
S.~Choudhury, {\em {Field Theoretic Approaches To Early Universe}}.
\newblock PhD thesis, Indian Statistical Inst., Calcutta, 2016.
\newblock \href{http://arxiv.org/abs/1603.08306}{{\ttfamily arXiv:1603.08306
  [hep-th]}}.

\bibitem{Choudhury:2016cso}
S.~Choudhury, S.~Panda, and R.~Singh, ``{Bell violation in the Sky},''
  \href{http://dx.doi.org/10.1140/epjc/s10052-016-4553-3}{{\em Eur. Phys. J. C}
  {\bfseries 77} no.~2, (2017) 60},
  \href{http://arxiv.org/abs/1607.00237}{{\ttfamily arXiv:1607.00237
  [hep-th]}}.

\bibitem{Choudhury:2017cos}
S.~Choudhury, ``{COSMOS-$e'$- soft Higgsotic attractors},''
  \href{http://dx.doi.org/10.1140/epjc/s10052-017-5001-8}{{\em Eur. Phys. J. C}
  {\bfseries 77} no.~7, (2017) 469},
  \href{http://arxiv.org/abs/1703.01750}{{\ttfamily arXiv:1703.01750
  [hep-th]}}.

\bibitem{Naskar:2017ekm}
A.~Naskar, S.~Choudhury, A.~Banerjee, and S.~Pal, ``{EFT of Inflation:
  Reflections on CMB and Forecasts on LSS Surveys},''
  \href{http://arxiv.org/abs/1706.08051}{{\ttfamily arXiv:1706.08051
  [astro-ph.CO]}}.

\bibitem{Choudhury:2017glj}
S.~Choudhury, ``{CMB from EFT},''
  \href{http://dx.doi.org/10.3390/universe5060155}{{\em Universe} {\bfseries 5}
  no.~6, (2019) 155}, \href{http://arxiv.org/abs/1712.04766}{{\ttfamily
  arXiv:1712.04766 [hep-th]}}.

\bibitem{Choudhury:2018glz}
S.~Choudhury, {\em {Quantum Field Theory approaches to Early Universe
  Cosmology}}.
\newblock LAP LAMBERT Academic Publishing, 5, 2018.

\bibitem{HosseiniMansoori:2023zop}
S.~A. Hosseini~Mansoori, F.~Felegary, M.~Roshan, O.~Akarsu, and M.~Sami,
  ``{$\mathbb{T}^{2}$- inflation: Sourced by energy-momentum squared
  gravity},'' \href{http://arxiv.org/abs/2306.09181}{{\ttfamily
  arXiv:2306.09181 [gr-qc]}}.

\bibitem{Geng:2015fla}
C.-Q. Geng, M.~W. Hossain, R.~Myrzakulov, M.~Sami, and E.~N. Saridakis,
  ``{Quintessential inflation with canonical and noncanonical scalar fields and
  Planck 2015 results},''
  \href{http://dx.doi.org/10.1103/PhysRevD.92.023522}{{\em Phys. Rev. D}
  {\bfseries 92} no.~2, (2015) 023522},
  \href{http://arxiv.org/abs/1502.03597}{{\ttfamily arXiv:1502.03597 [gr-qc]}}.

\bibitem{WaliHossain:2014usl}
M.~Wali~Hossain, R.~Myrzakulov, M.~Sami, and E.~N. Saridakis, ``{Unification of
  inflation and dark energy \`a la quintessential inflation},''
  \href{http://dx.doi.org/10.1142/S0218271815300141}{{\em Int. J. Mod. Phys. D}
  {\bfseries 24} no.~05, (2015) 1530014},
  \href{http://arxiv.org/abs/1410.6100}{{\ttfamily arXiv:1410.6100 [gr-qc]}}.

\bibitem{Hossain:2014coa}
M.~W. Hossain, R.~Myrzakulov, M.~Sami, and E.~N. Saridakis, ``{Class of
  quintessential inflation models with parameter space consistent with
  BICEP2},'' \href{http://dx.doi.org/10.1103/PhysRevD.89.123513}{{\em Phys.
  Rev. D} {\bfseries 89} no.~12, (2014) 123513},
  \href{http://arxiv.org/abs/1404.1445}{{\ttfamily arXiv:1404.1445 [gr-qc]}}.

\bibitem{Hossain:2014xha}
M.~W. Hossain, R.~Myrzakulov, M.~Sami, and E.~N. Saridakis, ``{Variable
  gravity: A suitable framework for quintessential inflation},''
  \href{http://dx.doi.org/10.1103/PhysRevD.90.023512}{{\em Phys. Rev. D}
  {\bfseries 90} no.~2, (2014) 023512},
  \href{http://arxiv.org/abs/1402.6661}{{\ttfamily arXiv:1402.6661 [gr-qc]}}.

\bibitem{Martin:2013tda}
J.~Martin, C.~Ringeval, and V.~Vennin, ``{Encyclop\ae{}dia Inflationaris},''
  \href{http://dx.doi.org/10.1016/j.dark.2014.01.003}{{\em Phys. Dark Univ.}
  {\bfseries 5-6} (2014) 75--235},
  \href{http://arxiv.org/abs/1303.3787}{{\ttfamily arXiv:1303.3787
  [astro-ph.CO]}}.

\bibitem{Benetti:2013cja}
M.~Benetti, ``{Updating constraints on inflationary features in the primordial
  power spectrum with the Planck data},''
  \href{http://dx.doi.org/10.1103/PhysRevD.88.087302}{{\em Phys. Rev. D}
  {\bfseries 88} (2013) 087302},
  \href{http://arxiv.org/abs/1308.6406}{{\ttfamily arXiv:1308.6406
  [astro-ph.CO]}}.

\bibitem{Martin:2013nzq}
J.~Martin, C.~Ringeval, R.~Trotta, and V.~Vennin, ``{The Best Inflationary
  Models After Planck},''
  \href{http://dx.doi.org/10.1088/1475-7516/2014/03/039}{{\em JCAP} {\bfseries
  03} (2014) 039}, \href{http://arxiv.org/abs/1312.3529}{{\ttfamily
  arXiv:1312.3529 [astro-ph.CO]}}.

\bibitem{Creminelli:2014oaa}
P.~Creminelli, D.~L\'opez~Nacir, M.~Simonovi\'c, G.~Trevisan, and
  M.~Zaldarriaga, ``{$\phi^2$ or Not $\phi^2$: Testing the Simplest
  Inflationary Potential},''
  \href{http://dx.doi.org/10.1103/PhysRevLett.112.241303}{{\em Phys. Rev.
  Lett.} {\bfseries 112} no.~24, (2014) 241303},
  \href{http://arxiv.org/abs/1404.1065}{{\ttfamily arXiv:1404.1065
  [astro-ph.CO]}}.

\bibitem{Dai:2014jja}
L.~Dai, M.~Kamionkowski, and J.~Wang, ``{Reheating constraints to inflationary
  models},'' \href{http://dx.doi.org/10.1103/PhysRevLett.113.041302}{{\em Phys.
  Rev. Lett.} {\bfseries 113} (2014) 041302},
  \href{http://arxiv.org/abs/1404.6704}{{\ttfamily arXiv:1404.6704
  [astro-ph.CO]}}.

\bibitem{Benetti:2016tvm}
M.~Benetti and J.~S. Alcaniz, ``{Bayesian analysis of inflationary features in
  Planck and SDSS data},''
  \href{http://dx.doi.org/10.1103/PhysRevD.94.023526}{{\em Phys. Rev. D}
  {\bfseries 94} no.~2, (2016) 023526},
  \href{http://arxiv.org/abs/1604.08156}{{\ttfamily arXiv:1604.08156
  [astro-ph.CO]}}.

\bibitem{Campista:2017ovq}
M.~Campista, M.~Benetti, and J.~Alcaniz, ``{Testing non-minimally coupled
  inflation with CMB data: a Bayesian analysis},''
  \href{http://dx.doi.org/10.1088/1475-7516/2017/09/010}{{\em JCAP} {\bfseries
  09} (2017) 010}, \href{http://arxiv.org/abs/1705.08877}{{\ttfamily
  arXiv:1705.08877 [astro-ph.CO]}}.

\bibitem{Keeley:2020rmo}
R.~E. Keeley, A.~Shafieloo, D.~K. Hazra, and T.~Souradeep, ``{Inflation Wars: A
  New Hope},'' \href{http://dx.doi.org/10.1088/1475-7516/2020/09/055}{{\em
  JCAP} {\bfseries 09} (2020) 055},
  \href{http://arxiv.org/abs/2006.12710}{{\ttfamily arXiv:2006.12710
  [astro-ph.CO]}}.

\bibitem{Vagnozzi:2020rcz}
S.~Vagnozzi, E.~Di~Valentino, S.~Gariazzo, A.~Melchiorri, O.~Mena, and J.~Silk,
  ``{The galaxy power spectrum take on spatial curvature and cosmic
  concordance},'' \href{http://dx.doi.org/10.1016/j.dark.2021.100851}{{\em
  Phys. Dark Univ.} {\bfseries 33} (2021) 100851},
  \href{http://arxiv.org/abs/2010.02230}{{\ttfamily arXiv:2010.02230
  [astro-ph.CO]}}.

\bibitem{Vagnozzi:2020dfn}
S.~Vagnozzi, A.~Loeb, and M.~Moresco, ``{Eppur \`e piatto? The Cosmic
  Chronometers Take on Spatial Curvature and Cosmic Concordance},''
  \href{http://dx.doi.org/10.3847/1538-4357/abd4df}{{\em Astrophys. J.}
  {\bfseries 908} no.~1, (2021) 84},
  \href{http://arxiv.org/abs/2011.11645}{{\ttfamily arXiv:2011.11645
  [astro-ph.CO]}}.

\bibitem{Vagnozzi:2023lwo}
S.~Vagnozzi, ``{Inflationary interpretation of the stochastic gravitational
  wave background signal detected by pulsar timing array experiments},''
  \href{http://arxiv.org/abs/2306.16912}{{\ttfamily arXiv:2306.16912
  [astro-ph.CO]}}.

\bibitem{Cabass:2022wjy}
G.~Cabass, M.~M. Ivanov, O.~H.~E. Philcox, M.~Simonovi\'c, and M.~Zaldarriaga,
  ``{Constraints on Single-Field Inflation from the BOSS Galaxy Survey},''
  \href{http://dx.doi.org/10.1103/PhysRevLett.129.021301}{{\em Phys. Rev.
  Lett.} {\bfseries 129} no.~2, (2022) 021301},
  \href{http://arxiv.org/abs/2201.07238}{{\ttfamily arXiv:2201.07238
  [astro-ph.CO]}}.

\bibitem{Cabass:2022ymb}
G.~Cabass, M.~M. Ivanov, O.~H.~E. Philcox, M.~Simonovi\'c, and M.~Zaldarriaga,
  ``{Constraints on multifield inflation from the BOSS galaxy survey},''
  \href{http://dx.doi.org/10.1103/PhysRevD.106.043506}{{\em Phys. Rev. D}
  {\bfseries 106} no.~4, (2022) 043506},
  \href{http://arxiv.org/abs/2204.01781}{{\ttfamily arXiv:2204.01781
  [astro-ph.CO]}}.

\bibitem{CMB-S4:2016ple}
{\bfseries CMB-S4} Collaboration, K.~N. Abazajian {\em et~al.}, ``{CMB-S4
  Science Book, First Edition},''
  \href{http://arxiv.org/abs/1610.02743}{{\ttfamily arXiv:1610.02743
  [astro-ph.CO]}}.

\bibitem{SimonsObservatory:2018koc}
{\bfseries Simons Observatory} Collaboration, P.~Ade {\em et~al.}, ``{The
  Simons Observatory: Science goals and forecasts},''
  \href{http://dx.doi.org/10.1088/1475-7516/2019/02/056}{{\em JCAP} {\bfseries
  02} (2019) 056}, \href{http://arxiv.org/abs/1808.07445}{{\ttfamily
  arXiv:1808.07445 [astro-ph.CO]}}.

\bibitem{SimonsObservatory:2019qwx}
{\bfseries Simons Observatory} Collaboration, M.~H. Abitbol {\em et~al.},
  ``{The Simons Observatory: Astro2020 Decadal Project Whitepaper},'' {\em
  Bull. Am. Astron. Soc.} {\bfseries 51} (2019) 147,
  \href{http://arxiv.org/abs/1907.08284}{{\ttfamily arXiv:1907.08284
  [astro-ph.IM]}}.

\bibitem{Kamionkowski:2015yta}
M.~Kamionkowski and E.~D. Kovetz, ``{The Quest for B Modes from Inflationary
  Gravitational Waves},''
  \href{http://dx.doi.org/10.1146/annurev-astro-081915-023433}{{\em Ann. Rev.
  Astron. Astrophys.} {\bfseries 54} (2016) 227--269},
  \href{http://arxiv.org/abs/1510.06042}{{\ttfamily arXiv:1510.06042
  [astro-ph.CO]}}.

\bibitem{Choudhury:2020yaa}
S.~Choudhury, ``{The Cosmological OTOC: Formulating new cosmological
  micro-canonical correlation functions for random chaotic fluctuations in
  Out-of-Equilibrium Quantum Statistical Field Theory},''
  \href{http://dx.doi.org/10.3390/sym12091527}{{\em Symmetry} {\bfseries 12}
  no.~9, (2020) 1527}, \href{http://arxiv.org/abs/2005.11750}{{\ttfamily
  arXiv:2005.11750 [hep-th]}}.

\bibitem{Choudhury:2021tuu}
S.~Choudhury, ``{The Cosmological OTOC: A New Proposal for Quantifying
  Auto-correlated Random Non-chaotic Primordial Fluctuations},''
  \href{http://dx.doi.org/10.20944/preprints202102.0616.v1}{{\em Symmetry}
  {\bfseries 13} no.~4, (2021) 599},
  \href{http://arxiv.org/abs/2106.01305}{{\ttfamily arXiv:2106.01305
  [physics.gen-ph]}}.

\bibitem{Adhikari:2021ked}
K.~Adhikari, S.~Choudhury, H.~N. Pandya, and R.~Srivastava, ``{PGW Circuit
  Complexity},'' \href{http://arxiv.org/abs/2108.10334}{{\ttfamily
  arXiv:2108.10334 [gr-qc]}}.

\bibitem{Akama:2023jsb}
S.~Akama and H.~W.~H. Tahara, ``{Imprints of primordial gravitational waves
  with non-Bunch-Davies initial states on CMB bispectra},''
  \href{http://arxiv.org/abs/2306.17752}{{\ttfamily arXiv:2306.17752 [gr-qc]}}.

\bibitem{Albayrak:2023hie}
S.~Albayrak, P.~Benincasa, and C.~D. Pueyo, ``{Perturbative Unitarity and the
  Wavefunction of the Universe},''
  \href{http://arxiv.org/abs/2305.19686}{{\ttfamily arXiv:2305.19686
  [hep-th]}}.

\bibitem{Choudhury:2022mch}
S.~Choudhury, ``{Entanglement negativity in de Sitter biverse from Stringy
  Axionic Bell pair: An analysis using Bunch-Davies vacuum},''
  \href{http://arxiv.org/abs/2301.05203}{{\ttfamily arXiv:2301.05203
  [hep-th]}}.

\bibitem{Colas:2022kfu}
T.~Colas, J.~Grain, and V.~Vennin, ``{Quantum recoherence in the early
  universe},'' \href{http://arxiv.org/abs/2212.09486}{{\ttfamily
  arXiv:2212.09486 [gr-qc]}}.

\bibitem{Aalsma:2022eru}
L.~Aalsma, M.~M. Faruk, J.~P. van~der Schaar, M.~Visser, and J.~de~Witte,
  ``{Late-Time Correlators and Complex Geodesics in de Sitter Space},''
  \href{http://arxiv.org/abs/2212.01394}{{\ttfamily arXiv:2212.01394
  [hep-th]}}.

\bibitem{Chapman:2022mqd}
S.~Chapman, D.~A. Galante, E.~Harris, S.~U. Sheorey, and D.~Vegh, ``{Complex
  geodesics in de Sitter space},''
  \href{http://dx.doi.org/10.1007/JHEP03(2023)006}{{\em JHEP} {\bfseries 03}
  (2023) 006}, \href{http://arxiv.org/abs/2212.01398}{{\ttfamily
  arXiv:2212.01398 [hep-th]}}.

\bibitem{Letey:2022hdp}
M.~I. Letey, Z.~Shumaylov, F.~J. Agocs, W.~J. Handley, M.~P. Hobson, and A.~N.
  Lasenby, ``{Quantum Initial Conditions for Curved Inflating Universes},''
  \href{http://arxiv.org/abs/2211.17248}{{\ttfamily arXiv:2211.17248 [gr-qc]}}.

\bibitem{Penna-Lima:2022dmx}
M.~Penna-Lima, N.~Pinto-Neto, and S.~D.~P. Vitenti, ``{New formalism to define
  vacuum states for scalar fields in curved spacetimes},''
  \href{http://dx.doi.org/10.1103/PhysRevD.107.065019}{{\em Phys. Rev. D}
  {\bfseries 107} no.~6, (2023) 065019},
  \href{http://arxiv.org/abs/2207.08270}{{\ttfamily arXiv:2207.08270 [gr-qc]}}.

\bibitem{Kanno:2022mkx}
S.~Kanno and M.~Sasaki, ``{Graviton non-gaussianity in
  \ensuremath{\alpha}-vacuum},''
  \href{http://dx.doi.org/10.1007/JHEP08(2022)210}{{\em JHEP} {\bfseries 08}
  (2022) 210}, \href{http://arxiv.org/abs/2206.03667}{{\ttfamily
  arXiv:2206.03667 [hep-th]}}.

\bibitem{Fumagalli:2021mpc}
J.~Fumagalli, G.~A. Palma, S.~Renaux-Petel, S.~Sypsas, L.~T. Witkowski, and
  C.~Zenteno, ``{Primordial gravitational waves from excited states},''
  \href{http://dx.doi.org/10.1007/JHEP03(2022)196}{{\em JHEP} {\bfseries 03}
  (2022) 196}, \href{http://arxiv.org/abs/2111.14664}{{\ttfamily
  arXiv:2111.14664 [astro-ph.CO]}}.

\bibitem{Sleight:2021plv}
C.~Sleight and M.~Taronna, ``{From dS to AdS and back},''
  \href{http://dx.doi.org/10.1007/JHEP12(2021)074}{{\em JHEP} {\bfseries 12}
  (2021) 074}, \href{http://arxiv.org/abs/2109.02725}{{\ttfamily
  arXiv:2109.02725 [hep-th]}}.

\bibitem{Chen:2010bka}
X.~Chen, ``{Folded Resonant Non-Gaussianity in General Single Field
  Inflation},'' \href{http://dx.doi.org/10.1088/1475-7516/2010/12/003}{{\em
  JCAP} {\bfseries 12} (2010) 003},
  \href{http://arxiv.org/abs/1008.2485}{{\ttfamily arXiv:1008.2485 [hep-th]}}.

\bibitem{Wang:2014kqa}
Y.~Wang and W.~Xue, ``{Inflation and Alternatives with Blue Tensor Spectra},''
  \href{http://dx.doi.org/10.1088/1475-7516/2014/10/075}{{\em JCAP} {\bfseries
  10} (2014) 075}, \href{http://arxiv.org/abs/1403.5817}{{\ttfamily
  arXiv:1403.5817 [astro-ph.CO]}}.

\bibitem{Ashoorioon:2014nta}
A.~Ashoorioon, K.~Dimopoulos, M.~M. Sheikh-Jabbari, and G.~Shiu,
  ``{Non-Bunch\textendash{}Davis initial state reconciles chaotic models with
  BICEP and Planck},''
  \href{http://dx.doi.org/10.1016/j.physletb.2014.08.038}{{\em Phys. Lett. B}
  {\bfseries 737} (2014) 98--102},
  \href{http://arxiv.org/abs/1403.6099}{{\ttfamily arXiv:1403.6099 [hep-th]}}.

\bibitem{Ashoorioon:2013eia}
A.~Ashoorioon, K.~Dimopoulos, M.~M. Sheikh-Jabbari, and G.~Shiu,
  ``{Reconciliation of High Energy Scale Models of Inflation with Planck},''
  \href{http://dx.doi.org/10.1088/1475-7516/2014/02/025}{{\em JCAP} {\bfseries
  02} (2014) 025}, \href{http://arxiv.org/abs/1306.4914}{{\ttfamily
  arXiv:1306.4914 [hep-th]}}.

\bibitem{Huang:2015gka}
Q.-G. Huang and S.~Wang, ``{No evidence for the blue-tilted power spectrum of
  relic gravitational waves},''
  \href{http://dx.doi.org/10.1088/1475-7516/2015/06/021}{{\em JCAP} {\bfseries
  06} (2015) 021}, \href{http://arxiv.org/abs/1502.02541}{{\ttfamily
  arXiv:1502.02541 [astro-ph.CO]}}.

\bibitem{Huang:2015gca}
Q.-G. Huang, S.~Wang, and W.~Zhao, ``{Forecasting sensitivity on tilt of power
  spectrum of primordial gravitational waves after Planck satellite},''
  \href{http://dx.doi.org/10.1088/1475-7516/2015/10/035}{{\em JCAP} {\bfseries
  10} (2015) 035}, \href{http://arxiv.org/abs/1509.02676}{{\ttfamily
  arXiv:1509.02676 [astro-ph.CO]}}.

\bibitem{Huang:2017gmp}
Q.-G. Huang and S.~Wang, ``{Optimistic estimation on probing primordial
  gravitational waves with CMB B-mode polarization},''
  \href{http://dx.doi.org/10.1093/mnras/sty3262}{{\em Mon. Not. Roy. Astron.
  Soc.} {\bfseries 483} no.~2, (2019) 2177--2184},
  \href{http://arxiv.org/abs/1701.06115}{{\ttfamily arXiv:1701.06115
  [astro-ph.CO]}}.

\bibitem{amaro2017laser}
P.~Amaro-Seoane, H.~Audley, S.~Babak, J.~Baker, E.~Barausse, P.~Bender,
  E.~Berti, P.~Binetruy, M.~Born, D.~Bortoluzzi, {\em et~al.}, ``Laser
  interferometer space antenna,'' {\em arXiv preprint arXiv:1702.00786} (2017)
  .

\bibitem{Kawamura:2011zz}
S.~Kawamura {\em et~al.}, ``{The Japanese space gravitational wave antenna:
  DECIGO},'' \href{http://dx.doi.org/10.1088/0264-9381/28/9/094011}{{\em Class.
  Quant. Grav.} {\bfseries 28} (2011) 094011}.

\bibitem{Crowder:2005nr}
J.~Crowder and N.~J. Cornish, ``{Beyond LISA: Exploring future gravitational
  wave missions},'' \href{http://dx.doi.org/10.1103/PhysRevD.72.083005}{{\em
  Phys. Rev. D} {\bfseries 72} (2005) 083005},
  \href{http://arxiv.org/abs/gr-qc/0506015}{{\ttfamily arXiv:gr-qc/0506015}}.

\bibitem{Punturo:2010zz}
M.~Punturo {\em et~al.}, ``{The Einstein Telescope: A third-generation
  gravitational wave observatory},''
  \href{http://dx.doi.org/10.1088/0264-9381/27/19/194002}{{\em Class. Quant.
  Grav.} {\bfseries 27} (2010) 194002}.

\bibitem{Reitze:2019iox}
D.~Reitze {\em et~al.}, ``{Cosmic Explorer: The U.S. Contribution to
  Gravitational-Wave Astronomy beyond LIGO},'' {\em Bull. Am. Astron. Soc.}
  {\bfseries 51} no.~7, (2019) 035,
  \href{http://arxiv.org/abs/1907.04833}{{\ttfamily arXiv:1907.04833
  [astro-ph.IM]}}.

\bibitem{KAGRA:2018plz}
{\bfseries KAGRA} Collaboration, T.~Akutsu {\em et~al.}, ``{KAGRA: 2.5
  Generation Interferometric Gravitational Wave Detector},''
  \href{http://dx.doi.org/10.1038/s41550-018-0658-y}{{\em Nature Astron.}
  {\bfseries 3} no.~1, (2019) 35--40},
  \href{http://arxiv.org/abs/1811.08079}{{\ttfamily arXiv:1811.08079 [gr-qc]}}.

\bibitem{VIRGO:2014yos}
{\bfseries VIRGO} Collaboration, F.~Acernese {\em et~al.}, ``{Advanced Virgo: a
  second-generation interferometric gravitational wave detector},''
  \href{http://dx.doi.org/10.1088/0264-9381/32/2/024001}{{\em Class. Quant.
  Grav.} {\bfseries 32} no.~2, (2015) 024001},
  \href{http://arxiv.org/abs/1408.3978}{{\ttfamily arXiv:1408.3978 [gr-qc]}}.

\bibitem{LIGOScientific:2014pky}
{\bfseries LIGO Scientific} Collaboration, J.~Aasi {\em et~al.}, ``{Advanced
  LIGO},'' \href{http://dx.doi.org/10.1088/0264-9381/32/7/074001}{{\em Class.
  Quant. Grav.} {\bfseries 32} (2015) 074001},
  \href{http://arxiv.org/abs/1411.4547}{{\ttfamily arXiv:1411.4547 [gr-qc]}}.

\bibitem{Guendelman:2024ezk}
E.~Guendelman, ``{Holomorphic gravity and its regularization of Locally Signed
  Coordinate Invariance},'' \href{http://arxiv.org/abs/2402.00140}{{\ttfamily
  arXiv:2402.00140 [gr-qc]}}.

\bibitem{Guendelman:2023spd}
E.~Guendelman, ``{Holomorphic general coordinate invariant modified measure
  gravitational theory},''
  \href{http://dx.doi.org/10.1016/j.aop.2023.169466}{{\em Annals Phys.}
  {\bfseries 458} (2023) 169466},
  \href{http://arxiv.org/abs/2308.09246}{{\ttfamily arXiv:2308.09246 [gr-qc]}}.

\bibitem{Guendelman:2023vso}
E.~Guendelman, ``{Signed Coordinate Invariance, invariant lagrangians and
  manifolds, the time problem in quantum cosmology, quantum space time,
  spacetimes and antispacetimes},''
  \href{http://arxiv.org/abs/2304.04056}{{\ttfamily arXiv:2304.04056 [gr-qc]}}.

\bibitem{Hossain:2014ova}
M.~W. Hossain, R.~Myrzakulov, M.~Sami, and E.~N. Saridakis, ``{Evading Lyth
  bound in models of quintessential inflation},''
  \href{http://dx.doi.org/10.1016/j.physletb.2014.08.051}{{\em Phys. Lett. B}
  {\bfseries 737} (2014) 191--195},
  \href{http://arxiv.org/abs/1405.7491}{{\ttfamily arXiv:1405.7491 [gr-qc]}}.

\bibitem{Piao:2004tq}
Y.-S. Piao and Y.-Z. Zhang, ``{Phantom inflation and primordial perturbation
  spectrum},'' \href{http://dx.doi.org/10.1103/PhysRevD.70.063513}{{\em Phys.
  Rev. D} {\bfseries 70} (2004) 063513},
  \href{http://arxiv.org/abs/astro-ph/0401231}{{\ttfamily
  arXiv:astro-ph/0401231}}.

\bibitem{Liu:2010dh}
Z.-G. Liu, J.~Zhang, and Y.-S. Piao, ``{Phantom Inflation with A Steplike
  Potential},'' \href{http://dx.doi.org/10.1016/j.physletb.2010.12.055}{{\em
  Phys. Lett. B} {\bfseries 697} (2011) 407--411},
  \href{http://arxiv.org/abs/1012.0673}{{\ttfamily arXiv:1012.0673 [gr-qc]}}.

\bibitem{Brandenberger:2008nx}
R.~H. Brandenberger, ``{String Gas Cosmology},''
\newblock 8, 2008.
\newblock \href{http://arxiv.org/abs/0808.0746}{{\ttfamily arXiv:0808.0746
  [hep-th]}}.

\bibitem{Brandenberger:2011et}
R.~H. Brandenberger, ``{String Gas Cosmology: Progress and Problems},''
  \href{http://dx.doi.org/10.1088/0264-9381/28/20/204005}{{\em Class. Quant.
  Grav.} {\bfseries 28} (2011) 204005},
  \href{http://arxiv.org/abs/1105.3247}{{\ttfamily arXiv:1105.3247 [hep-th]}}.

\bibitem{Khoury:2001wf}
J.~Khoury, B.~A. Ovrut, P.~J. Steinhardt, and N.~Turok, ``{The Ekpyrotic
  universe: Colliding branes and the origin of the hot big bang},''
  \href{http://dx.doi.org/10.1103/PhysRevD.64.123522}{{\em Phys. Rev. D}
  {\bfseries 64} (2001) 123522},
  \href{http://arxiv.org/abs/hep-th/0103239}{{\ttfamily arXiv:hep-th/0103239}}.

\bibitem{Lehners:2008vx}
J.-L. Lehners, ``{Ekpyrotic and Cyclic Cosmology},''
  \href{http://dx.doi.org/10.1016/j.physrep.2008.06.001}{{\em Phys. Rept.}
  {\bfseries 465} (2008) 223--263},
  \href{http://arxiv.org/abs/0806.1245}{{\ttfamily arXiv:0806.1245
  [astro-ph]}}.

\bibitem{Brandenberger:2016vhg}
R.~Brandenberger and P.~Peter, ``{Bouncing Cosmologies: Progress and
  Problems},'' \href{http://dx.doi.org/10.1007/s10701-016-0057-0}{{\em Found.
  Phys.} {\bfseries 47} no.~6, (2017) 797--850},
  \href{http://arxiv.org/abs/1603.05834}{{\ttfamily arXiv:1603.05834
  [hep-th]}}.

\bibitem{Brandenberger:2023wtd}
R.~Brandenberger and G.~A. Mitchell, ``{A bouncing cosmology from VECROs},''
  \href{http://dx.doi.org/10.1140/epjc/s10052-023-11501-2}{{\em Eur. Phys. J.
  C} {\bfseries 83} no.~4, (2023) 308},
  \href{http://arxiv.org/abs/2302.12924}{{\ttfamily arXiv:2302.12924
  [hep-th]}}.

\bibitem{Koehn:2015vvy}
M.~Koehn, J.-L. Lehners, and B.~Ovrut, ``{Nonsingular bouncing cosmology:
  Consistency of the effective description},''
  \href{http://dx.doi.org/10.1103/PhysRevD.93.103501}{{\em Phys. Rev. D}
  {\bfseries 93} no.~10, (2016) 103501},
  \href{http://arxiv.org/abs/1512.03807}{{\ttfamily arXiv:1512.03807
  [hep-th]}}.

\bibitem{Lehners:2015mra}
J.-L. Lehners and E.~Wilson-Ewing, ``{Running of the scalar spectral index in
  bouncing cosmologies},''
  \href{http://dx.doi.org/10.1088/1475-7516/2015/10/038}{{\em JCAP} {\bfseries
  10} (2015) 038}, \href{http://arxiv.org/abs/1507.08112}{{\ttfamily
  arXiv:1507.08112 [astro-ph.CO]}}.

\bibitem{Ijjas:2018qbo}
A.~Ijjas and P.~J. Steinhardt, ``{Bouncing Cosmology made simple},''
  \href{http://dx.doi.org/10.1088/1361-6382/aac482}{{\em Class. Quant. Grav.}
  {\bfseries 35} no.~13, (2018) 135004},
  \href{http://arxiv.org/abs/1803.01961}{{\ttfamily arXiv:1803.01961
  [astro-ph.CO]}}.

\bibitem{Bhargava:2020fhl}
P.~Bhargava, S.~Choudhury, S.~Chowdhury, A.~Mishara, S.~P. Selvam, S.~Panda,
  and G.~D. Pasquino, ``{Quantum aspects of chaos and complexity from bouncing
  cosmology: A study with two-mode single field squeezed state formalism},''
  \href{http://dx.doi.org/10.21468/SciPostPhysCore.4.4.026}{{\em SciPost Phys.
  Core} {\bfseries 4} (2021) 026},
  \href{http://arxiv.org/abs/2009.03893}{{\ttfamily arXiv:2009.03893
  [hep-th]}}.

\bibitem{Agullo:2016tjh}
I.~Agullo and P.~Singh, {\em {Loop Quantum Cosmology}},
  \href{http://dx.doi.org/10.1142/9789813220003_0007}{pp.~183--240}.
\newblock WSP, 2017.
\newblock \href{http://arxiv.org/abs/1612.01236}{{\ttfamily arXiv:1612.01236
  [gr-qc]}}.

\bibitem{Bojowald:2005epg}
M.~Bojowald, ``{Loop quantum cosmology},''
  \href{http://dx.doi.org/10.12942/lrr-2005-11}{{\em Living Rev. Rel.}
  {\bfseries 8} (2005) 11},
  \href{http://arxiv.org/abs/gr-qc/0601085}{{\ttfamily arXiv:gr-qc/0601085}}.

\bibitem{Bojowald:1999tr}
M.~Bojowald, ``{Loop quantum cosmology. I. Kinematics},''
  \href{http://dx.doi.org/10.1088/0264-9381/17/6/312}{{\em Class. Quant. Grav.}
  {\bfseries 17} (2000) 1489--1508},
  \href{http://arxiv.org/abs/gr-qc/9910103}{{\ttfamily arXiv:gr-qc/9910103}}.

\bibitem{Koshelev:2022olc}
A.~S. Koshelev, K.~S. Kumar, and A.~A. Starobinsky, ``{Generalized non-local
  $R^2$-like inflation},'' \href{http://arxiv.org/abs/2209.02515}{{\ttfamily
  arXiv:2209.02515 [hep-th]}}.

\bibitem{Koshelev:2020foq}
A.~S. Koshelev, K.~Sravan~Kumar, A.~Mazumdar, and A.~A. Starobinsky,
  ``{Non-Gaussianities and tensor-to-scalar ratio in non-local R$^{2}$-like
  inflation},'' \href{http://dx.doi.org/10.1007/JHEP06(2020)152}{{\em JHEP}
  {\bfseries 06} (2020) 152}, \href{http://arxiv.org/abs/2003.00629}{{\ttfamily
  arXiv:2003.00629 [hep-th]}}.

\bibitem{Choudhury:2018rjl}
S.~Choudhury, A.~Mukherjee, P.~Chauhan, and S.~Bhattacherjee, ``{Quantum
  Out-of-Equilibrium Cosmology},''
  \href{http://dx.doi.org/10.1140/epjc/s10052-019-6751-2}{{\em Eur. Phys. J. C}
  {\bfseries 79} no.~4, (2019) 320},
  \href{http://arxiv.org/abs/1809.02732}{{\ttfamily arXiv:1809.02732
  [hep-th]}}.

\bibitem{Choudhury:2018bcf}
S.~Choudhury and A.~Mukherjee, ``{Quantum randomness in the Sky},''
  \href{http://dx.doi.org/10.1140/epjc/s10052-019-7072-1}{{\em Eur. Phys. J. C}
  {\bfseries 79} no.~7, (2019) 554},
  \href{http://arxiv.org/abs/1812.04107}{{\ttfamily arXiv:1812.04107
  [physics.gen-ph]}}.

\bibitem{Planck:2018jri}
{\bfseries Planck} Collaboration, Y.~Akrami {\em et~al.}, ``{Planck 2018
  results. X. Constraints on inflation},''
  \href{http://dx.doi.org/10.1051/0004-6361/201833887}{{\em Astron. Astrophys.}
  {\bfseries 641} (2020) A10},
  \href{http://arxiv.org/abs/1807.06211}{{\ttfamily arXiv:1807.06211
  [astro-ph.CO]}}.

\bibitem{Zhao:2013bba}
W.~Zhao, Y.~Zhang, X.-P. You, and Z.-H. Zhu, ``{Constraints of relic
  gravitational waves by pulsar timing arrays: Forecasts for the FAST and SKA
  projects},'' \href{http://dx.doi.org/10.1103/PhysRevD.87.124012}{{\em Phys.
  Rev. D} {\bfseries 87} no.~12, (2013) 124012},
  \href{http://arxiv.org/abs/1303.6718}{{\ttfamily arXiv:1303.6718
  [astro-ph.CO]}}.

\bibitem{Planck:2018vyg}
{\bfseries Planck} Collaboration, N.~Aghanim {\em et~al.}, ``{Planck 2018
  results. VI. Cosmological parameters},''
  \href{http://dx.doi.org/10.1051/0004-6361/201833910}{{\em Astron. Astrophys.}
  {\bfseries 641} (2020) A6}, \href{http://arxiv.org/abs/1807.06209}{{\ttfamily
  arXiv:1807.06209 [astro-ph.CO]}}. [Erratum: Astron.Astrophys. 652, C4
  (2021)].

\bibitem{Kristiano:2022maq}
J.~Kristiano and J.~Yokoyama, ``{Ruling Out Primordial Black Hole Formation
  From Single-Field Inflation},''
  \href{http://arxiv.org/abs/2211.03395}{{\ttfamily arXiv:2211.03395
  [hep-th]}}.

\bibitem{Kristiano:2023scm}
J.~Kristiano and J.~Yokoyama, ``{Response to criticism on ''Ruling Out
  Primordial Black Hole Formation From Single-Field Inflation'': A note on
  bispectrum and one-loop correction in single-field inflation with primordial
  black hole formation},'' \href{http://arxiv.org/abs/2303.00341}{{\ttfamily
  arXiv:2303.00341 [hep-th]}}.

\bibitem{Choudhury:2023vuj}
S.~Choudhury, M.~R. Gangopadhyay, and M.~Sami, ``{No-go for the formation of
  heavy mass Primordial Black Holes in Single Field Inflation},''
  \href{http://arxiv.org/abs/2301.10000}{{\ttfamily arXiv:2301.10000
  [astro-ph.CO]}}.

\bibitem{Choudhury:2023jlt}
S.~Choudhury, S.~Panda, and M.~Sami, ``{No-go for PBH formation in EFT of
  single field inflation},'' \href{http://arxiv.org/abs/2302.05655}{{\ttfamily
  arXiv:2302.05655 [astro-ph.CO]}}.

\bibitem{Choudhury:2023rks}
S.~Choudhury, S.~Panda, and M.~Sami, ``{Quantum loop effects on the power
  spectrum and constraints on primordial black holes},''
  \href{http://arxiv.org/abs/2303.06066}{{\ttfamily arXiv:2303.06066
  [astro-ph.CO]}}.

\bibitem{Choudhury:2023hvf}
S.~Choudhury, S.~Panda, and M.~Sami, ``{Galileon inflation evades the no-go for
  PBH formation in the single-field framework},''
  \href{http://arxiv.org/abs/2304.04065}{{\ttfamily arXiv:2304.04065
  [astro-ph.CO]}}.

\bibitem{Choudhury:2023kdb}
S.~Choudhury, A.~Karde, S.~Panda, and M.~Sami, ``{Primordial non-Gaussianity
  from ultra slow-roll Galileon inflation},''
  \href{http://arxiv.org/abs/2306.12334}{{\ttfamily arXiv:2306.12334
  [astro-ph.CO]}}.

\bibitem{Riotto:2023hoz}
A.~Riotto, ``{The Primordial Black Hole Formation from Single-Field Inflation
  is Not Ruled Out},'' \href{http://arxiv.org/abs/2301.00599}{{\ttfamily
  arXiv:2301.00599 [astro-ph.CO]}}.

\bibitem{Riotto:2023gpm}
A.~Riotto, ``{The Primordial Black Hole Formation from Single-Field Inflation
  is Still Not Ruled Out},'' \href{http://arxiv.org/abs/2303.01727}{{\ttfamily
  arXiv:2303.01727 [astro-ph.CO]}}.

\bibitem{Firouzjahi:2023aum}
H.~Firouzjahi, ``{One-loop Corrections in Power Spectrum in Single Field
  Inflation},'' \href{http://arxiv.org/abs/2303.12025}{{\ttfamily
  arXiv:2303.12025 [astro-ph.CO]}}.

\bibitem{Motohashi:2023syh}
H.~Motohashi and Y.~Tada, ``{Squeezed bispectrum and one-loop corrections in
  transient constant-roll inflation},''
  \href{http://arxiv.org/abs/2303.16035}{{\ttfamily arXiv:2303.16035
  [astro-ph.CO]}}.

\bibitem{Firouzjahi:2023ahg}
H.~Firouzjahi and A.~Riotto, ``{Primordial Black Holes and Loops in
  Single-Field Inflation},'' \href{http://arxiv.org/abs/2304.07801}{{\ttfamily
  arXiv:2304.07801 [astro-ph.CO]}}.

\bibitem{Franciolini:2023lgy}
G.~Franciolini, A.~Iovino, Junior., M.~Taoso, and A.~Urbano, ``{One loop to
  rule them all: Perturbativity in the presence of ultra slow-roll dynamics},''
  \href{http://arxiv.org/abs/2305.03491}{{\ttfamily arXiv:2305.03491
  [astro-ph.CO]}}.

\bibitem{Tasinato:2023ukp}
G.~Tasinato, ``{A large $|\eta|$ approach to single field inflation},''
  \href{http://arxiv.org/abs/2305.11568}{{\ttfamily arXiv:2305.11568
  [hep-th]}}.

\bibitem{Cheng:2023ikq}
S.-L. Cheng, D.-S. Lee, and K.-W. Ng, ``{Primordial perturbations from
  ultra-slow-roll single-field inflation with quantum loop effects},''
  \href{http://arxiv.org/abs/2305.16810}{{\ttfamily arXiv:2305.16810
  [astro-ph.CO]}}.

\bibitem{Choudhury:2023hfm}
S.~Choudhury, A.~Karde, S.~Panda, and M.~Sami, ``{Scalar induced gravity waves
  from ultra slow-roll Galileon inflation},''
  \href{http://arxiv.org/abs/2308.09273}{{\ttfamily arXiv:2308.09273
  [astro-ph.CO]}}.

\bibitem{Bhattacharya:2023ysp}
G.~Bhattacharya, S.~Choudhury, K.~Dey, S.~Ghosh, A.~Karde, and N.~S. Mishra,
  ``{Evading no-go for PBH formation and production of SIGWs using Multiple
  Sharp Transitions in EFT of single field inflation},''
  \href{http://arxiv.org/abs/2309.00973}{{\ttfamily arXiv:2309.00973
  [astro-ph.CO]}}.

\bibitem{Choudhury:2023fwk}
S.~Choudhury, K.~Dey, A.~Karde, S.~Panda, and M.~Sami, ``{Primordial
  non-Gaussianity as a saviour for PBH overproduction in SIGWs generated by
  Pulsar Timing Arrays for Galileon inflation},''
  \href{http://arxiv.org/abs/2310.11034}{{\ttfamily arXiv:2310.11034
  [astro-ph.CO]}}.

\bibitem{Choudhury:2023fjs}
S.~Choudhury, K.~Dey, and A.~Karde, ``{Untangling PBH overproduction in
  $w$-SIGWs generated by Pulsar Timing Arrays for MST-EFT of single field
  inflation},'' \href{http://arxiv.org/abs/2311.15065}{{\ttfamily
  arXiv:2311.15065 [astro-ph.CO]}}.

\bibitem{Choudhury:2024one}
S.~Choudhury, A.~Karde, S.~Panda, and M.~Sami, ``{Realisation of the ultra-slow
  roll phase in Galileon inflation and PBH overproduction},''
  \href{http://arxiv.org/abs/2401.10925}{{\ttfamily arXiv:2401.10925
  [astro-ph.CO]}}.

\end{thebibliography}\endgroup

\end{document}